\documentclass{JHEP3}

\usepackage{graphicx}
\usepackage{mathrsfs}
\usepackage{amsmath, amsthm, amssymb}
\newcommand{\be}{\begin{equation}}
\newcommand{\ee}{\end{equation}}
\newcommand{\beq}{\begin{equation}}
\newcommand{\eeq}{\end{equation}}
\newcommand{\bea}{\begin{eqnarray}}
\newcommand{\eea}{\end{eqnarray}}

\def\simgt{\stackrel{>}{{}_\sim}}
\newcommand{\gsim}{\lower.7ex\hbox{$\;\stackrel{\textstyle>}{\sim}\;$}}
\newcommand{\lsim}{\lower.7ex\hbox{$\;\stackrel{\textstyle<}{\sim}\;$}}

\newcommand\etal{{\it {et al.}}}

\newcommand\tanb{tan \beta}

\newcommand\ps{\mbox{ ps}}
\newcommand{\mev}{\mbox{ MeV}}
\newcommand{\gev}{\mbox{ GeV}}

\newcommand{\cl}{\text{CL}}

\newcommand{\msbar}{\overline{MS}} 

\newcommand{\mtpole}{M_t}
\newcommand{\alphas}{\alpha_s(M_Z)^{\overline{MS}}}
\newcommand{\alphaemmz}{\alpha_{\text{em}}(M_Z)^{\overline{MS}}}
\newcommand{\sineff}{\sin^2 \theta_{\rm{eff}}}
\newcommand{\mbmbmsbar}{m_b(m_b)^{\msbar} }

\newcommand{\hl}{h}

\newcommand{\BR}{BR}

\newcommand{\brbsgamma}{\BR(\overline{B}\rightarrow X_s\gamma)}

\newcommand{\delmbs}{\Delta M_{B_s}}

\newcommand\brbsmumu{\BR(\overline{B}_s\to\mu^+\mu^-)}

\newcommand\RBtaunu{\frac{\BR(B_u \to \tau \nu)}{\BR(B_u \to \tau \nu)_{SM}}}
\newcommand\DeltaO{\Delta_{0-}}
\newcommand\RBDtaunuBDenu{\frac{\BR(B \to D \tau \nu)}{\BR(B \to D e \nu)}}
\newcommand\Rl{R_{l23}}

\newcommand\Dstaunu{\BR(D_s \to \tau \nu)}
\newcommand\Dsmunu{\BR(D_s\to \mu \nu)} 

\newcommand{\abundchi}{\Omega_\chi h^2}

\newcommand{\mhl}{m_h}
\newcommand{\zetah}{\zeta_h}

\newcommand{\amusm}{a_{\mu}^{\text{SM}}}
\newcommand{\gmtwo}{(g-2)_{\mu}}

\title{MSSM Forecast for the LHC}

\author{Maria Eugenia Cabrera\\
        Instituto de F\'isica Te\'orica, IFT-UAM/CSIC \\
        U.A.M., Cantoblanco, \\
        28049 Madrid, Spain\\
        E-mail: \email{maria.cabrera@uam.es}}

\author{J. Alberto Casas\\
        Instituto de F\'isica Te\'orica, IFT-UAM/CSIC \\
        U.A.M., Cantoblanco, \\
        28049 Madrid, Spain \\
        E-mail: \email{alberto.casas@uam.es}}

\author{Roberto Ruiz de Austri\\
        Instituto de F\'isica Corpuscular, IFIC-UV/CSIC \\
        Valencia, Spain \\
        E-mail: \email{rruiz@ific.uv.es}}

\abstract{\small
We perform a forecast of the MSSM with universal soft terms (CMSSM) for the LHC, based on an improved Bayesian analysis. We do not incorporate ad hoc measures of the fine-tuning to penalize unnatural possibilities: such penalization arises from the Bayesian analysis itself when the experimental value of $M_Z$ is considered. This allows to scan the whole parameter space, allowing arbitrarily large soft terms. Still the low-energy region is statistically favoured (even before including dark matter or g-2 constraints). Contrary
to other studies, the results are almost unaffected by changing the upper limits taken for the soft terms. The results are also remarkable stable when using flat or logarithmic priors, a fact that arises from the larger statistical weight of the low-energy region in both cases. Then we incorporate all the important experimental constrains to the analysis, obtaining a map of the probability density of the MSSM parameter space, i.e. the forecast of the MSSM. Since not all the experimental information is equally robust, 
we perform separate analyses depending on the group of observables used. When only the most robust ones are used, the favoured region of the parameter space contains a significant portion outside the LHC reach. This effect gets reinforced if the Higgs mass is not close to its present experimental limit and persits when dark matter constraints are included. Only when the g-2 constraint (based on $e^+e^-$ data) is considered, the preferred region (for $\mu>0$) is well inside the LHC scope. We also perform a Bayesian comparison of the positive- and negative-$\mu$ possibilities. 
}

\keywords{Supersymmetric Effective Theories, Beyond Standard Model, Supersymmetry Phenomenology}
\preprint{IFT-UAM/CSIC-09-59}

\begin{document}

\section{Introduction}

The idea of an LHC forecast for the Minimal Supersymmetric Standard Model (MSSM) is to use all the present (theoretical and experimental) information available to determine the relative probability of the different regions of the MSSM parameter space. This includes theoretical constraints (and perhaps prejudices) and experimental constraints, such as electroweak precision tests. For recent work on this subject see refs.
\cite{Allanach:2005kz, Allanach:2006jc, deAustri:2006pe, Allanach:2007qk, Roszkowski:2007fd, Buchmueller:2008qe, Trotta:2008bp, Ellis:2008di, AbdusSalam:2009qd, Buchmueller:2009fn}. An appropriate framework to perform such forecast is 
the Bayesian approach (for a review see ref. \cite{D'Agostini:1995fv}), which allows a sensible statistical analysis and to separate in a neat way the objective and subjective pieces of information. 

The probability density of a particular point in the parameter space, say $\{\theta_i^0\}$, given a certain set of {\em data}, is the so-called posterior probability density function (pdf), $p(\theta_i^0|{\rm data})$, which is given by the fundamental Bayesian relation 
\bea
\label{Bayes}
p(\theta_i^0|{\rm data})\ =\ p({\rm data}|\theta_i^0)\ p(\theta_i^0)\ \frac{1}{p({\rm data})}\ .
\eea
Here $p({\rm data}|\theta_i^0)$ is the likelihood (sometimes denoted by ${\cal L}$), i.e. the probability density of measuring the given data for the chosen point in the parameter space\footnote{Frequentist approaches, which are an alternative to Bayesian ones, are based on the analysis of the likelihood function in the parameter space; see ref.~\cite{Buchmueller:2009fn} for a recent frequentist analysis of the MSSM.}. $p(\theta_i^0)$ is the prior, i.e. the ``theoretical" probability density that we assign a priori to the point in the parameter space. Finally, $p({\rm data})$ is a normalization factor which plays no role unless one wishes to compare different classes of models, so for the moment it can be dropped from the previous formula. 
One can say that in eq.~(\ref{Bayes}) the first factor (the likelihood) is objective, while the second (the prior) contains our prejudices about 
how the probability is distributed a priori in the parameter space, given all our previous knowledge about the model. 

Ignoring the prior factor is not necessarily the most reasonable or ``free of prejudices" attitude. Such procedure amounts to an implicit choice for the prior, namely 
a completely flat prior in the parameters. However, choosing e.g. $\theta_i^2$ as initial parameters 
instead of $\theta_i$, the previous flat prior becomes non-flat. So one needs some theoretical basis to establish, at least, the parameters whose prior can be reasonably taken as flat.

Besides, note that a choice for the allowed ranges of the various parameters is necessary in order to make statistical statements. Often one is interested in showing the probability density of one (or several) of the initial parameters, say $\theta_i,\ i=1,...,N_1$, but not in the others, $\theta_i,\ i=N_1+1,...,N$.
Then, one has to {\em marginalize} the latter, i.e. integrate in the parameter space:
\bea
\label{marg}
p(\theta_i,\ i=1,...,N_1|{\rm data})\ = \int d\theta_{N_1+1},...,d\theta_N\ p(\theta_i,\ i=1,...,N|{\rm data})\ 
 .
\eea
This procedure is very useful and common to make predictions about the values of particularly interesting parameters. Now, in order to perform the marginalization, we need an input for the prior functions {\em and} for the range of allowed values of the parameters, which determines the range of the definite integration (\ref{marg}). A choice for these ingredients is therefore inescapable in trying to make LHC forecasts.

In the present paper we apply these concepts to the study of the MSSM \cite{Martin:1997ns}. More precisely, we will consider a standard framework, often called CMSSM or MSUGRA, in which the soft parameters are assumed universal at a high scale ($M_X$), where the supersymmetry (SUSY) breaking is transmitted to the observable sector, as happens e.g. in the gravity-mediated SUSY breaking scenario. Hence, our parameter-space is defined by the following parameters:
\bea
\label{MSSMparameters}
\{\theta_i\}\ =\ \{m,M,A,B,\mu,s\}
\ .
\eea
Here $m$, $M$ and $A$ are the universal scalar mass, gaugino mass and trilinear scalar coupling; $B$ is the bilinear scalar coupling; $\mu$ is the usual Higgs mass term in the superpotential; and $s$ stands for the SM-like parameters of the MSSM. The latter include the $SU(3)\times SU(2)\times U(1)_Y$ gauge couplings, $g_3,g,g'$, and the Yukawa couplings, which in turn determine the fermion masses and mixing angles.

In sect.~2 we explain the set up for this study. In our opinion we have improved previous analyses in several aspects. Namely, we have not made ad hoc assumptions to penalize fine-tuned regions. A nice consequence is the absence of dependences on the initial ranges for the MSSM parameters. Besides, we have done a rigorous treatment of the nuisance variables (in particular Yukawa couplings) and we have made a satisfactory choice of priors for the initial parameters (actually two different choices to evaluate the dependence on the prior). 
In sect.~3 we compare the relative probability of the high- and low-energy (i.e. accessible to LHC) regions of the MSSM parameter space. We show that, for any reasonable prior, the low-energy region is statistically favoured only after properly incorporating the information about the scale of electroweak breaking. Sections 4 and 5 are devoted to include all the important experimental constrains into the analysis, for positive and negative $\mu$-parameter respectively. In this way, we obtain a map of the probability density of the MSSM parameter space, i.e. the MSSM forecast for the LHC. We distinguish between the most robust experimental data (EW observables, limits on masses of supersymmetric particles, etc.) and more controversial data ($g_\mu-2$) or model-dependent constraints (Dark Matter), performing separate analyses depending on the group of observables used. The comparison between the positive- and negative-$\mu$ cases is done in sect.~5. In sect.~6 we present a summary of the analysis and the main conclusions.

\section{The set up for the scan of the MSSM}

\subsection{Electroweak breaking}

The main motivation of low-energy SUSY is the nice implementation of the electroweak (EW) breaking, where the EW scale (or, equivalently, the $Z$ mass) does not suffer from enormous (quadratic) radiative corrections. Actually, in the MSSM the EW breaking occurs naturally in a substantial part of the parameter space. This success is greatly due to the SUSY radiative contributions to the Higgs potential. Of course, in our analysis, the points of the parameter space that do not have a correct EW breaking are to be discarded, as usual.

It is common lore that the parameters of the MSSM, $\{m,M,A,B,\mu\}$, should not be far from the experimental EW scale in order to avoid unnatural fine-tunings to obtain the correct size of the E.W breaking. There is a rich literature \cite{Ellis:1986yg, Barbieri:1987fn, deCarlos:1993yy, Giusti:1998gz, Strumia:1999fr} about the best way to quantify this fine-tuning. It is normally understood that regions of the parameter space with large fine-tuning (typically at large values of the soft parameters) are to be considered unnatural and thus uninteresting. The exception to this rule are landscape-like scenarios, which we do not consider in detail in this paper. Previous Bayesian studies of the MSSM have attempted to incorporate this criterion by implementing some penalization of the fine-tuned regions, e.g. using a conveniently modified prior for the MSSM parameters \cite{Allanach:2006jc, Giusti:1998gz, Allanach:2007qk}.
Another (more usual) practice has been to restrict the range of the soft parameters to $\lsim $ few TeV, but this makes the results dependent on the actual ranges considered.

However, since the naturalness arguments are deep down statistical arguments, one might expect that an effective penalization of fine-tunings
should arise from the Bayesian analysis itself, with no need of introducing 
``naturalness priors"  or restricting the soft terms to the low-energy scale. It was shown in ref. \cite{Cabrera:2008tj} that this is indeed the case (see also \cite{Strumia:1999fr}
for a previous observation in this sense). Let us briefly recall the argument. The key point is to consider $M_Z^{\rm exp}$ as experimental data on a similar foot to the others, entering the total likelihood, ${\cal L}$. For the sake of simplicity let us approximate the likelihood associated to the $Z$ mass as a Dirac delta, so
\bea
\label{likelihood}
p({\rm data}|s, m, M, A, B, \mu)\ \simeq\ \delta(M_Z-M_Z^{\rm exp})\ {\cal L}_{\rm rest}\ ,
\eea
where ${\cal L}_{\rm rest}$ is the likelihood associated to all the physical observables, except $M_Z$. Now, we can take advantage of this Dirac delta to marginalize the pdf
in one of the initial parameters, e.g. $\mu$, performing a change of variable $\mu\rightarrow M_Z$:
\bea
\label{marg_mu}
%{}\hspace{-1cm}
p(s, m, M, A, B| \ {\rm data} )& = &\int d\mu\ p(s, m, M, A, B, \mu | 
{\rm data} )
\nonumber \\
&=& \ \int dM_Z \left[\frac{d\mu}{d M_Z}\right]p({\rm data}|s, m, M, A, B, \mu)\ 
\nonumber\\ 
&\simeq&\ {\cal L}_{\rm rest}  \left[\frac{d\mu}{d M_Z}\right]_{\mu_Z}
p(s, m, M, A, B, \mu_Z)\ .
\eea
where $\mu_Z$ is the value of $\mu$ that reproduces the experimental value
of $M_Z$ for the given values of $\{s, m, M, A, B\}$, and $p(s, m, M, A, B, \mu)$ is the prior in the initial parameters (still undefined). This marginalized pdf can be written as
\bea
\label{pcmu}
p(s, m, M, A, B| \ {\rm data} )\
=\ 2\ {\cal L}_{\rm rest}\  \frac{\mu_0}{M_Z}\ \frac{1}{c_\mu}\
p(s, m, M, A, B, \mu_0)\ .
\eea
where $c_\mu = \left|\frac{\partial \ln M_Z^2}{\partial \ln \mu}\right|$ is the conventional Barbieri-Giudice measure \cite{Ellis:1986yg, Barbieri:1987fn} of the degree of fine-tuning. Thus, the presence of this fine-tuning parameter in the denominator penalizes the regions of the parameter space with large fine-tuning, as desired. 

As we will see in full detail in sect.~3, this is enough to make the high-scale region of the parameter space, say soft terms $\gsim$ few TeV, statistically insignificant; which allows in turn to consider a wide range for the soft parameters (up to the very $M_X$). In consequence, the results of our analysis are essentially independent on the upper limits of the MSSM parameters, in contrast with previous studies.

\subsection{Treatment of nuisance variables and Yukawa couplings}

It is common in statistical problems that not all the parameters that define the system are of the same interest. In the problem at hand we are interested in determining the probability maps 
for the MSSM parameters that describe the new physics, i.e. $\{m,M,A,B,\mu\}$, but not (or not at the same level) for the SM-like parameters, denoted by $\{s\}$. However, the {\em nuisance parameters} $\{s\}$ play an important role in extracting experimental consequences from the MSSM. The usual technique to eliminate nuisance parameters is simply marginalizing them, i.e. integrating the pdf (\ref{pcmu}) in the $\{s\}$ variables (for a review see ref. \cite{Berger:1999}). When the value of a nuisance parameter is in one-to-one correspondence to a high-quality experimental piece of information (included in ${\cal L}_{\rm rest}$), this integration simply selects the ``experimental" value of the nuisance parameter, which thus becomes (basically) a constant with no further statistical significance in the analysis. Note that in that case the prior on such nuisance parameter becomes irrelevant. In the MSSM, nuisance parameters of this class are the gauge couplings, $\{g_3,g,g'\}$,
%\footnote{Strictly speaking, the initial theoretical inputs are the gauge couplings at high energy, which are related to the experimental (low-energy) ones by the renormalization-group running. This running depends on the other MSSM parameters through the position of thresholds associated with different particles. Hence, two viable MSSM models have slightly different values of the gauge couplings at high energy, and thus the theoretical prior on the couplings would play an (almost insignificant) role in the statistical comparison of the two models.}
which thus can be extracted from the analysis (for an extended discussion see \cite{Cabrera:2008tj}).

In the pure SM a similar argument can be used to eliminate the Yukawa couplings, since they  are in one-to-one correspondence to the quark and lepton masses. However, in the MSSM these masses depend also on the relative value of the two expectation values of the two Higsses, i.e. $\tan \beta\ \equiv\ v_2/v_1$. At tree-level, the up-type-quark, down-type-quark and charged lepton masses go like $m_u \sim y_u v \sin \beta$, $m_d \sim y_d v \cos \beta$ and $m_e \sim y_e v \cos \beta$ respectively; where $v^2=2(v_1^2 + v_2^2) = (246\ {\rm GeV})^2$ is proportional to the $Z$ mass squared. Note that $\tan\beta$ is a derived quantity, which is obtained upon minimization of the scalar potential $V(H_1, H_2)$, and thus takes different values at different points of the MSSM parameter space. 
This means that two viable MSSM models (with the same fermion masses) will have in general very different values of the Yukawa couplings, and thus the theoretical prior, $p(y)$, will play a relevant and non-ignorable role in their relative probability. Any Bayesian analysis of the MSSM amounts to an explicit or implicit assumption about the prior in the Yukawa couplings.

In previous Bayesian analyses of the MSSM the role of the Yukawa couplings was basically ignored: their values were just taken as needed to reproduce the experimental fermion masses, within uncertainties. Even if this procedure can be seen as ``sensible", it is worth wondering which kind of prior $p(y)$ corresponds to. As shown in ref. \cite{Cabrera:2008tj} this can be worked out by marginalizing the Yukawa couplings, using the experimental information about the fermion masses. Let us discuss briefly see how this works.

Just for the sake of the argument, let us approximate the associated likelihood as a product of Dirac deltas
\bea
\label{likelihood_mt}
{\cal L}_{\rm fermion\ masses}\ =\ \delta(m_t-m_t^{\rm exp})\ \delta(m_b-m_b^{\rm exp})\ ....
\eea
and the fermion masses by the tree-level expressions
\bea
\label{mt}
m_t= \frac{1}{\sqrt{2}}y_t^{\rm low} v s_\beta, \;\;
m_b= \frac{1}{\sqrt{2}}y_b^{\rm low} v c_\beta,\;\;\;\;\;
\rm{etc.}
\eea
where $s_\beta\equiv \sin\beta$, $c_\beta\equiv \cos\beta$ and $y_i^{\rm low}$ are the low-energy Yukawa couplings. Suppose further that $y_i^{\rm low} = R_i y_i$, where $y_i$ are the high-energy Yukawa couplings (and thus the input parameters) and the renormalization-group (RG) factor $R_i$ does not depend on $y_i$ itself. Now, it is easy to work out the factor introduced in the pdf when the $y_i$ variables are marginalized:
\bea
\label{marg_y}
&&\hspace{-1cm}\int [dy_t\ dy_b\cdots]\ p(y, m, M, A, B| \ {\rm data} )\
=\ \int [dy_t\ dy_b\cdots]\ p(y) 
\delta(m_t-m_t^{\rm exp})\ \delta(m_b-m_b^{\rm exp})\cdots
\nonumber\\
&&\hspace{3cm}\sim\  p(y)\ \left|\frac{d y_t}{d m_t}\right|\left|\frac{d y_b}{d m_b}\right|\cdots\ 
=\ p(y)\ s_\beta^{-1}\ c_\beta^{-1}\cdots
\eea
where $p(y)$ denotes the prior in the Yukawa couplings (which we assume that factorizes from the other priors). Eq.~(\ref{marg_y}) represents the footprint of the Yukawa couplings in the pdf. Now, taking logarithmically flat priors in the Yukawas, i.e. $p(y_i)\propto 1/y_i$, then the $s_\beta^{-1}\ c_\beta^{-1}\cdots$ factors get cancelled.
This is therefore the prior implicitly assumed in the previous Bayesian analyses, and the one we will adopt in this paper. Remarkably, for independent reasons, we find the logarithmically flat prior for Yukawa couplings a most sensible choice. Certainly there is no convincing origin for the experimental pattern of fermion masses, and thus of Yukawa couplings. However it is a fact that these come in very assorted orders of magnitude (from ${\cal O}(10 ^{-6})$ for the electron to ${\cal O}(1)$ for the top), suggesting that the underlying mechanism may produce Yukawa couplings of different orders with similar efficiency; and this is the meaning of a logarithmic prior.

Of course the above discussion is oversimplified. The physical (pole) masses include radiative corrections. Besides, the RG factor $R_i$ for the top Yukawa coupling has an important dependence on the Yukawa itself. These subtleties have been incorporated to the full analysis.

\subsection{Variables for the MSSM scan}

Although our initial set of variables is $\{m,M,A,B,\mu,s\}$, and this is the one on which we have to set our theoretical prior, for the purposes of scanning the MSSM parameter space it is much more convenient to trade some of them by other parameters with more direct phenomenological significance. We have already seen that it is worth to trade $\mu$ by $M_Z$, which is automatically integrated out. Similarly, we have seen that the Yukawa couplings are nuisance variables that are profitably traded by the physical fermion masses and easily integrated out. Besides all this, it is highly advantageous to trade the initial $B-$parameter by the derived $\tan\beta$ parameter. The main reason is that for a given viable choice of $\{m,M,A,\tan\beta\}$, there are exactly two values of $\mu$ (with opposite sign and the same absolute value at low energy) leading to the correct value of $M_Z$. Thus working in one of the two (positive and negative) branches of $\mu$, each
point in the $\{m,M,A,\tan\beta\}$ space corresponds exactly to one model, whereas a point
in the $\{m,M,A,B\}$ space may correspond to several models, introducing a conceptual and technical complication in the analysis. 

Consequently we should compute the whole Jacobian, $J$, of the transformation
\bea
\label{change_3}
\{\mu,y_t,B\}\ \rightarrow\  \{M_Z,m_t,t\},\;\;\;\;\;\;t\equiv\tan\beta\ .
\eea
Then the {\em effective prior} in the new variables becomes
\bea
\label{eff_prior}
p_{\rm eff}(g_i, m_t, m, M, A, \tan\beta)\  \equiv\ 
J|_{\mu=\mu_Z}\  p(g_i, y_t, m, M, A, B, \mu=\mu_Z)\
\eea
where we have already marginalized $M_Z$ using the associated likelihood $\sim \delta(M_Z-M_Z^{\rm exp})$ (recall that $\mu_Z$ is the value of $\mu$ that reproduces the experimental $M_Z$.) In eqs.(\ref{change_3}, \ref{eff_prior}) we have made explicit the dependence just on the top Yukawa coupling and mass, but for other fermions goes the same. 

So, to prepare the scan in the new variables we need an explicit evaluation of the Jacobian factor. For that we must know the dependence of the old variables on the new ones. This dependence is extracted from the minimization equations of the Higgs scalar potential, $V(H_1, H_2)$ (which connect $\{\mu, B\}$ with $\{M_Z, \tan\beta\}$), and from the relation between the yukawa couplings and fermion masses. This dependences, even when radiative corrections are included, have the form
\bea
\label{muyB}
\mu=f(M_Z,y,t),\;\;\; y=g(M_Z,m_t,t),\;\;\; B=h(\mu,y,t)\ ,
\eea
where $f$, $g$, $h$ are well defined functions, for which we give approximate analytical expressions below. Here we have made explicit only the dependence on the variables involved in the change of variables (\ref{change_3}). Note that $y$ depends on $M_Z$ since $v\propto M_Z$. In consequence
\bea
\label{J}
J =  \frac{\partial f}{\partial M_Z}\ \frac{\partial g}{\partial m_t}
\ \frac{\partial h}{\partial t} \ \ .
\eea
where the factor $\partial f/\partial M_Z$ carries essentially the fine-tuning penalization discussed in subsect.~2.1. 

For the numerical analysis we have evaluated $J$ using the \texttt{SoftSusy} code \cite{softsusy} which implements the full one-loop contributions and leading two-loop terms to the tadpoles for the electroweak breaking conditions with parameters running at two-loops. This essentially corresponds to the next-to-leading log approximation. 

However, it is possible to give an analytical and quite accurate expression of $J$ by working with the tree-level potential with parameters running at one-loop (i.e. essentially the leading log approximation). Then
\bea
\label{mu}
\mu_{\rm low}^2= \frac{m_{H_1}^2 - m_{H_2}^2t^2}{t^2-1} - \frac{M_Z^2}{2}
\eea
\bea
\label{B}
B_{\rm low}= \frac{s_{2\beta}}{2\mu_{\rm low}}(m_{H_1}^2 + m_{H_2}^2+2\mu_{\rm low}^2)
\eea
\bea
\label{y}
y_{\rm low}=\frac{m_t}{v\ s_\beta}\ .
\eea
Here the ``low'' subscript indicates that the quantity is evaluated at low scale (more precisely, at a representative supersymmetric mass, such as the geometric average of the stop masses). The soft masses $m_{H_i}^2$ are also understood at low scale. For notational simplicity, we have dropped the subscript $t$ from the Yukawa coupling. 
Note that all these low-energy quantities contain an implicit dependence on the top Yukawa coupling through the corresponding RG equations. Besides,
\bea
\label{muBLH}
\mu_{\rm low}= R_\mu(y)\mu,\;\;\;\; B_{\rm low}=B + \Delta_{RG}B(y),\;\;\;\; y_{\rm low}\simeq \frac{y E(Q_{\rm low})}{1+6yF(Q_{\rm low})}\ ,
\eea
where $R_\mu(y), \Delta_{RG}B(y)$ are definite functions of $y$; $Q$ is the renormalization scale, $F = \int_{Q_{\rm high}}^{Q_{\rm low}} E \ln Q$, and $E(Q)$ is a definite function that depends just on the gauge couplings \cite{Ibanez:1983di}. Plugging
eqs.(\ref{mu}--\ref{muBLH}) into eqs.(\ref{muyB}, \ref{J})
it is straightforward to get an explicit expression for $J$. Plugging the latter back into eq.~(\ref{eff_prior}) we get an approximate form for the effective prior
\bea
\label{approx_eff_prior}
p_{\rm eff}(m_t, m, M, A, \tan\beta)\ \ \propto \   \left[\frac{E}{R_\mu^2}\right]\ 
\frac{y}{y_{\rm low}} \frac{t^2-1}{t(1+t^2)} 
\frac{B_{\rm low}}{\mu_Z}  
p(m, M, A, B, \mu=\mu_Z)\ 
,
\eea
where we have taken a logarithmically flat prior for the Yukawa couplings (i.e. $p(y_i)\propto y_i^{-1}$), as discussed above. 
This is the prior to be used when the MSSM parameter space is scanned in the usual variables $\{m, M, A, \tan\beta)\}$ and $\mu$ is taken as required to reproduce the correct EW breaking (the information about $M_Z^{\rm exp}$ is thus automatically incorporated). Let us stress that its form stems just from the relation between the initial variables and the phenomenological ones, indicated in eq.(\ref{change_3}), and it is not ``subjective"
at all. Besides, the prefactor in the r.h.s. of eq.(\ref{approx_eff_prior}) (which is essentially the Jacobian) is valid for any MSSM, not just the CMSSM. The subjectivity lies in the $p(m, M, A, B, \mu)$ piece, i.e. the prior in the initial parameters, for which we have still to make a choice. 
Furthermore, the prefactor in eq.(\ref{approx_eff_prior}) contains the above-discussed penalization of fine-tuned regions, something that may be not so obvious, but that will become clear in sect.~3. Finally, the form of the prefactor implies an effective penalization of large $\tan\beta$, reflecting the smaller statistical weight of this possibility. Actually, the implicit fine-tuning associated to a large $\tan\beta$ was already noted in ref.~\cite{Hall:1993gn, Nelson:1993vc}, where it was estimated to be of order $1/\tan\beta$, in agreement with eq.(2.15). This is logical. From eq.(2.12) we see that
\bea
\label{rango_MS}
\frac{1}{\tan\beta}=\frac{\mu_{\rm low} B_{\rm low}}{m_{H_1}^2+m_{H_2}^2+2\mu_{\rm low}^2}
\eea
The denominator of this expression has the size of the typical soft terms (which we will call $M_S$). Therefore a large $\tan\beta$ requires abnormally small $\mu_{\rm low} B_{\rm low}$. As a matter of fact, $\mu$ cannot be very small, otherwise the mass of the lightest chargino would be below the experimental limit. Therefore $\tan\beta$ requires very small $B_{\rm low}$. But this cannot be naturally arranged since the radiative contributions to $B$ (i.e. its RG evolution from high to low scale) are sizeable (of order $M_S$) \cite{Martin:1997ns}. Thus small $B_{\rm low}$ requires a tuning between its initial (high scale) value and the radiative corrections.

\subsection{Priors in the initial parameters}

The choice of the prior in the initial parameters, $\{m, M, A, B, \mu\}$  must reflect our knowledge about them, before consideration of the experimental data (to be included in the likelihood piece). In our case, we have already made some non-trivial, though quite reasonable, assumptions about them, namely the hypothesis of universality of the soft terms (which is supported by the strong constraints from FCNC processes) at a very high scale (this restricts the analysis to scenarios where the transfer of SUSY breaking is suppressed by a high scale, as happens e.g. in models with gravity-mediated SUSY breaking). 

To go further we must consider the dynamical origin of the parameters. Four of them, $\{m, M, A, B\}$, are soft SUSY-breaking parameters. They typically go like $\sim F/\Lambda$, where $F$ is the SUSY breaking scale, which corresponds to the dominant VEV among the auxiliary fields in the SUSY breaking sector (it can be an $F-$term or a $D-$term) and $\Lambda$ is the messenger scale, associated to the interactions that transmit the breaking to the observable sector. Since the soft-breaking terms share a common origin it is logical to assume that their sizes are also similar. Of course, there are several contributions to a particular soft term, which depend on the details of the superpotential, the K\"ahler potential and the gauge kinetic function of the complete theory (see e.g. ref. \cite{Kaplunovsky:1993rd}). So, it is reasonable to assume that a particular soft term can get any value (with essentially flat probability) of the order of the typical size of the soft terms or below it. There are special cases, like split SUSY scenarios, where the soft terms can be classified in two groups that feel differently the breaking of SUSY. In
  those instances, the priors should also be considered in two separate groups. But those cases are out of the scope of the present analysis, which is focussed on the simplest, most conventional and less baroque framework, which consists of a common SUSY breaking origin and transmission for all the soft terms. The $\mu-$parameter is not a soft term, but a parameter of the superpotential. However, it is desirable that its size is related (e.g. through the Giudice-Masiero mechanism \cite{Giudice:1988yz}) to the SUSY breaking scale. Otherwise, one has to face the so-called $\mu-$problem, i.e. why should be the size of $\mu$ similar to the soft terms', as is required for a correct electroweak breaking (see eq.(\ref{mu})). Thus, concerning the prior, we can consider $\mu$ on a similar foot to the other soft terms.

Now, we are going to make the previous discussion more quantitative. Let us call $M_S$ the typical size of the soft terms in the observable sector, $M_S\sim F/\Lambda$. Then, we define the ranges of variation of the initial parameters as
\bea
\label{rangos_mMABmu}
-qM_S \leq &B&\leq qM_S
\nonumber\\
-qM_S \leq &A&\leq qM_S
\nonumber\\
0 \leq  &m&\leq qM_S
\nonumber\\
0 \leq &M&\leq qM_S
\nonumber\\
0 \leq &\mu&\leq qM_S
\eea
where $q$ is an ${\cal O}(1)$ factor. We have considered here the branch of positive $\mu$. For the negative one we simply replace $\mu\rightarrow -\mu$.
We have taken the same $q$ for all the parameters, since we find no reason to make distinctions among them. Note that we can take $q=1$ with no loss of generality, provided $M_S$ is allowed to vary in the range $0 \leq M_S\leq \infty$.
In practice, to avoid divergences in the priors, we have to take a finite range for $M_S$, say
\bea
\label{rango_MS}
M_S^0 \leq M_S\leq M_X\,, \;\;\;\; M_S^0 \sim 10\; {\rm GeV}
\eea
Nevertheless, the values of the upper and lower limits of the $M_S$ range are going to be irrelevant, as it will become clear soon. Consequently, we can still take $q=1$.

We have discussed the ranges of the parameters, but not the shape of the priors. As already stated, we find reasonable to assume (conveniently normalized) flat priors for the soft parameters inside the ranges (\ref{rangos_mMABmu}), i.e. 
\bea
\label{init_priors}
p(m)=p(M)=p(\mu)=\frac{1}{M_S}\;,\;\;\;\;p(A)=p(B)=\frac{1}{2M_S}
\eea
Still we have to decide what is the prior in $M_S$, and it is at this point where we have to take the decision of assuming a flat or logarithmic prior in the scale of SUSY breaking. We have considered the two possibilities throughout the paper. The comparison of the results from both choices will give us a measure of the prior-dependence of the analysis.

\vspace{0.2cm}
\noindent {\bf Logarithmic prior}
\vspace{0.2cm}

Let us start assuming a logarithmic prior in $M_S$, which we consider the most reasonable option, since it amounts to consider all the possible orders of magnitude of the SUSY breaking in the observable sector on the same foot (this occurs e.g. in conventional SUSY breaking by gaugino condensation in a hidden sector). Then, 
\bea
\label{prior_MS_log}
p(M_S) = N_{M_S}\frac{1}{M_S}\ ,
\eea
where $N_{M_S}$ is a normalization constant, which turns out to be completely irrelevant. Now, we can marginalize $M_S$, which thus disappears completely from the subsequent analysis, leaving a prior which depends just on the $\{m, M, A, B, \mu\}$ parameters\footnote{This procedure is a ``hierarchical Bayesian technique", first used in ref. \cite{Allanach:2007qk}, but using complicated functions that were not possible to integrate analytically.}:
\bea
\label{MS_marg}
p(m, M, A, B,\mu) &=& \frac{N_{M_S}}{4}\int^{M_X}_{\max\{m, M, |A|, |B|,\mu, M_S^0\}} 
\frac{1}{M_S^6}\ dM_S
\nonumber\\
&=& \frac{N_{M_S}}{20}\ \left[\frac{1}{[\max\{m, M, |A|, |B|,\mu, M_S^0\}]^5} - \frac{1}{M_X^5}\right]
\nonumber\\
&\simeq& \frac{N_{M_S}}{20}\ \frac{1}{[\max\{m, M, |A|, |B|,\mu, M_S^0\}]^5}
\eea
Of course, the prefactor is just an irrelevant normalization constant. Note that we have neglected the ${1}/{M_X^5}$ term, which simply forced the prior to strictly vanish in the $M_X$ limit. This effect is only appreciable when one of the parameters is close to $M_X$, otherwise it is completely negligible (note the fifth power in the denominators). 
On the other hand, as mentioned in subsect. 2.1 and will become clear soon, once the EW breaking is incorporated to the analysis, regions of very large initial parameters become irrelevant. In consequence, eq.(\ref{MS_marg}) is an excellent approximation. Note that the value of $M_X$ disappears from the analysis. The value of the lower limit on $M_S$, i.e. $M_S^0$, is also irrelevant. Note that its presence in the denominator of eq.(\ref{MS_marg}) avoids the prior to diverge when the parameters are very small. This ``regulating" effect is only felt when {\em all} the parameters are below $M_S^0$. However, we know that this region will be killed by the experimental data once they are taken into account (through the likelihood piece in the pdf). E.g. the upper bounds on chargino masses require $|\mu|\gsim 100$ GeV. Hence, the value of $M_S^0$ plays no relevant role, apart from the formal regularization of the prior. Let us recall that the above prior (\ref{MS_marg}) is the one to be plugged in eq.(\ref{eff_prior}) (or in the approximated expression (\ref{approx_eff_prior})) to get the effective prior in the scan parameters.

It is funny to compare the prior of eq.(\ref{MS_marg}) with a ``more conventional" logarithmic prior, i.e. $p(m, M, A, B,\mu)\propto 1/(m, M, A, B,\mu)$. First of all, the ``more conventional" prior is not regulated unless one imposes that the parameters should not go below some low-scale (or that the prior does not behave logarithmically flat in that region). But then the results are sensitive to the cut-off scale chosen.
Note that the prior for phenomenologically viable points, with e.g. very small $A$ and large $\mu$ (thus avoiding the constraints from chargino masses), will depend on the precise treatment of this region. Apart from this annoyance, the conventional logarithmic prior treats the parameters as uncorrelated objects. This produces non-realistic distortions. E.g. a point of the parameter space where some parameters are very large, but the others are very small, can have a value of the prior (i.e. an assigned probability) larger than another point where all the parameters are ${\cal O}({\rm TeV})$. However, this goes against the expectative that all the initial parameters are likely to have similar sizes, as they share a common dynamical origin. In other words, it is not sensible to increase the prior probability (in a very significant amount) just because one of the parameters is abnormally small, compared to the others. These problems are nicely avoided by the simple prior (\ref{MS_marg}), reflecting the way it has been constructed. 

To finish this discussion, let us note that the prior (\ref{MS_marg}) does not have the form of a product of individual priors defined for each parameter. Still, we can get the form of the prior for just one parameter, marginalizing the others before including any experimental information. For instance, the prior in the gaugino mass, $M$, is obtained by marginalizing in  $m, A, B,\mu$, leading to
\bea
\label{M_marg}
{\cal P}(M) \propto \frac{1}{\max\{M, M_S^0\}} 
\eea
where we have neglected $\sim 1/M_X$ contributions. This has indeed the form of a logarithmically flat prior. Of course, similar individual priors are obtained for the other parameters.

\vspace{0.2cm}
\noindent {\bf Flat prior}
\vspace{0.2cm}

We can now repeat the previous analysis, assuming a flat prior for $M_S$, which amounts to consider all the values of the SUSY breaking in the observable sector on the same foot.
Hence we maintain the ranges for the parameters, eqs.~(\ref{rangos_mMABmu}), (\ref{rango_MS}), and the flat priors inside those ranges, eq.~(\ref{init_priors}). We just replace the logarithmically flat prior in $M_S$, eq.~(\ref{prior_MS_log}), by a flat prior
\bea
\label{prior_MS_flat}
p(M_S) = N_{M_S}
\eea
where $N_{M_S}$ is an irrelevant normalization constant $\sim 1/M_X$. Again, to obtain the prior in the $\{m,M,A,B,\mu\}$ variables, we marginalize in $M_S$. The previous result
(\ref{MS_marg}) becomes now

\bea
\label{MS_marg_2}
p(m, M, A, B,\mu) \propto \ \frac{1}{[\max\{m, M, |A|, |B|,\mu, M_S^0\}]^4}
\eea
where, once more (and for the same reasons) we have neglected a $1/M_X^4$ contribution. 
The difference with eq.(\ref{MS_marg}) is that now we have one power less in the denominator. Again, the prior (\ref{MS_marg_2}) is the one to be plugged back into eq.(\ref{eff_prior}) in order to get the effective prior in the scan parameters.

We can also repeat the exercise of obtaining the prior for an individual parameter, say $M$, by marginalizing the others. In this case, the previous equation (\ref{M_marg}) becomes
\bea
\label{M_marg_2}
{\cal P}(M) \sim \ln \frac{M_X}{\max\{M, M_S^0\}} \ ,
\eea
In essence this is a flat prior in $M$, as it does not change much along orders of magnitude. E.g. in the $100\ {\rm GeV}\leq M\leq 4\ {\rm TeV}$ range it just changes a 13\%. Again, the other parameters go in a similar way.

\section{High-energy vs Low-energy regions and the EW breaking}

At first sight it may seem that the assumption of a logarithmic prior, see eqs.(\ref{MS_marg}, \ref{M_marg}), amounts  to a strong preference for the low-energy region of the parameter space, i.e. for $\{m,M,A,B,\mu\}$ not far from the EW scale. However, this is not true. We may ask the following question: What is a priori the relative probability that a parameter, say $M$, 
lies in the low-energy (accessible to the LHC) region, $100\ {\rm Gev} \lsim M\lsim 2\ {\rm TeV}$ versus the chance that it lies at a higher scale, $2\ {\rm TeV} \lsim M\lsim M_X$.
Using eq.(\ref{M_marg}), it is clear that this relative probability is 
\bea
\label{comp_zonas}
\frac{{\cal P}(100\ {\rm GeV}\leq M\leq 2\ {\rm TeV})}
{{\cal P}(2{\rm TeV}\leq M\leq M_X)}\simeq \frac{1}{12}
\eea
(in an obvious notation). I.e. in the initial set-up the most probable situation is that SUSY escapes LHC detection, even with logarithmic prior. Note that this is not so if one cuts-off the ranges of the parameters at a few TeV, as is very usually done, but we allow them to vary all the way up to $M_X$. Of course, the situation is much more dramatic for a flat prior. Using eq.(\ref{M_marg_2}) we see that in that case
\bea
\label{comp_zonas_2}
\frac{{\cal P}(100\ {\rm GeV}\leq M\leq 2\ {\rm TeV})}
{{\cal P}(2{\rm TeV}\leq M\leq M_X)}\simeq 3\times 10^{-12}\ .
\eea
Hence, the flat-prior set-up assigns a negligible initial probability to LHC detection.

Fortunately things change for better as soon as we incorporate the experimental information about the size of the EW breaking, i.e. $M_Z^{\rm exp}$. We have already discussed in subsect. 2.1 how $M_Z^{\rm exp}$ can be used to marginalize $\mu$, leaving a footprint of fine-tuning penalization. Now we are going to be more precise. We will take the effective prior in the scan variables, given by eq.(\ref{eff_prior}) or by the approximate expression (\ref{approx_eff_prior}). Recall that these expressions already incorporate the experimental information on $M_Z$ and the marginalization of $\mu$. 

Now, we can evaluate once more the relative probability between the low- and high-energy regions (say for the $M$ parameter again), but with this effective prior, i.e. incorporating the information about the EW breaking scale. For the sake of clarity we present now an analytical discussion, with some approximations, that gives correctly the essential results and allows to show the physical reasons behind them. At the end we will present the numerical results.

Hence, we have to marginalize the $\{m,A,\tan\beta\}$ parameters, since the $\mu$-parameter has already been marginalized. Let us first perform the integration in $\{m,A\}$. Note that for a given value $M=M_1$, and $\tan\beta$ fixed, only a portion of the $\{m,A\}$ plane will be able to accommodate, by adjusting the $\mu$-parameter, the required EW breaking. Let us call this region $R_1$. Therefore the integration is only extended to
$R_1$
\be
\label{marg_Am}
{\cal P}(M_1,\tan\beta)\sim\int_{R_1}dm\ dA\ p_{\rm eff}(m_t, m, M, A, \tan\beta)
\ee
where $p_{\rm eff}(m_t, m, M, A, \tan\beta)$ is given by eq.(\ref{eff_prior}) or eq.(\ref{approx_eff_prior}).
We have to compare this probability with the one for a different gaugino mass, say
$M_2>M_1$ (we keep $\tan\beta$ fixed). The expression for ${\cal P}(M_2,\tan\beta)$ is completely analogous to eq.(\ref{marg_Am}). The only subtle point is what is the new allowed region, $R_2$. An approximate way to determine it is the following. In almost all the parameter space the squared SUSY-parameters, $\{m^2, M^2, A^2, B^2, \mu^2\}$ are much larger than $M_Z^2$. Therefore, the combination of them producing the correct value of $M_Z$ is almost identical to the one producing $M_Z=0$. So we can approximate $R_1$ and $R_2$ as the regions giving $M_Z=0$. Now, if we neglect for a moment RG effects, it is clear that for each point $\{m_1, A_1\}\in R_1$ there is another point $\{m_2, A_2\}\in R_2$, producing the same breaking, given by $m_2=\frac{M_2}{M_1}m_1$, $A_2=\frac{M_2}{M_1}A_1$, $\mu_2=\frac{M_2}{M_1}\mu_1$, $B_2=\frac{M_2}{M_1}B_1$. In other words, $R_2\sim \frac{M_2}{M_1}R_1$. RG effects do not in principle modify this relation since they are proportional to the very soft terms. However, there is a residual effect: since the running goes from $M_X$ (where the SUSY parameters are defined) until the scale where the EW breaking is evaluated ($\sim$ stop masses), there is logarithmic correction, $\propto\log (M_2/M_1)$, which would slightly modify the shape of $R_2$. On the other hand, there will be points in $R_2$ that will go out of the allowed ranges of the parameters. I.e., $R_2$ will be slightly smaller than $\frac{M_2}{M_1}R_1$. This means that we are slightly overestimating the weight of the high-scale parameter space, which is a conservative attitude for this discussion. 
Now we can express the integration in the $R_2$ region in terms of the $R_1$ one,
\bea
\label{marg_Am_2}
{\cal P}(M_2,\tan\beta)&\sim&
\int_{R_2} dm_2\ dA_2\ 
p_{\rm eff}(m_2, M_2, A_2, \mu_2, \tan\beta)\ 
\nonumber\\
&=& 
\ \int_{R_1} dm_1\ dA_1\ \left(\frac{M_1}{M_2}\right)^3\ 
p_{\rm eff}(m_1, M_1, A_1, \mu_1, \tan\beta)\ 
\nonumber\\
&=& 
\ \left(\frac{M_1}{M_2}\right)^3\ 
{\cal P}(M_1,\tan\beta)\ ,
\eea
Here have used the fact that, when we assume a logarithmic initial prior, the effective prior, $p_{\rm eff}$, scales as
\bea
\label{escal_peff}
p_{\rm eff}(m_2, M_2, A_2, \mu_2, \tan\beta)\ =\ \left(\frac{M_1}{M_2}\right)^5\ 
p_{\rm eff}(m_1, M_1, A_1, \mu_1, \tan\beta)\ . 
\eea
This can be easily noticed from the approximate expression of $p_{\rm eff}$ in 
eq.(\ref{approx_eff_prior}). This relation is exact, essentially, up to small RG-effects in the $\frac{B_{\rm low}}{\mu_Z}$ factor involved in (\ref{approx_eff_prior}). 

The last step is to marginalize $\tan \beta$. But this will not affect the relative probability we are interested in, since the factor obtained by integrating $\tan\beta$ is identical for $M_1$ and for $M_2$. Alternatively, we can leave $\tan \beta$ fixed at some arbitrary value.
The important point is that the relative probability goes like
\bea
\label{escal_pM}
\frac{{\cal P}(M_2,\tan\beta)}{{\cal P}(M_1,\tan\beta)}\ \sim\ \frac{M_1^3}{M_2^3}\ .
\eea
In other words, 
\bea
\label{escal_pM_2}
{\cal P}(M,\tan\beta)\ \sim\ \frac{1}{M^3}
\eea
This should be compared with the ${\cal P}(M)\ \sim\ \frac{1}{M}$ behaviour obtained in eq.(\ref{M_marg}), when the experimental $M_Z$ was not taken into account. We see that the pdf in $M$ has gained two powers of $M$ in the denominator. This is the fine-tuning penalization that arises on its own from the Bayesian analysis. Now the relative probability of the low-energy (accessible to the LHC) region versus the probability of a higher scale becomes
\bea
\label{comp_zonas_Z}
\frac{{\cal P}(100\ {\rm GeV}\leq M\leq 2\ {\rm TeV})}
{{\cal P}(2{\rm TeV}\leq M\leq M_X)}\simeq \left(\frac{2\ {\rm TeV}}{100\ {\rm GeV}}\right)^2\sim 10^2
\eea
to be compared with eq.(\ref{comp_zonas}) before including the EW breaking in the analysis. In conclusion, once the EW breaking is correctly incorporated in the Bayesian analysis (but not before!), 99\% of the probability lives in the low-energy (LHC-relevant) region of the parameter space. Note that this is achieved without invoking other kinds of constraints (like Dark Matter or g-2 constraints) that are often used to set the scale of the soft terms not far from the EW scale. Hence, the main reason to believe that SUSY should be accessible at LHC scales comes from the EW breaking itself, i.e. the original motivation for phenomenological SUSY. We find this result very satisfactory. Needless to say that for the other parameters the results go in a completely analogous way.

We have discussed the impact of the EW breaking scale when a logarithmic prior is used. For a flat prior, it is straightforward to repeat the above discussion, taking into account that the prior $p(m,M,A,B,\mu)$ [and thus $p_{\rm eff}(m, M, A, \mu, \tan\beta)$] has one power of mass less in the denominator.
Therefore at the end of the day we arrive at
\bea
\label{escal_pM_2}
{\cal P}(M,\tan\beta)\ \sim\ \frac{1}{M^2}
\eea
to be compared with the almost flat behaviour before including the experimental EW information, see eq.(\ref{MS_marg_2}). Consequently, the relative probability of the low-energy and high-energy regions of the parameter space, for a flat prior, becomes
\bea
\label{comp_zonas_Z_2}
\frac{{\cal P}(100\ {\rm GeV}\leq M\leq 2\ {\rm TeV})}
{{\cal P}(2{\rm TeV}\leq M\leq M_X)}\simeq \frac{2\ {\rm TeV}}{100\ {\rm GeV}} \simeq 10
\eea
So, even assuming a flat prior for the typical size of the soft breaking terms, up to the $M_X$ scale, we see that the EW breaking is sufficient to put 90\% of the total probability in the LHC-interesting region. This contrasts strongly with previous analysis, and, again, we consider it very satisfactory.

We have checked the previous arguments by performing the analysis in a numerical 
way. For the posterior samples we adopt the MultiNest \cite{Feroz:2007kg} 
algorithm as implemented in the \texttt{SuperBayeS} code \cite{superbayes}. It 
is based on the framework of Nested Sampling, recently invented by 
Skilling \cite{SkillingNS,Skilling:2006}. MultiNest has been developed
in such a way as to be an extremely efficient sampler even for
likelihood functions defined over a parameter space of large
dimensionality with a very complex structure as it is the case of the 
CMSSM. The main purpose of the Multinest is the computation of the  
Bayesian evidence and its uncertainty but it produces posterior inferences 
as a by--product. For the marginalization procedure we have used the above-discussed ranges
for our priors, i.e. from $0$ to $M_X$ for 
$m$, $M$ and $|A|$. Besides, we have used $2 < \tan\beta < 62$. The lower limit comes from present bound on the Higgs mass \cite{Schael:2006cr}. The upper one comes from imposing 
Yukawa couplings in the perturbative regime \cite{Casas:1998vh}. The precise value $\tan\beta< 62$ has been chosen to allow an strict comparison with previous analyses in refs.~\cite{deAustri:2006pe, Allanach:2007qk}
However, the precise value of this upper bound turns out to be irrelevant, as the region of very large $\tan\beta$ is strongly suppressed (see the discussion after eq.~(\ref{approx_eff_prior}) and ref.~\cite{Ellis:2001msa}).

In Fig. 1 the red line shows the prior in $M$ (upper panels) and $m$ (lower panels), when the other parameters are marginalized, using logarithmic (left panels) or flat (right panels) initial prior for the scale of SUSY breaking in the observable sector ($M_S$). The blue bar distributions show the pdf once the EW breaking is incorporated in the analysis, i.e. the effective prior in the scan variables, $p_{\rm eff}$, see eq.(\ref{eff_prior}) and the approximate form (\ref{approx_eff_prior}). The logarithmic scale in the horizontal axes allows to see that most of the probability, which initially lies in the high-energy region ($M, m$ above the TeV scale), flows dramatically into the low-energy region once the EW breaking is considered. 
Actually, most of the probability falls inside the LHC discovery reach (even with just 1 fb$^{-1}$ \cite{Ball:2007zza, Baer:2009dn}).
Quantitatively, the results are in good agreement with the previous discussion. Although the distribution of probability above 1 TeV is almost invisible in the plots (especially for log priors), it is actually different from zero and follows from the approximate law of eq.(\ref{escal_pM_2}). Notice also that, at this stage, all the points are equally ``best-fit points" (even at extremely large $M,m$), since they are equally in reproducing $M_Z$, the only experimental information so far considered. 

\begin{figure}[t]
\begin{center}
\label{}
\includegraphics[angle=0,width=0.4\linewidth]{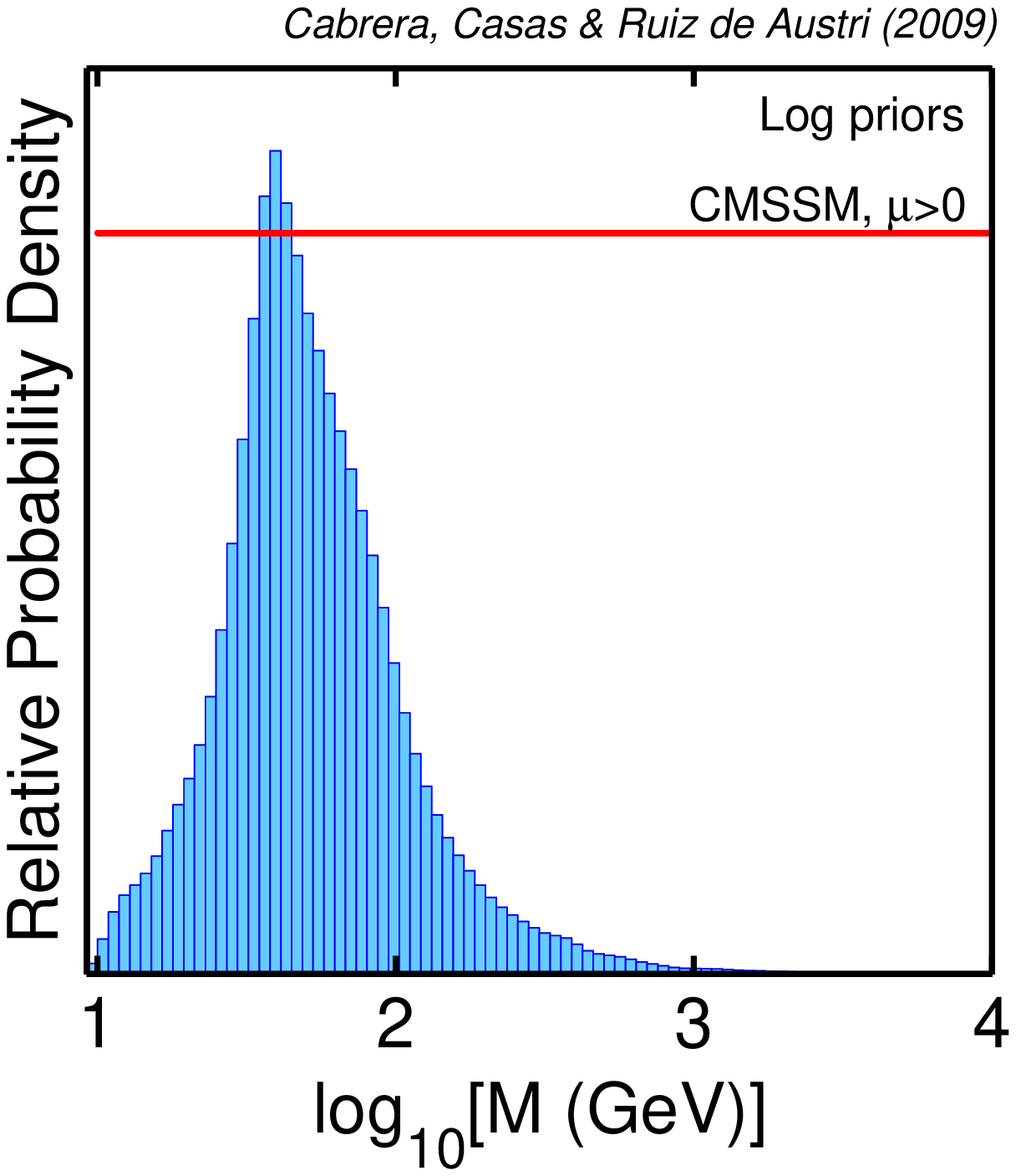} \hspace{1.2cm}
\includegraphics[angle=0,width=0.4\linewidth]{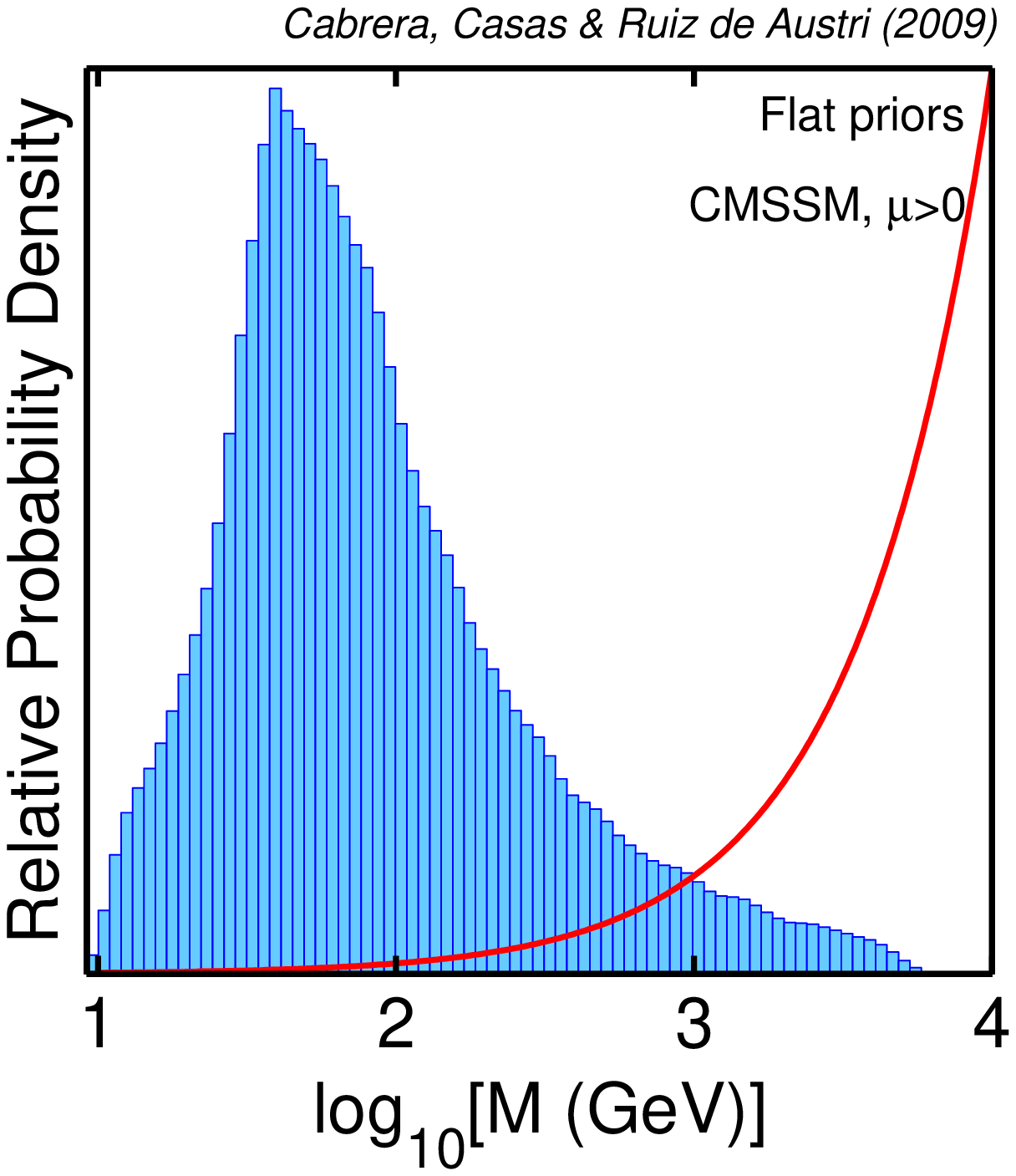}  \\ 
\vspace{1.0cm} 
\includegraphics[angle=0,width=0.4\linewidth]{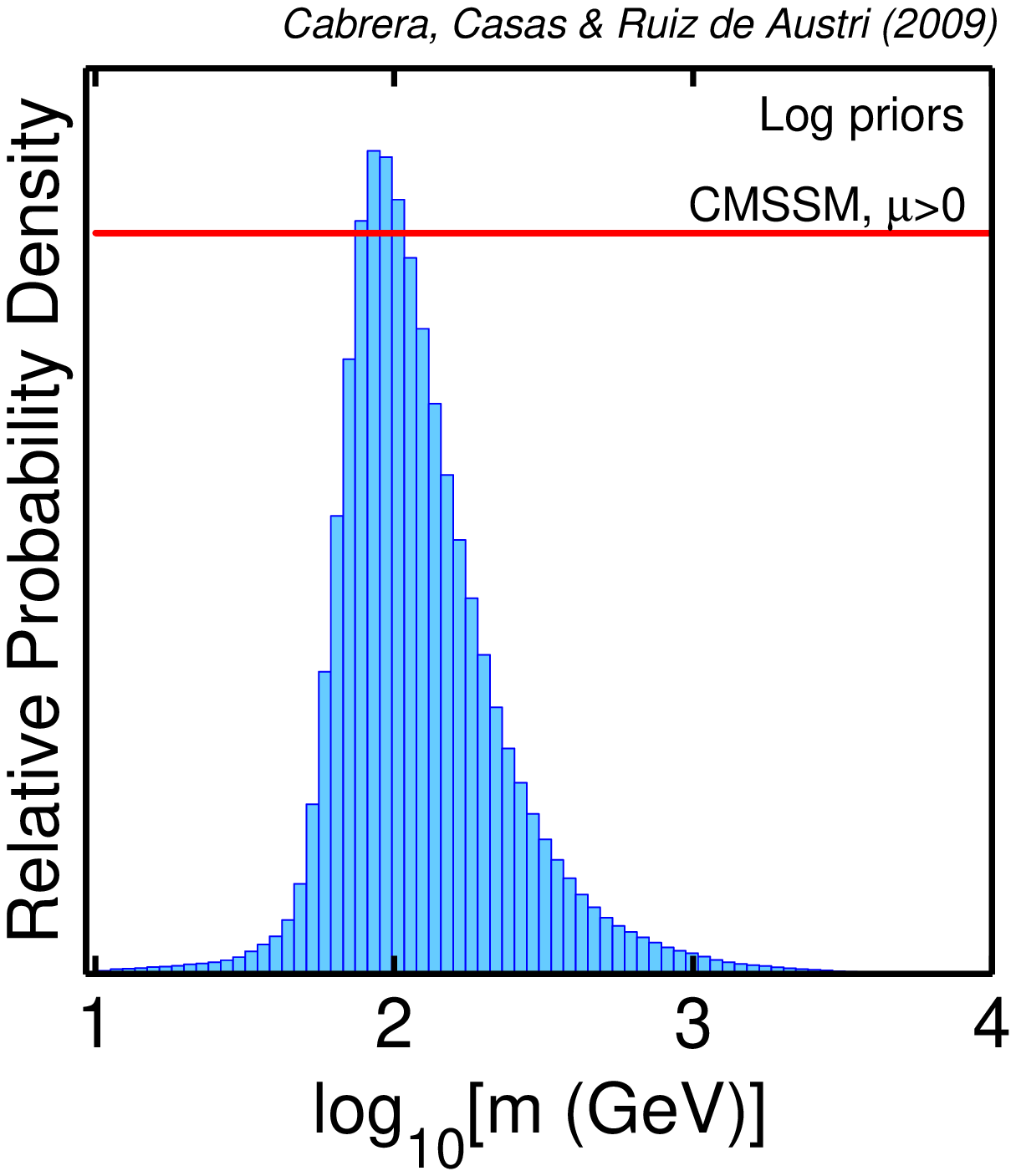} \hspace{1.2cm}
\includegraphics[angle=0,width=0.4\linewidth]{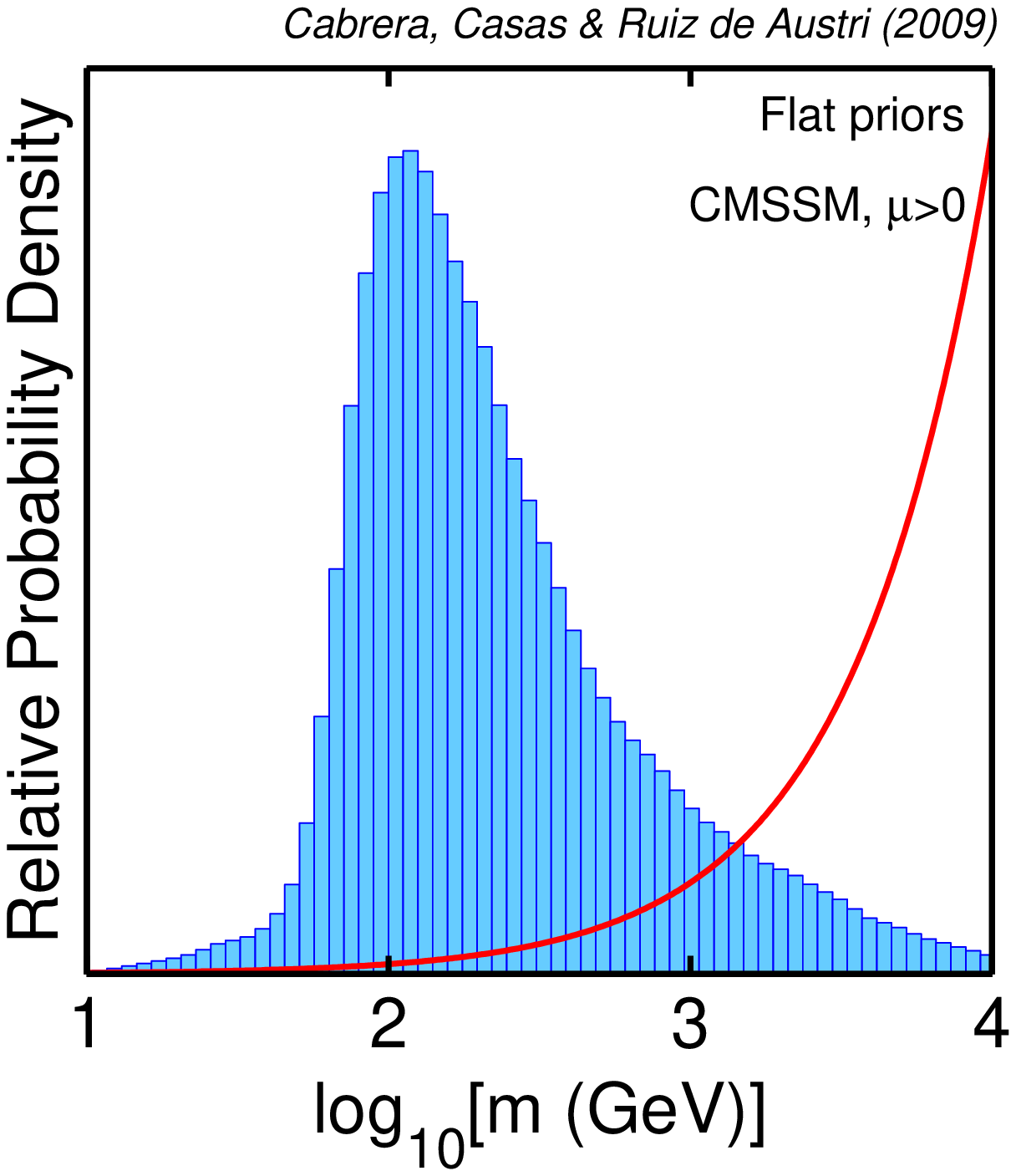}
\caption[text]{1D marginalized posterior probability distribution of the 
$M$ and $m$ parameters (upper and lower panels respectively) for logarithmic (left panels) 
and flat (right panels) priors in the $\mu>0$ case, for a scan including the information about the EW 
breaking ($M_Z^{\rm exp}$). 
The red lines represent the marginalized prior. All given in arbitrary units.} 
\end{center}
\end{figure}

Besides  making the high-energy parameter space quite irrelevant, the EW breaking has another dramatic effect, which is visible in Fig. 1. Namely, the probability distributions (pdfs) based on a logarithmic or on a flat prior are quite similar, {\em after} the incorporation of the EW scale. That is, the favoured regions of the parameter space are quite independent of the choice of the prior. Normally, a behaviour of this kind is attributed to the fact that the data are powerful enough to select a region of the parameter space, so that the general expression of the pdf, eq.(\ref{Bayes}), is dominated by the likelihood piece. However this is not the case here. As a matter of fact, concerning the likelihood, there are points with arbitrary large parameters that are as good as the low-energy ones, since they correctly reproduce $M_Z^{\rm exp}$, the only data so far considered. The low-energy region is preferred because it is statistically much more significant, as we have discussed above. But this is a {\em Bayesian} effect, non-existent in a frequentist analysis. Therefore the situation is very good from the Bayesian point of view: the results are quite independent from the type of prior, but to see the preferred regions we need the Bayesian procedures.

To finish this section, let us note that the previous statistical argument supports low-energy supersymmetry breaking (in the observable sector), even in a landscape scenario. In other words, even if there were many more vacua with supersymmetry breaking at large scale, most of realistic vacua would correspond to low-energy supersymmetry breaking, for rather generic a-priori distributions of all possible vacua (for related work in this line see \cite{Susskind:2004uv}).

\section{Experimental Constraints}

In this section we will incorporate all the relevant experimental information to the likelihood piece of the probability distribution (all but $M_Z^{\rm exp}$, which has already been taken into account). This amounts to include many experimental observables and bounds, with their error bars, and to calculate the predictions for them in the MSSM.

As originally demonstrated in \cite{Allanach:2005kz,deAustri:2006pe}, the values of the 
relevant SM-like parameters ({\em nuisance parameters}) can strongly influence some of the 
CMSSM predictions. For our analysis we take the set
\be \label{nuipars:eq} \left\{\mtpole,\ \mbmbmsbar,\ \alphaemmz,\ \alphas \right\}, 
\ee
where $\mtpole$ is the pole top quark mass, while the other three parameters (the bottom mass, the electromagnetic and the strong coupling constants) are all evaluated in the $\msbar$ scheme at the indicated scales. The constraints on the SM nuisance parameters are given in 
Table~\ref{tab:nuisance}. 

On the other hand, there are the experimental values of accelerator and 
cosmological observables, which are listed in Table~\ref{tab:obs}. 
Instead of including all this information at once and show the results, we find more illustrative to do it in several steps. This will allow to show the effect of the various types of data on the probability distributions (which are sometimes opposite). On the other hand, not all the data are on the same foot of quality and reliability and it is convenient not to mix them from the beginning.

%%%%%%%%%%
\begin{table}[t]
\centering    .

\begin{tabular}{|l | l l | l|}
\hline
SM (nuisance)  &   Mean value  & \multicolumn{1}{c|}{Uncertainty} & Ref. \\
 parameter &   $\mu$      & ${\sigma}$ (exper.)  &  \\ \hline
$\mtpole$           &  172.6 GeV    & 1.4 GeV&  \cite{topmass:mar08} \\
$m_b (m_b)^{\overline{MS}}$ &4.20 GeV  & 0.07 GeV &  \cite{pdg07} \\
$\alphas$       &   0.1176   & 0.002 &  \cite{pdg07}\\
$1/\alphaemmz$  & 127.955 & 0.03 &  \cite{Hagiwara:2006jt} \\ \hline
\end{tabular}
\caption{Experimental mean $\mu$ and standard deviation $\sigma$
 adopted for the likelihood function for SM (nuisance) parameters,
 assumed to be described by a Gaussian distribution.
\label{tab:nuisance}}
\end{table}
%%%%%%%%%%%%%

In order to avoid a proliferation of plots we examine first the positive $\mu$ branch. In the next section we will show the relevant plots and results for negative $\mu$ and perform a comparison of the relative probability of the two possibilities.

\subsection{EW and B-physics observables, and limits on particle masses}

We start by considering the most reliable and robust pieces of experimental information: EW and B(D)-physics observables and lower bounds on the masses of supersymmetric particles and the Higgs mass. The complete list of the observables of this kind used in our analysis is given in Table 2 (all the entries except those concerning $a_\mu$ and dark matter constraints).

To calculate the MSSM spectrum we use \texttt{SoftSusy} \cite{softsusy}, where SUSY masses are computed at full one-loop level and the Higgs 
sector includes two-loop leading corrections \cite{Dedes:2003km}. We discard 
points suffering from unphysicalities: no self-consistent solutions to the 
RGEs, no EW breaking and tachyonic states. Furthermore, we require the neutralino 
to be the lightest supersymmetric particle (LSP) in order to be an 
acceptable dark matter candidate. The latter condition might be relaxed, as discussed in subsect.~4.3 below. In our treatment of the radiative corrections to
the electroweak observables $M_W$ and $\sineff$ we include full two-loop and 
known higher order SM corrections as computed in 
ref.~\cite{awramik-acfw04}, as well as
gluonic two-loop MSSM corrections obtained in~\cite{dghhjw97}.
%For details on how the EW observables are computed the reader is directed to 
%\cite{deAustri:2006pe}.

Roughly speaking, the MSSM parameter space is quite unconstrained by EW (LEP) observables, except for quite small values of the SUSY soft-terms (i.e. when the SUSY corrections are sizeable) \cite{deBoer:2003xm, Heinemeyer:2004gx}.
This is logical. As it is well known, the MSSM is free from the Little Hierarchy problem, understood as the tension between LEP observables and the need of new physics at ${\cal O}({\rm TeV})$ scales to avoid the hierarchy problem \cite{Barbieri:2000gf}. This is because R-parity prevents from tree-level SUSY contributions to higher order SM operators. In consequence, unless supersymmetric masses are quite small, the effect of SUSY on LEP observables is not important. 

Concerning B-physics observables, the branching ratio for the 
$B \rightarrow X_s \gamma$ decay (the most important one) has been computed with the numerical code 
\texttt{SusyBSG} \cite{Degrassi:2007kj} using the full NLO QCD contributions, 
including the two-loop calculation of the gluino contributions presented 
in \cite{Degrassi:2006eh} and the results of \cite{D'Ambrosio:2002ex} for the 
remaining non-QCD 
$\tan\beta$-enhanced contributions. The supersymmetric contributions to $b\rightarrow s \gamma$ grow with decreasing masses of the supersymmetric particles {\em and} with increasing $\tan\beta$. For $\mu>0$ they have the ``wrong sign", so larger supersymmetric masses are preferred. 
However the SUSY contribution  is never dramatic for masses around 1 TeV or larger. For the determination of $\delmbs$ we use expressions from ref. \cite{for1} which include dominant large $\tanb$-enhanced beyond-LO SUSY contributions from Higgs penguin diagrams.
The other B(D)-physics observables summarized in Table 2 
have been computed with the code \texttt{SuperIso} (for details
on the computation of the observables see \cite{Mahmoudi:2008tp} and references 
therein). Both codes have been integrated into \texttt{SuperBayes}.

%%%%%%%%%%%%%
\begin{table}[t]
\centering
\begin{tabular}{|l | l l l | c|}
\hline
Observable &   Mean value & \multicolumn{2}{c|}{Uncertainties} & ref.  \\
 &   $\mu$      & ${\sigma}$ (exper.)  & $\tau$ (theor.) &  (exper.)  \\\hline
 $M_W$     &  $80.398\gev$   & $27\mev$ & $15\mev$ & \cite{lepwwg}  \\
$\sineff{}$    &  $0.23149$      & $17\times10^{-5}$
               & $15\times10^{-5}$ &  \cite{lepwwg}    \\
$a_\mu^{\rm exp} \times 10^{10}$ &  11659208.9  & 6.33  &  -   & \cite{pdg07} \\
$\delta a_\mu \times 10^{10}\; (e^+ e^-)$ &  29.5 & 8.8 &  2.0 & \cite{Miller:2007kk}  \\
$\delta a_\mu \times 10^{10}\; (\tau)$  &  14.8  & 8.2  &  2.0 & \cite{Davier:2009ag}  \\
$\delmbs$     &  $17.77\ps^{-1}$  & $0.12\ps^{-1}$  & $2.4\ps^{-1}$
& \cite{cdf-deltambs}   \\
$\brbsgamma \times 10^{4}$ & 3.52 & 0.33 & 0.3 & \cite{hfag}  \\
$\RBtaunu$   &  1.28  & 0.38  & - & \cite{hfag}  \\
$\DeltaO  \times 10^{2}$   &  3.6  & 2.65  & - &  \cite{:2008cy}  \\
$\RBDtaunuBDenu \times 10^{2}$ & 41.6 & 12.8 & 3.5   &  \cite{Aubert:2007dsa}  \\
$\Rl$ & 1.004 & 0.007 & -  &  \cite{Antonelli:2008jg}  \\
$\Dstaunu \times 10^{2}$ & 5.7 & 0.4 & 0.2  &  \cite{Akeroyd:2009tn}  \\
$\Dsmunu  \times 10^{3}$ & 5.8 & 0.4 & 0.2  & \cite{Akeroyd:2009tn}  \\
$\abundchi$ &  0.1099 & 0.0062 & $0.1\,\abundchi$& \cite{wmap5yr}  \\\hline\hline
   &  Limit (95\%~\cl)  & \multicolumn{2}{c|}{$\tau$ (theor.)} & ref. \\ \hline
%   &                    &                            & (exper.) & (theor.) \\
$\brbsmumu$ &  $ <5.8\times 10^{-8}$
& \multicolumn{2}{c|}{14\%}  & \cite{cdf-bsmumu}  \\
$\mhl$  & $>114.4\gev$\ (SM-like Higgs)  & \multicolumn{2}{c|}{$3 \gev$}
& \cite{lhwg}   \\
$\zetah^2$
& $f(m_h)$  & \multicolumn{2}{c|}{negligible}  & \cite{lhwg}    \\
$m_{\tilde{q}}$ & $>375$ GeV   & \multicolumn{2}{c|}{5\%} & \cite{pdg07}  \\
$m_{\tilde{g}}$ & $>289$ GeV  & \multicolumn{2}{c|}{5\%} & \cite{pdg07}  \\
other sparticle masses  &  \multicolumn{3}{c|}{As in table~4 of
  ref. \cite{deAustri:2006pe}.}  &  \\ \hline
\end{tabular}
\caption{Summary of the observables used in the analysis to constrain the CMSSM
parameter space. Upper part:
Observables for which a positive measurement has been made. $\delta a_\mu= a_\mu^{\rm exp}-\amusm$ denotes the discrepancy between
the experimental value and the SM prediction of the anomalous magnetic
moment of the muon $\gmtwo$.
As explained in the text, for each quantity we use a
likelihood function with mean $\mu$ and standard deviation $s =
\sqrt{\sigma^2+ \tau^2}$, where $\sigma$ is the experimental
uncertainty and $\tau$ represents our estimate of the theoretical
uncertainty. Lower part: Observables for which only limits currently
exist.  The likelihood function is given in
ref.~\cite{deAustri:2006pe}, including in particular a smearing out of
experimental errors and limits to include an appropriate theoretical
uncertainty in the observables. $\mhl$ stands for the light Higgs mass
while $\zetah^2\equiv g^2(\hl ZZ)_{\text{MSSM}}/g^2(\hl ZZ)_{\text{SM}}$,
where $g$ stands for the Higgs coupling to the $Z$ and $W$ gauge boson
pairs. The references for the theoretical calculations are given in the text.
\label{tab:obs}}       
\end{table}

Experimental bounds used in the analysis are indicated in the second part 
of Table 2. 
These include bounds on supersymmetric masses (squarks, sleptons, gluinos, 
charginos, neutralinos) and the Higgs mass. 
In general, the constraints on supersymmetric masses tend obviously to cut 
off the region of the parameter space with too small values of $m, M$. 
On top of this, the bound on the Higgs mass is most relevant, and deserves 
special attention, as we are about to see. For details on how the likelihood 
is computed we refer to ref. \cite{deAustri:2006pe}.

For the quantities for which positive measurements have been made (as
listed in the upper part of Table~\ref{tab:obs}), we assume a
Gaussian likelihood function with a variance given by the sum of the
theoretical and experimental variances, as motivated by eq.~(3.3) in
ref.~\cite{deAustri:2006pe}. For the observables for which only lower or 
upper limits are available (as listed in the bottom part of
Table~\ref{tab:obs}) we use a smoothed-out version of the
likelihood function that accounts for the theoretical error in the
computation of the observable, see eq.~(3.5) and fig.~1 in
ref.~\cite{deAustri:2006pe}.  In particular, in applying a lower mass bound 
from LEP-II on the Higgs boson $h$ we take into account its dependence
on its coupling to the $Z$ boson pairs $\zetah^2$, 
as described in detail in
ref.~\cite{rrt2}. When $\zetah^2\simeq 1$, the LEP-II lower bound of
$114.4\gev$ (95\%~\cl)~\cite{lhwg} applies. For arbitrary values of
$\zetah$, we apply the LEP-II 95\%~\cl\ bounds on $\mhl$,
which we translate into the corresponding 95\%~\cl\ bound in the
$(\mhl, \zetah^2)$ plane. We then add a conservative theoretical
uncertainty $\tau(\mhl) = 3\gev$, following eq.~(3.5) in
ref.~\cite{deAustri:2006pe}. The best fit is then defined as the 
maximum value of the joint likelihood function.

\begin{figure}[t]
\begin{center}
\label{}
\includegraphics[angle=0,width=0.4\linewidth]{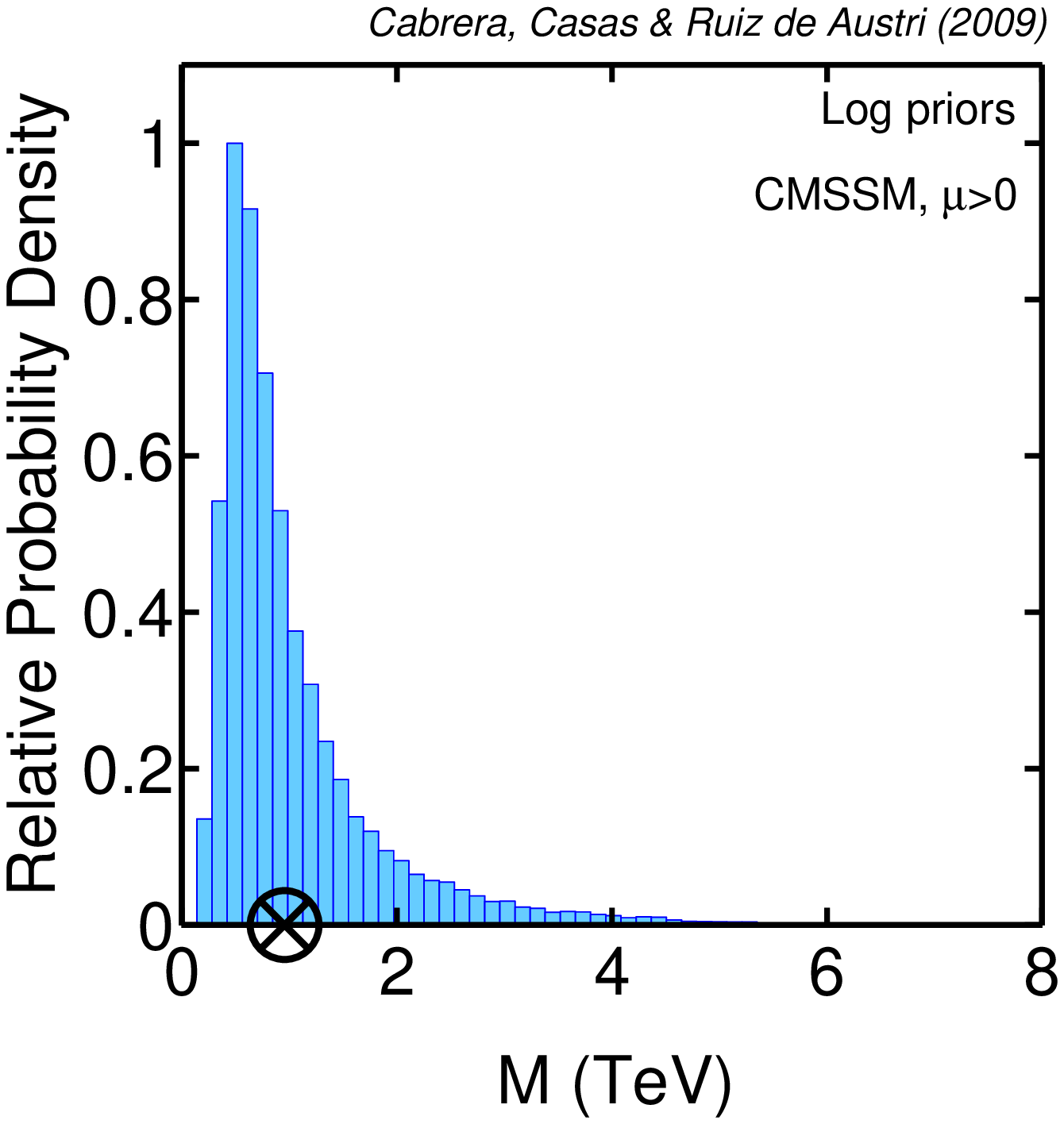} \hspace{1.2cm}
\includegraphics[angle=0,width=0.4\linewidth]{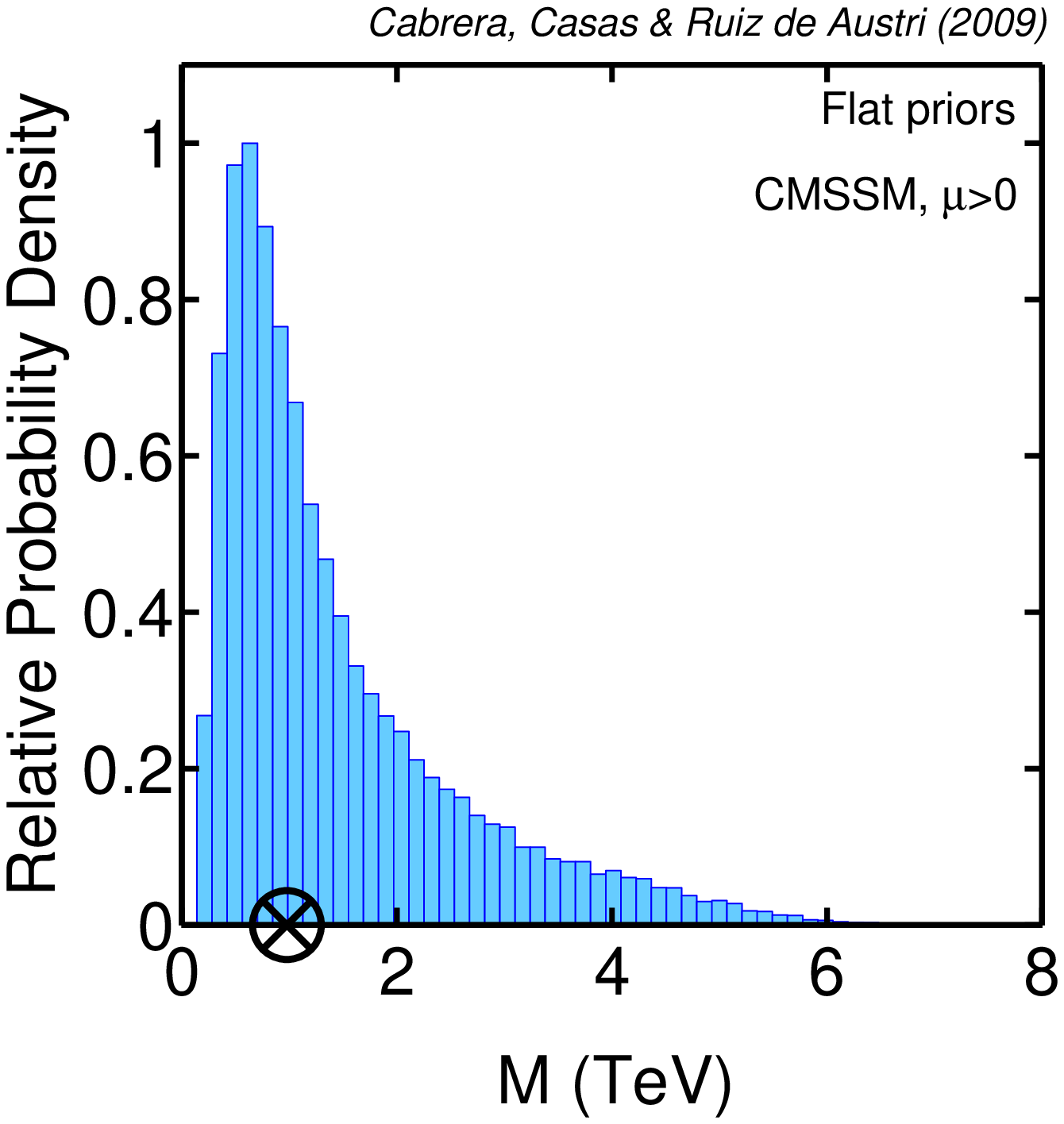} \\ \vspace{1.0cm}
\includegraphics[angle=0,width=0.4\linewidth]{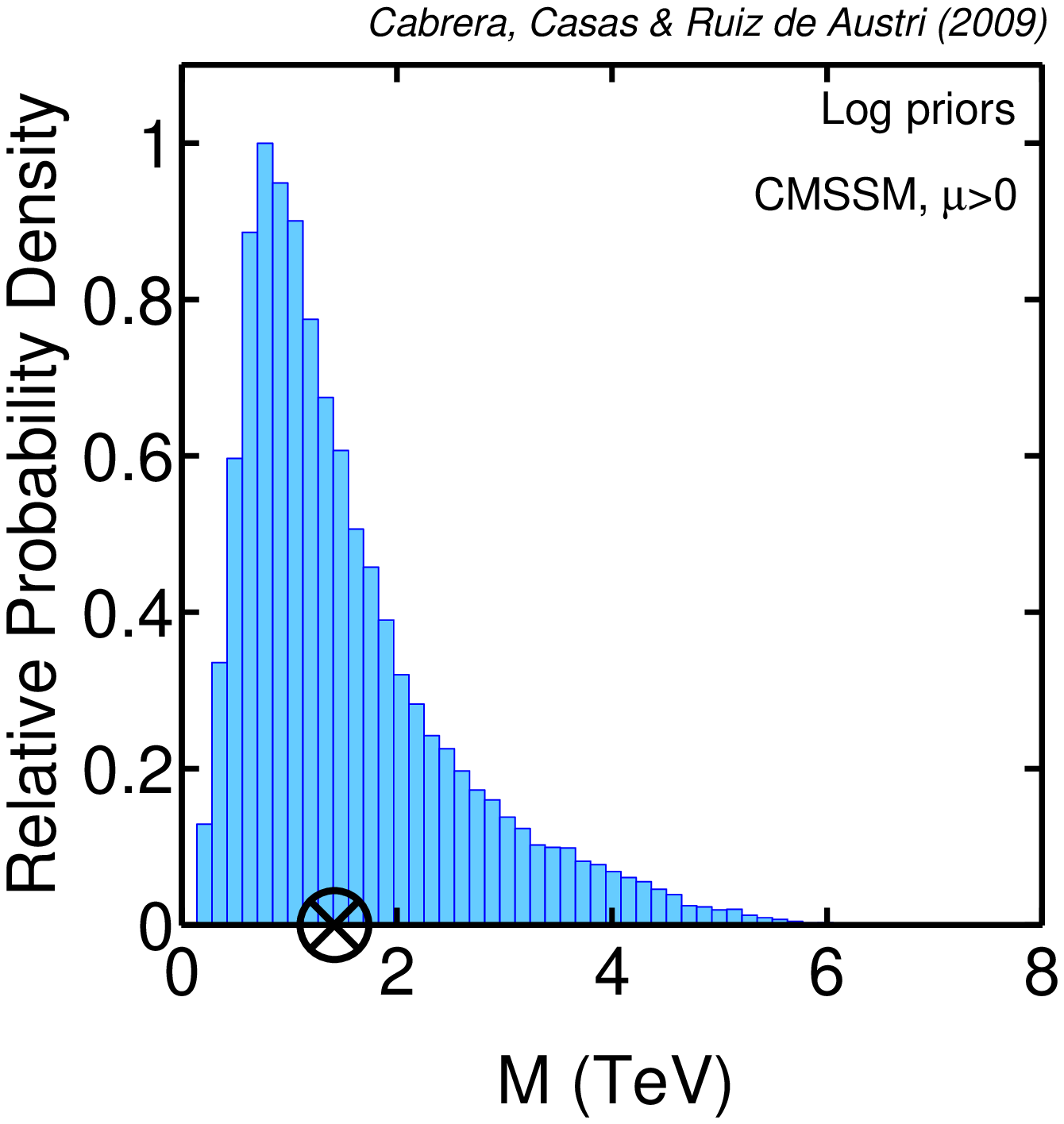} \hspace{1.2cm}
\includegraphics[angle=0,width=0.4\linewidth]{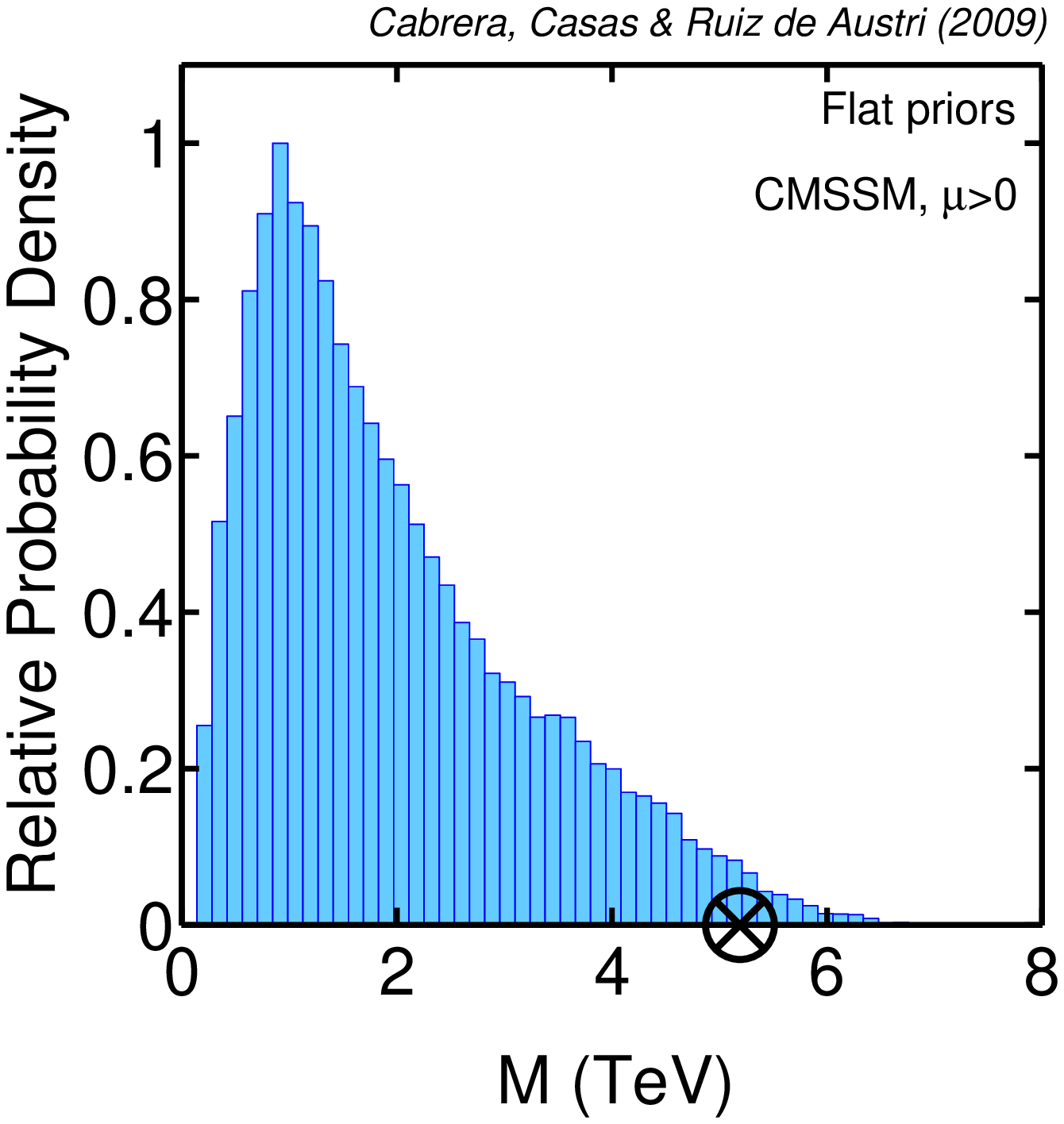}
\caption[text]{Upper panels show the 1D marginalized posterior probability 
distribution of the $M$  parameter for logarithmic (left panel) and 
flat (right panel) priors in the $\mu>0$ case for a scan including 
SM nuisance parameters constraints, EW breaking ($M_Z^{\rm exp}$), collider limits on Higgs and 
superpartner masses, and EW and B(D)-physics observables. 
Lower panels show the same but imposing a bound for the Higgs mass 
of $m_h \geq 120$ GeV. The cross corresponds to the best-fit point, defined as the one with highest likelihood.}
\end{center}
\end{figure}

Fig. 2 (upper panels) show the pdf for the gaugino mass parameter, $M$, once all this experimental information is incorporated. Again, the left (right) panels correspond to a
logarithmic (flat) initial prior for the scale of SUSY breaking in the observable sector ($M_S$). The reason to show the pdf of $M$ is to facilitate the comparison with the analogous 
probability distribution before the inclusion of the new pieces of experimental information (Fig. 1). Clearly, the bulk of the probability is now pushed into the high-energy region. This effect is basically due to the Higgs mass bound. As discussed above, concerning the other observables, everything works fine, as long as SUSY is not at too low scale. 
On the other hand, it is well known that in the MSSM the tree-level Higgs mass is bounded from above by $M_Z$, so radiative corrections (which grow logarithmically with the stop masses) are needed.

It is possible to be more quantitative by considering the dominant 1-loop correction \cite{Ellis:1991zd} to the theoretical upper bound on $m_h$ in the MSSM:
\bea
\label{mhMSSM}
m_h^2\leq M_Z^2 \cos^2 2 \beta + {3 m_t^4 \over 2\pi^2 v^2}
\log{M_{\tilde t}^2\over m_t^2} + ...  
\eea
where $m_t$ is the (running) top mass and $M_{\tilde t}$ is an average of stop masses. 
Hence, for a given lower bound on the Higgs mass, $m_h^{\rm min}$, one needs
\bea
\label{MSUSYMSSM}
M_{\tilde t}\;\simgt\;  e^{-2.1\cos^2 2\beta}
 e^{\left({m_h^{\rm min}}/{62\ {\rm GeV}}\right)^2} m_t\ .
\eea
Thus, an increase $\Delta m_h^2$ on the lower bound of the Higss mass squared approximately translates into a multiplicative factor for $M_{\tilde t}$ :
\bea
\label{MSUSYMSSM2}
M_{\tilde t}\;\rightarrow\; M_{\tilde t}\ e^{{\Delta m_h^2}/{(62\ {\rm GeV})^2}} \ ,
\eea
and a similar increase can be expected in the initial parameters $m$, $M$.

\begin{figure}[t]
\begin{center}
\label{}
\includegraphics[angle=0,width=0.35\linewidth]{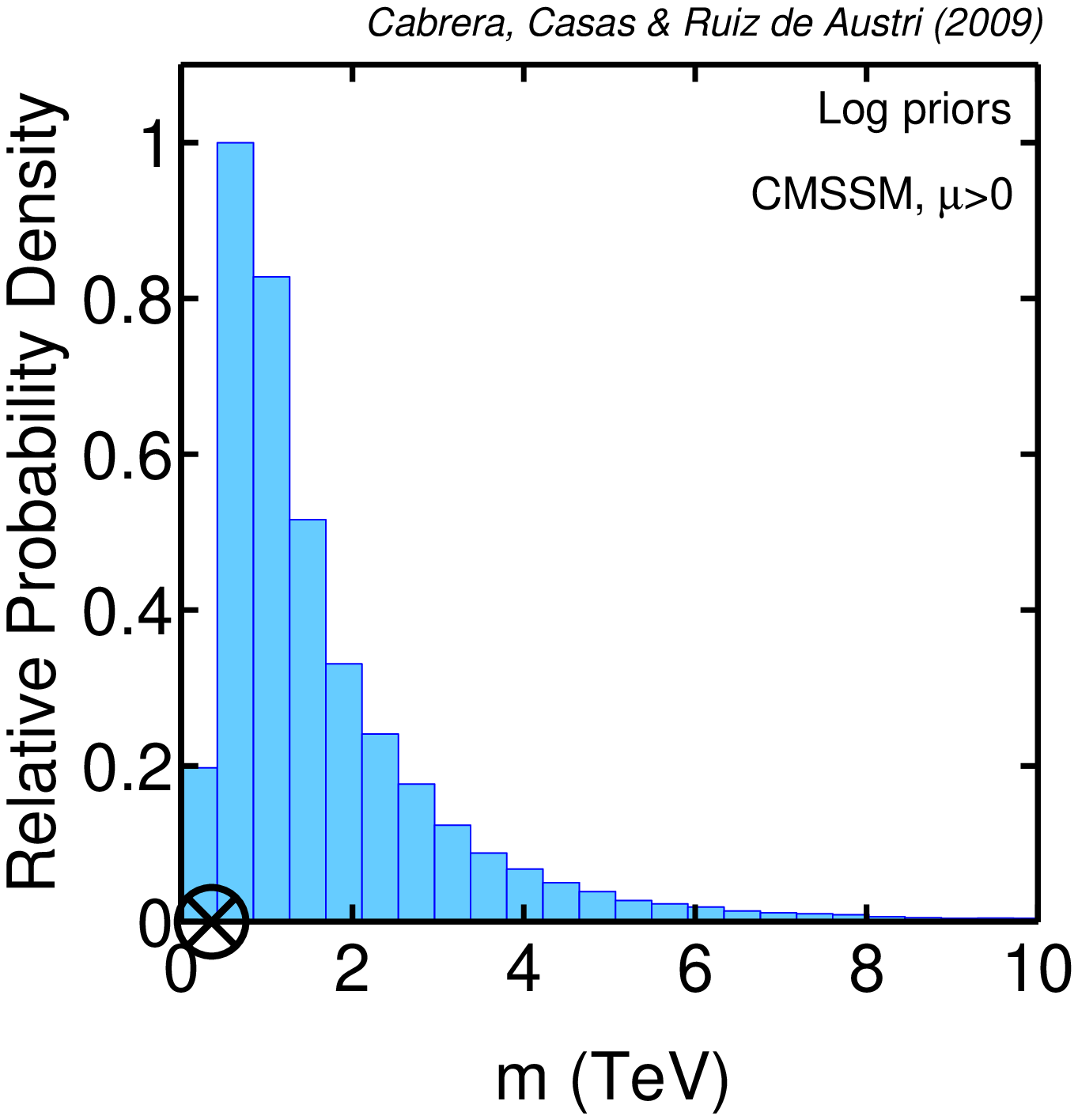} \hspace{1.2cm}
\includegraphics[angle=0,width=0.35\linewidth]{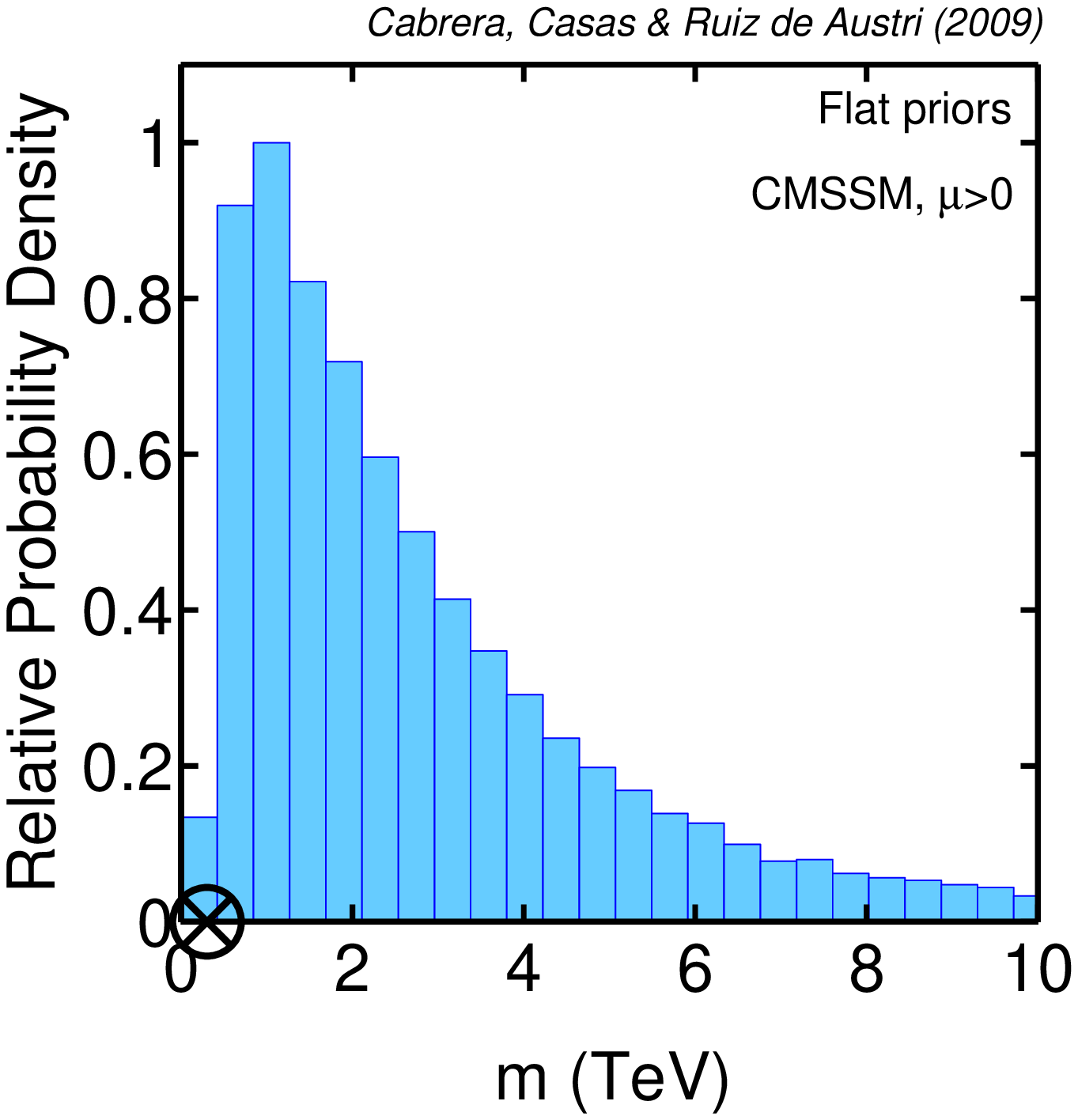} \\ \vspace{0.6cm}
\includegraphics[angle=0,width=0.35\linewidth]{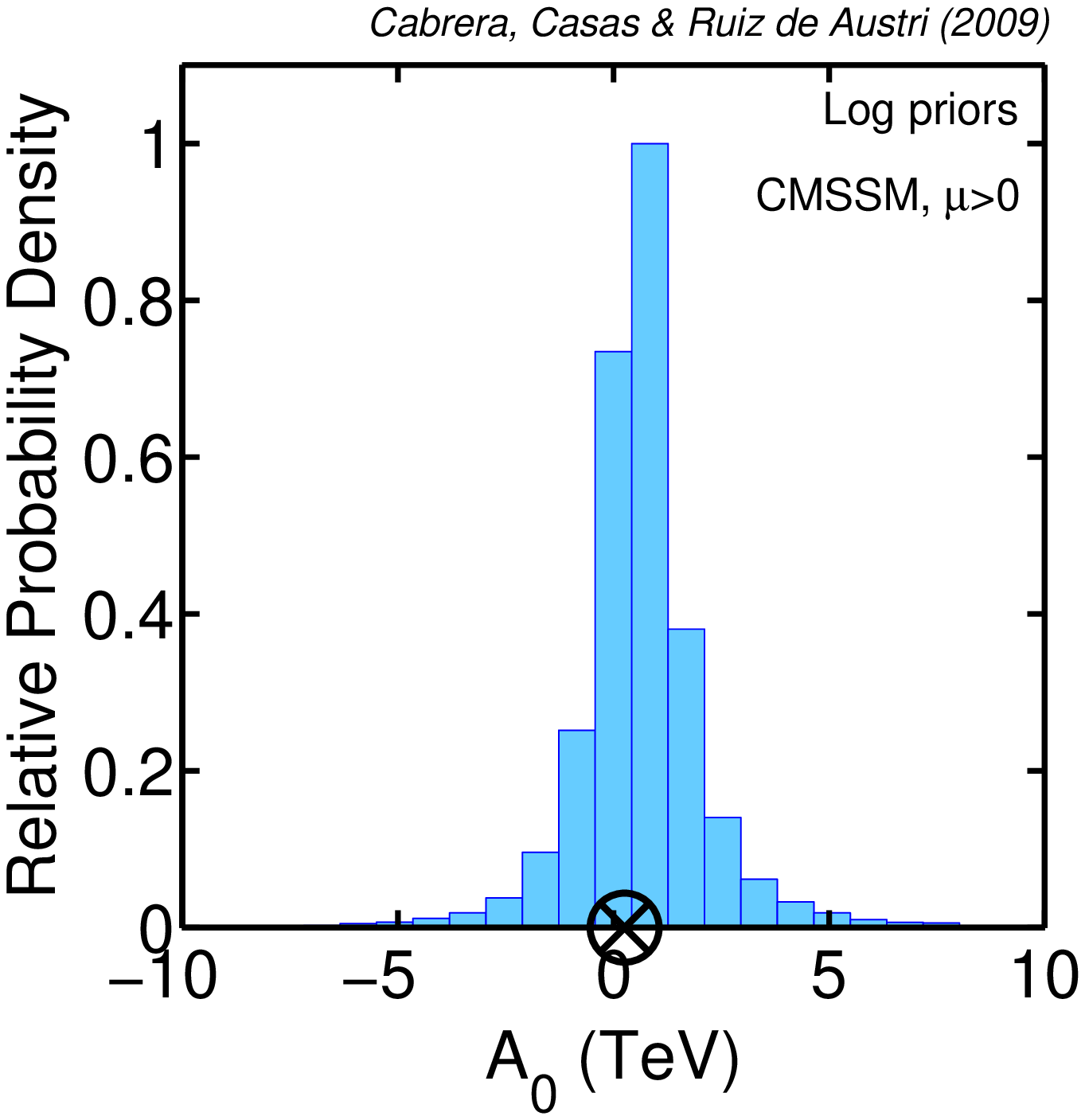} \hspace{1.2cm}
\includegraphics[angle=0,width=0.35\linewidth]{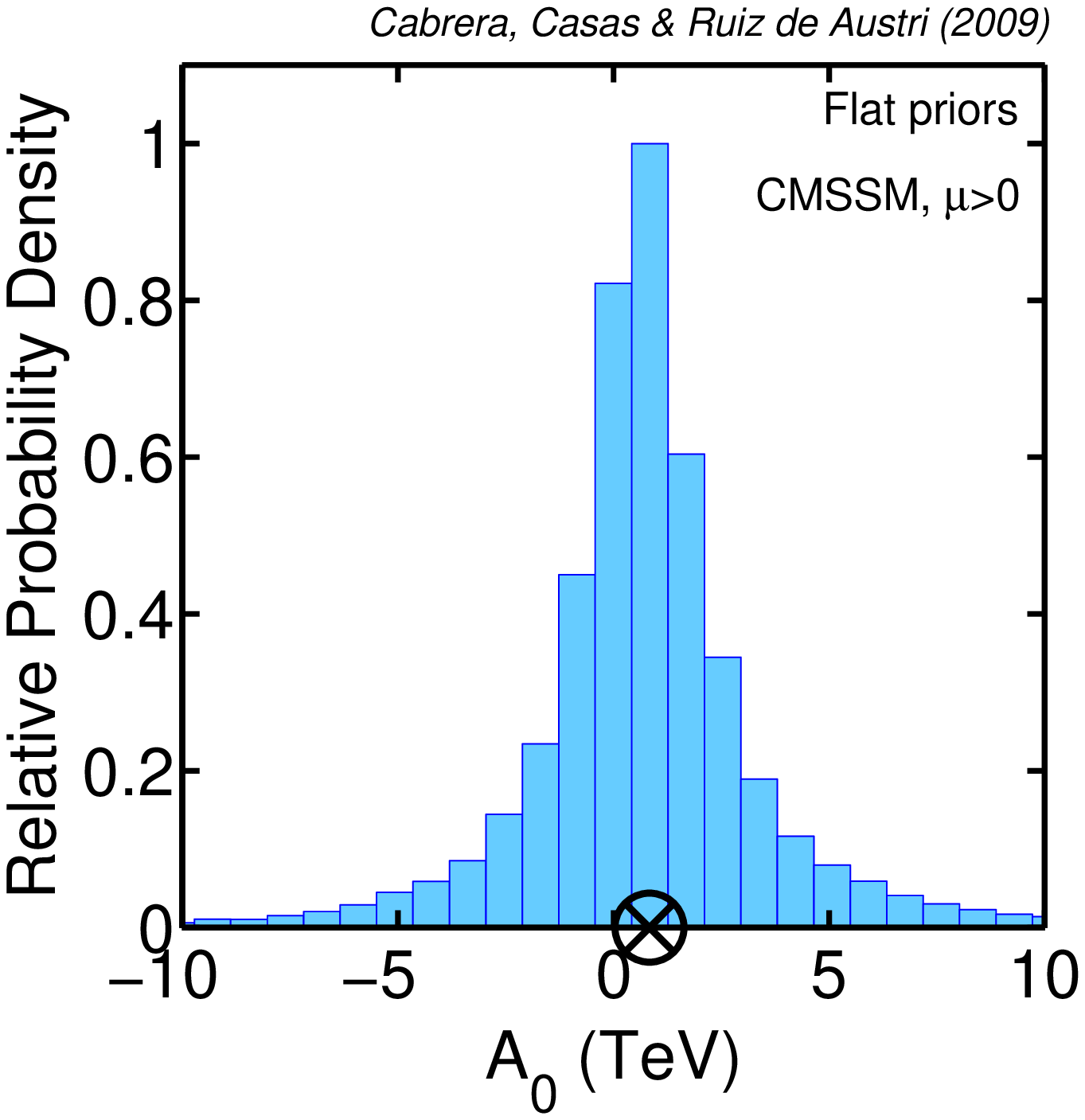} \\ \vspace{0.6cm}
\includegraphics[angle=0,width=0.35\linewidth]{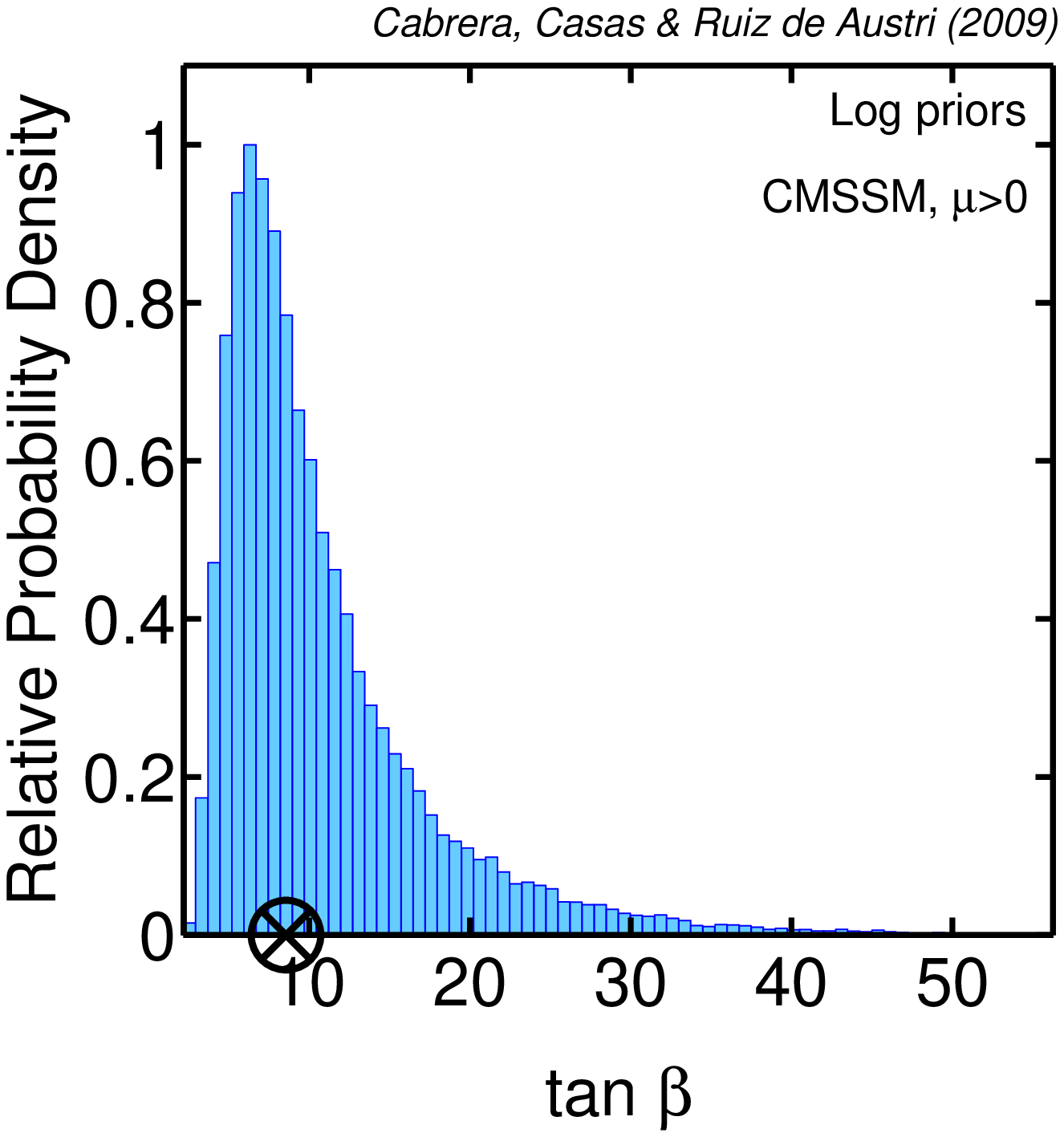} \hspace{1.2cm}
\includegraphics[angle=0,width=0.35\linewidth]{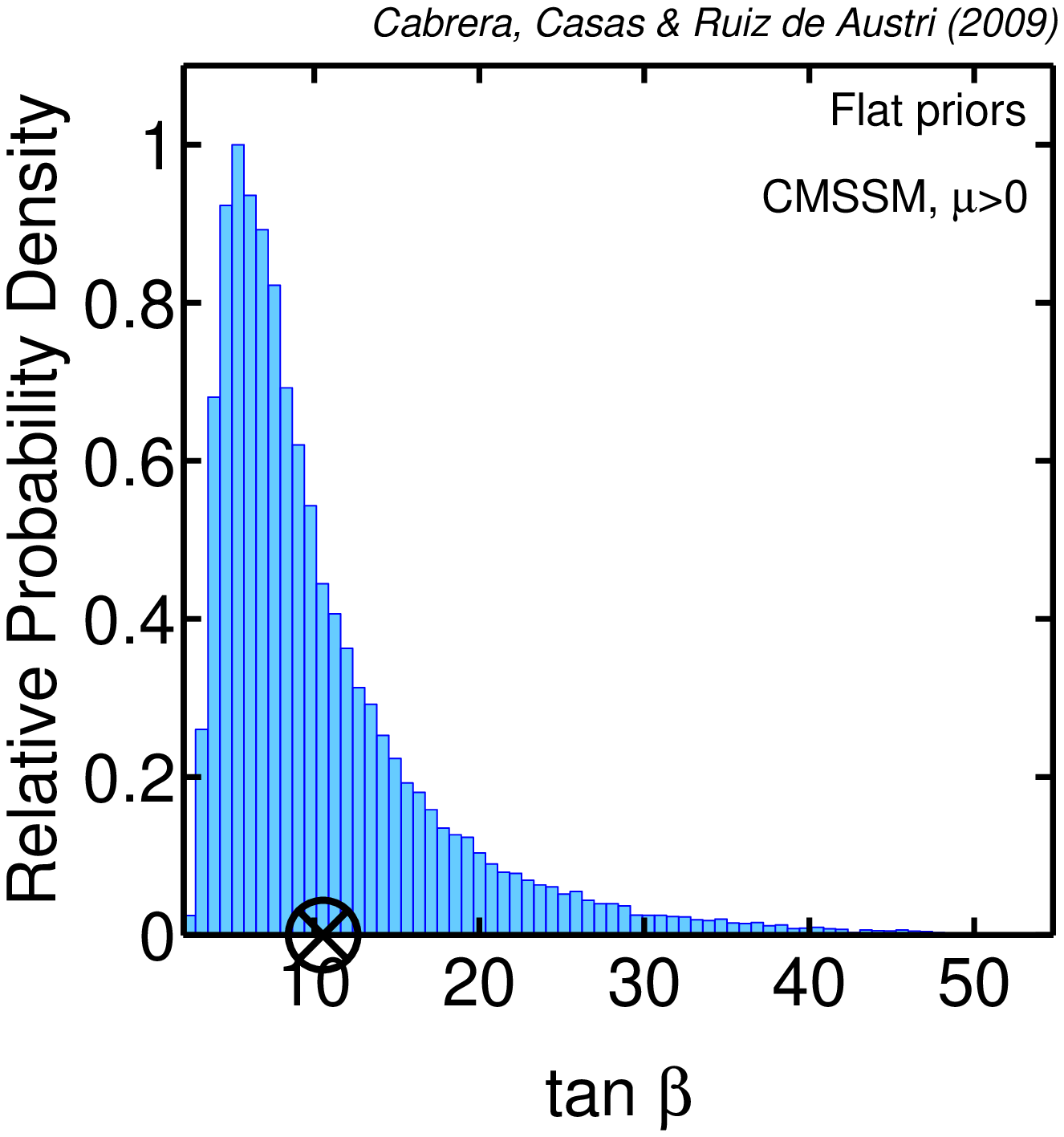}
\caption[test]{1D marginalized posterior probability distribution of the CMSSM parameters 
for logarithmic (left panels) and flat (right panels) priors in the $\mu>0$ 
case for a scan including SM nuisance parameters constraints, EW breaking ($M_Z^{\rm exp}$), collider limits 
on Higgs and superpartner masses, and EW and B(D)-physics observables. 
The cross corresponds to the best-fit point.}
\end{center}
\end{figure}

To illustrate these facts, we have re-done the pdfs assuming a different value of the Higgs mass bound, say $m_h\geq 120$ GeV. Of course this would correspond to the real situation if the Higgs mass turns out finally to lie in this range. According to the previous argument, we can expect now a longer push of the probability distribution into the high-energy region. And this is what happens, as it is shown in Fig.2 (lower panels). The effect is very important,
given the modest increase in the Higgs mass bound. Larger shifts in $m_h$ have an exponentially larger effect, as discussed above. So, if the MSSM is true and we wish to detect it at LHC, let us hope that $m_h$ is close to the present experimental limit\footnote{Certainly, it is well-known that a Higgs above 125 GeV is not easy to arrange in the MSSM, and that is at the origin of the difficulties. What the present analysis shows, in a more direct way, is how improbable is to arrange a large $m_h$ (see also Fig.~4 below) and the implications for the discovery of SUSY at the LHC.}.

Fig.3 shows some representative probability distributions for individual (initial and derived) parameters, i.e. once all the rest are marginalized. The dimension-full parameters ($m,A$) follow a trend similar to that of the gaugino mass, $M$ (which was already shown in Fig. 2). On the other hand, large values of $\tan\beta$ are penalized, mainly due to the Jacobian factor in the probability distribution, see eq.(\ref{approx_eff_prior}) and the subsequent discussion. It is worth to remark that this penalization of $\tan\beta$ contrasts with other Bayesian analyses, where the prior for $\tan\beta$ was taken as flat. Here it arises from the above-mentioned Jacobian factor and therefore has nothing to do with a particular choice of priors.
Fig.~4 shows the probability distribution for the Higgs mass. 
One can see that there are a significant number of points which 
evade the LEP-II $114.4\gev$ lower bound for the SM Higgs. This reflects the 
fact that we have employed the full likelihood function in the 
$(\mhl, \zetah^2)$ plane as described above and which allowes points with 
low Higgs masses where $\zetah^2=sin^2 (\beta -\alpha) \ll 1$. The corresponding Bayesian credibility intervals, representing the 68\% and 95\% of the total probability, are given in Table~\ref{tab:Higgs}. The central value for the Higgs mass is at 117--118 GeV. From that table one can see the robustness of the results under changes of the prior. Notice also the little discrepancy among the mean value of the posterior pdf and the best fit. 

%, which suggests that the posterior is dominated by the likelihood.
%This is confirmed by the little discrepancy among the mean value of
%the posterior pdf and the best fit. 

\begin{table}
\begin{center}
\begin{tabular}{|c|c|c|c|}
\hline
Parameter & Mean value & Best fit & 68\% (95\%) range \\ 
\hline\hline
$m_h$ (GeV)  & 240.8 &  & [235.5 : 246.8] ([222.6 : 251.3]) \\
$R^{-1}$ (GeV) & 589.3 &  & [511.5 : 668.6] ([467 : 781.9]) \\ \hline
\end{tabular}
\end{center}
\caption{Higgs mass mean value and best fit for logarithmic and flat
priors, and the $68\%$ and $95\%$ Bayesian equal-tails credibility intervals. 
All numbers are given in $\gev$ units.}
\label{tab:Higgs}
\end{table}

\begin{figure}[t]
\begin{center}
\label{}
\includegraphics[angle=0,width=0.4\linewidth]{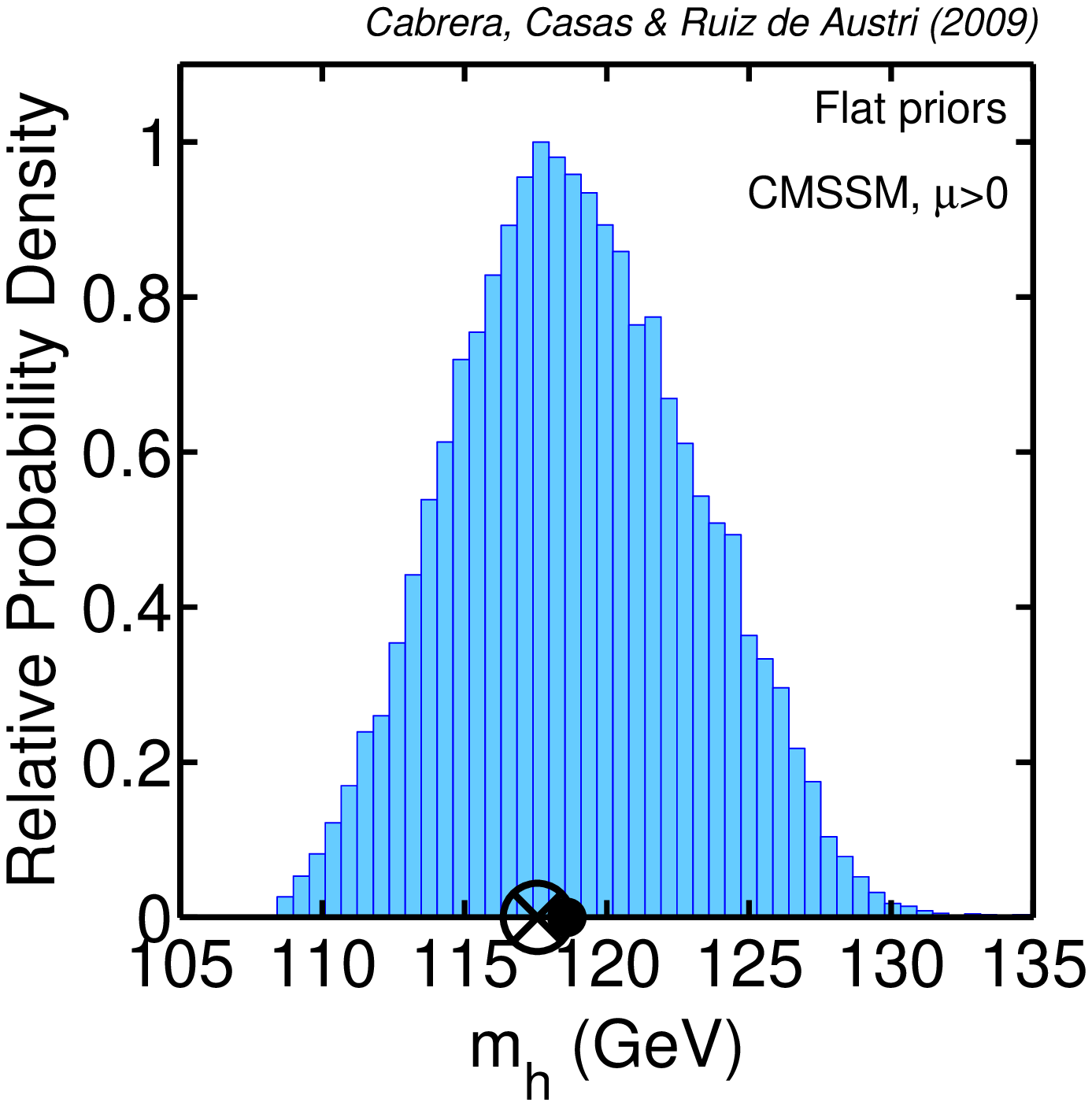} \hspace{1.2cm}
\includegraphics[angle=0,width=0.4\linewidth]{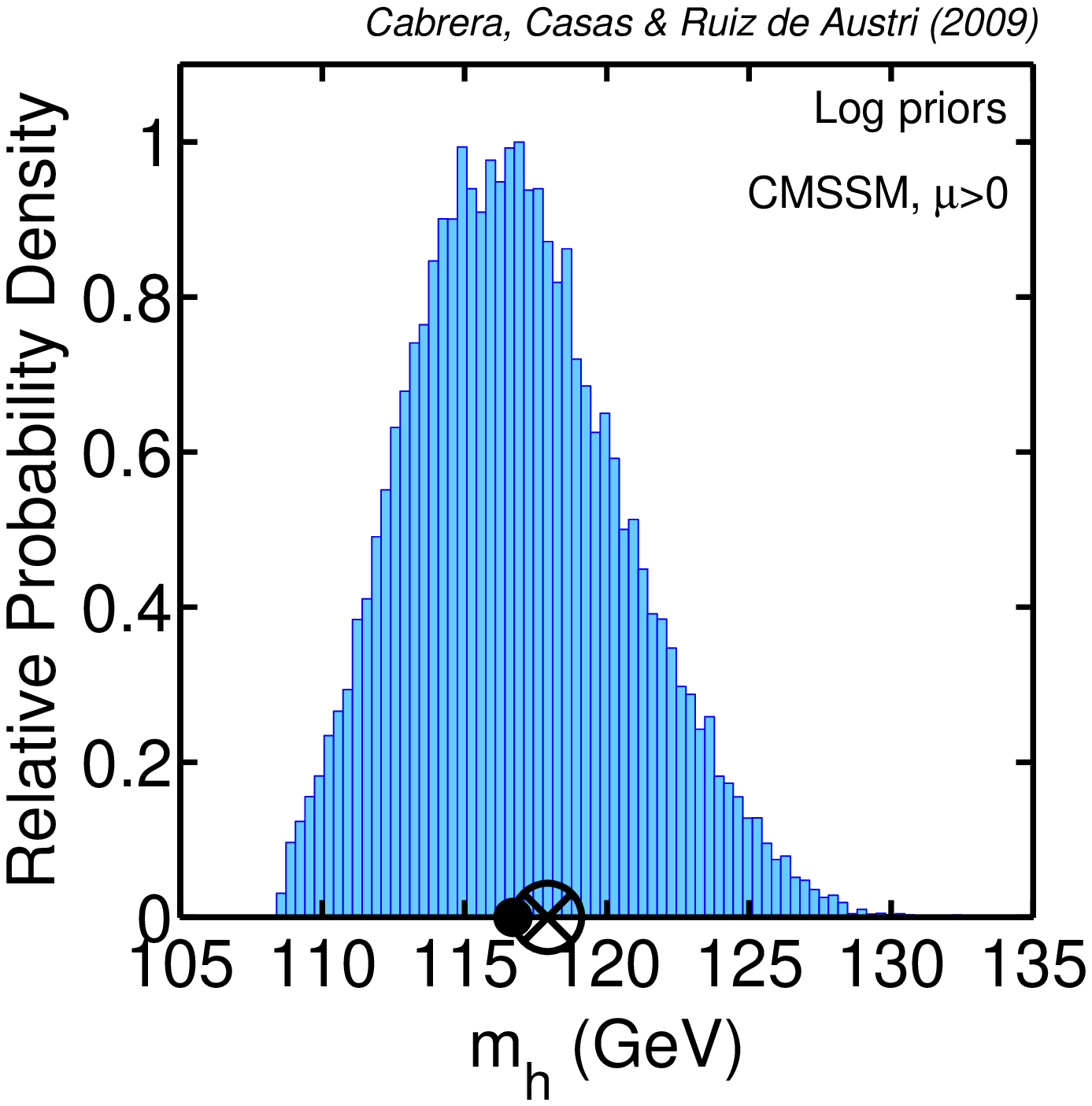}
\caption[text]{As Fig. 3, for the Higgs mass. 
The small filled circle represents the mean value of the posterior pdf and the cross corresponds to the best-fit point.} 
\end{center}
\end{figure}

Fig.5 shows the probability distribution in the $\{M,m\}$ and $\{\tan\beta,M\}$ planes (i.e. when all the parameters but two are marginalized). The results of Figs. 3--5 are shown all for logarithmic (left panels) and flat (right panels) priors, exhibiting a remarkable stability, which has already been discussed. In the $\{M,m\}$ plots we have shown also the discovery reach of LHC for 1 fb$^{-1}$ and 100 fb$^{-1}$ (with a center-of-mass energy of 14 TeV). These lines have taken from ref.\cite{Baer:2009dn}. They arise from a study of events with $\rm{N}_{\rm jets}\geq 2$ and an optimization of the cuts on $E_T^{\rm missing}$. (For a more detailed explanation of the procedure used see [29]). Strictly speaking, the lines correspond
to $A=0$, $\tan\beta=45$, but they provide a good indication of the LHC discovery potential in the short and medium term (for similar analyses see \cite{Ball:2007zza}). Now, it is clear that a substantial (though still non-dominant) part of the probability falls out of the LHC reach, an effect that it is more important for flat prior. This means that if we are unlucky, supersymmetry could evade LHC detection in the short, or even the long, term. On top of this, let us recall that if the Higgs mass is not close to its present experimental value, the preferred regions of the parameter space are quickly pushed to high-energy (see discussion about Fig.2), thus jeopardizing the discovery of supersymmetry.

\begin{figure}[t]
\begin{center}
\label{}
\includegraphics[angle=0,width=0.4\linewidth]{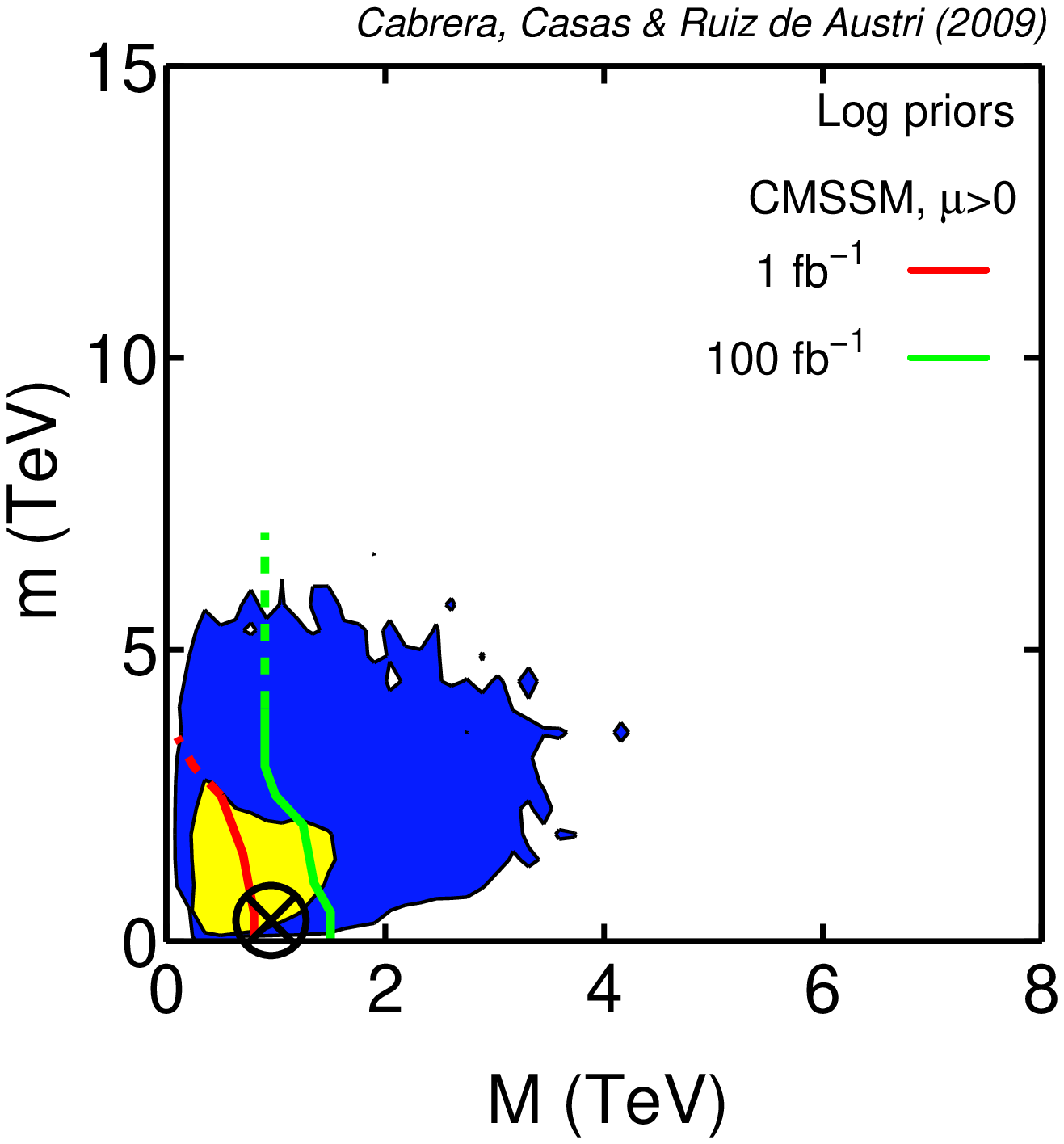} \hspace{1.2cm}
\includegraphics[angle=0,width=0.4\linewidth]{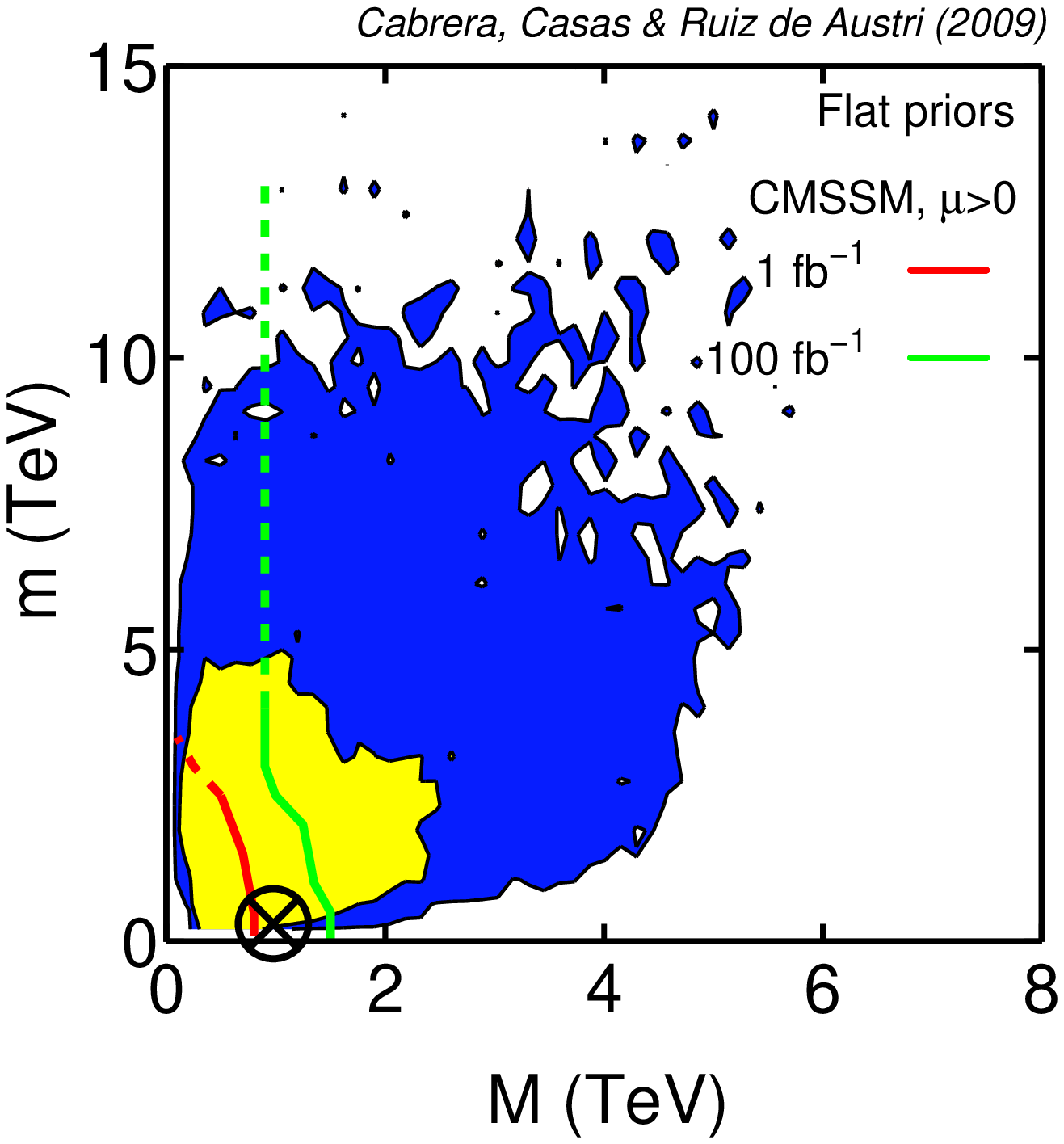} \\ \vspace{1.0cm}
\includegraphics[angle=0,width=0.4\linewidth]{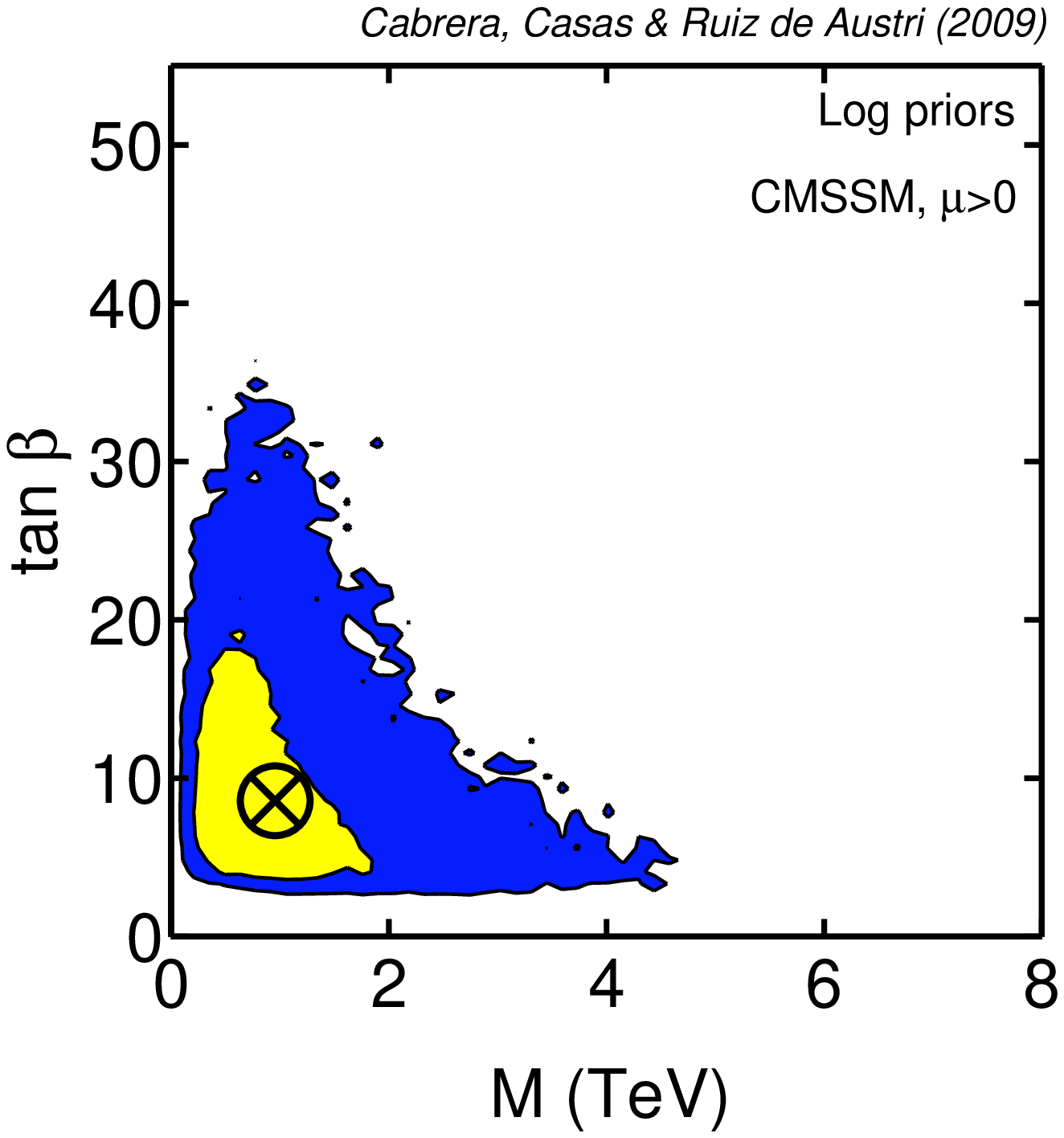} \hspace{1.2cm}
\includegraphics[angle=0,width=0.4\linewidth]{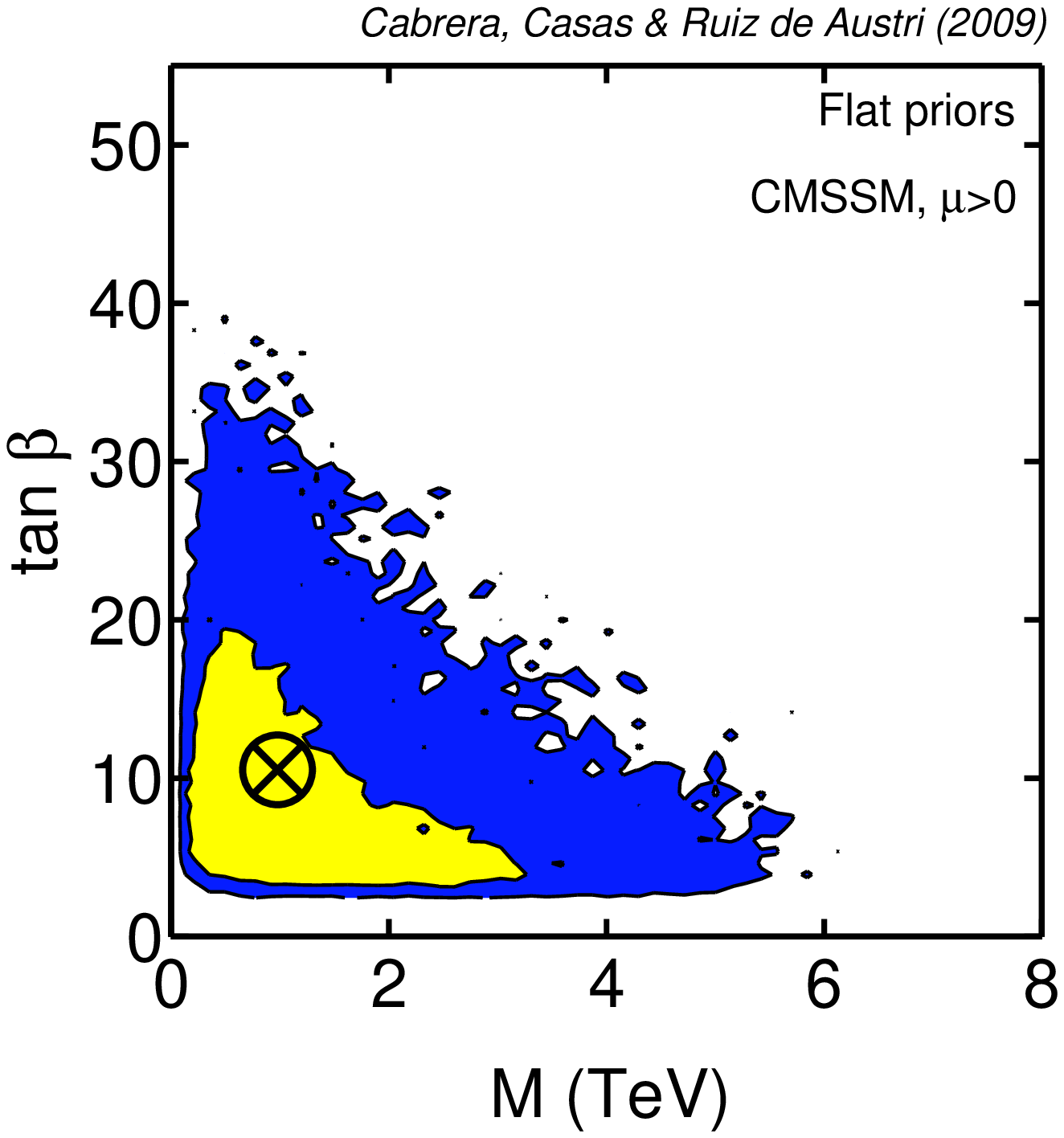}
\caption[test]{2D marginalized posterior probability distribution for 
logarithmic (left panels) and flat (right panels) priors in the 
$\mu>0$ case for a scan including SM nuisance parameters 
constraints, EW breaking ($M_Z^{\rm exp}$), collider limits on Higgs and superpartner masses, and EW and 
B(D)-physics observables. The inner and outer
contours enclose respective 68\% and 95\% joint regions.
The red(green) lines show discovery reach of LHC with 1(100) fb$^{-1}$.  
The cross corresponds to the best-fit point.}
\end{center}
\end{figure}

\subsection{Constraints from $(g-2)_\mu$}

The magnetic anomaly of the muon, $a_\mu= \frac{1}{2}(g-2)_\mu$ has been a classical and powerful test for new physics. At present, the experimental uncertainties in the experimental and theoretical determinations are on the verge of strongly constraining, or even giving a positive signal, of new physics. However, the situation is still somewhat uncertain, due essentially to inconsistencies between alternative determinations of the SM hadronic contribution, more precisely the contribution coming from the hadronic vacuum polarization diagram, say $\delta_{\rm had}^{\rm SM}a_\mu$. 

This contribution can be expressed in terms of the total hadronic cross section $e^+ e^-\rightarrow$ hadrons. Using direct experimental data for this cross section, one obtains a final result for $a_\mu$, which is at more than 3$\sigma$ from the current experimental determination, namely $\delta a_\mu=a_\mu^{\rm exp}-a_\mu^{\rm SM}=29.5\pm8.8\times 10^{-10}$. This has been often claimed as a signal of new physics. Obviously, if one accepts this point of view, the discrepancy should be cured by contributions of new physics, in our case MSSM contributions. The immediate implication is that supersymmetric masses should be brought to quite small values, in order to produce a large enough contribution, $\delta^{\rm MSSM}a_\mu $, to reconcile theory and experiment. Hence, SUSY should live at low-energy (accessible to LHC), mainly because of $a_\mu$. This is an independent argument from the the one based on the size of the EW scale, which has been discussed in sect.3.

The previous statement is quite strong. History has taught us that many experimental observables, in apparent disagreement with the SM prediction, have eventually converged with it. This occurred due to both experimental and theoretical subtleties and difficulties, that sometimes had not been fully understood or taken into account. Although, obviously, $a_\mu$ is a most relevant test for the SM, and hopefully will be a first indication of physics beyond the SM, it is perhaps prudent not taking for granted that this is so indeed. As a matter of fact, the experimental $e^+e^-\rightarrow$ hadrons cross section shows some inconsistencies between different groups of experimental data (see \cite{Davier:2009zi} for a recent account). This is especially notorious if one considers the hadronic $\tau-$decay data, which are theoretically related to the $e^+e^-$ hadronic cross section. Using the $\tau$--data, the 3.3$\sigma$ disagreement becomes 1.8$\sigma$, i.e. one comes back to the SM realm. Although the more direct $e^+e^-$ data are usually preferred to evaluate $a_\mu^{\rm SM}$, this discrepancy is warning us to be cautious about this procedure.

To illustrate this situation, we have performed two alternative analyses. In the first one we use the evaluation of $\delta_{\rm had}^{\rm SM}a_\mu$ based on $e^+e^-$ data. In the second, we use the one based on $\tau$-data. 
We compute $\delta_{\rm had}^{\rm SM}a_\mu$ at full one-loop level adding the 
logarithmic piece of the quantum electro-dynamics two-loop calculation plus  
two-loop contributions from both stop-Higgs and chargino-stop/sbottom 
\cite{Heinemeyer:2004yq}.
The effective two-loop effect due to a shift in the muon Yukawa coupling 
proportional to $\tan^2\beta$ has been added as well \cite{Marchetti:2008hw}.

\vspace{0.2cm}
\noindent {\bf Using $\delta_{\rm had}^{\rm SM}a_\mu$ from $e^+e^-$ data}
\vspace{0.2cm}

In this case, the inclusion of the $a_\mu$ constraint has a dramatic effect, as mentioned above. The preferred values of the soft terms are pushed into the low-energy region. Actually, the push is so strong that the predictions for other observables, in particular $b\rightarrow s\ \gamma$, start to be too large. This tension has been pointed out in ref.\cite{Feroz:2009dv}, and we would like to illustrate it here presenting some representative plots. Fig.6 shows the (non-normalized) pdf for the $m-$parameter in three different cases (taking always a logarithmic prior): a) using EW + Bounds + B-physics, as in subsect 4.1 (blue solid line); b) using EW + Bounds + $a_\mu$ (red dashed line); c) using EW + Bounds + B-physics + $a_\mu$ (green dashed-dotted line).

\begin{figure}[t]
\begin{center}
\label{}
\includegraphics[angle=0,width=0.4\linewidth]{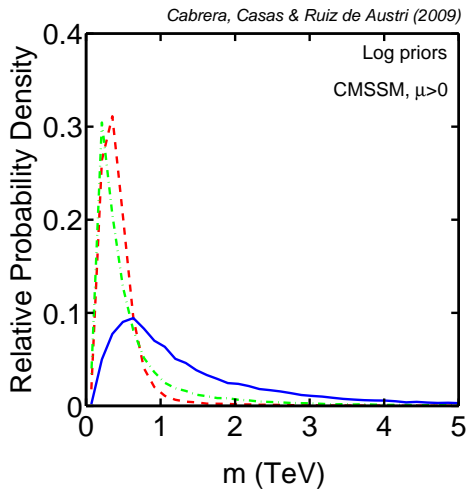} 
\caption{Non-normalized 1D marginalized posterior probability distribution 
of the $m$ parameter for logarithmic prior and $\mu>0$ including SM 
nuisance parameters constraints, EW breaking ($M_Z^{\rm exp}$), EW observables, collider limits on Higgs and superpartner 
masses: + B(D)-physics observables (blue solid line); +  $a_\mu$ (red dashed line); 
+ B(D)-physics and  $a_\mu$ (green dashed-dotted line).} 
\end{center}
\end{figure}

Clearly, the effect of just $a_\mu$ is to bring the preferred region for the soft terms from $\sim 1$ TeV to $\sim 300$ GeV. This effect is remarkably stable against variations of the type of prior, indicating that the data are now powerful enough to essentially select a region of the parameter space. Let us also mention that large values of $\tan\beta$ become now much more likely, being normally associated to the region of larger soft masses (recall that $\delta^{\rm MSSM} a_\mu $ grows with decreasing masses and increasing $\tan\beta$). When both $b\rightarrow s, \gamma$ and $a_\mu$ are taken into account, there is almost no region of the parameter space able to reproduce both experimental results within $2\sigma$. Therefore the likelihood factor gets suppressed, and the ``preferred" region of the parameter space (illustrated here by the green line) is somehow an average of the two previous cases. This tension between $b\rightarrow s, \gamma$ and $a_\mu$  can also be noticed by looking at Fig.7, where the left and right panels show the pdfs of $BR(B\rightarrow X_s \gamma)$ and $\delta^{\rm MSSM} a_\mu$ respectively, with  the same code for the lines as in Fig. 6. Besides, we have include a gaussian in each panel (solid black line), proportional to the likelihood, and thus centered at the experimental value with the experimental uncertainty. Comparing the position of the bulk of the probability distribution with the likelihood, it is clear that the most favourable cases are not really satisfactory reproducing the two measurements simultaneously, even though we have not attempted to quantify this tension in a rigorous way.

\begin{figure}[t]
\begin{center}
\label{}
\includegraphics[angle=0,width=0.4\linewidth]{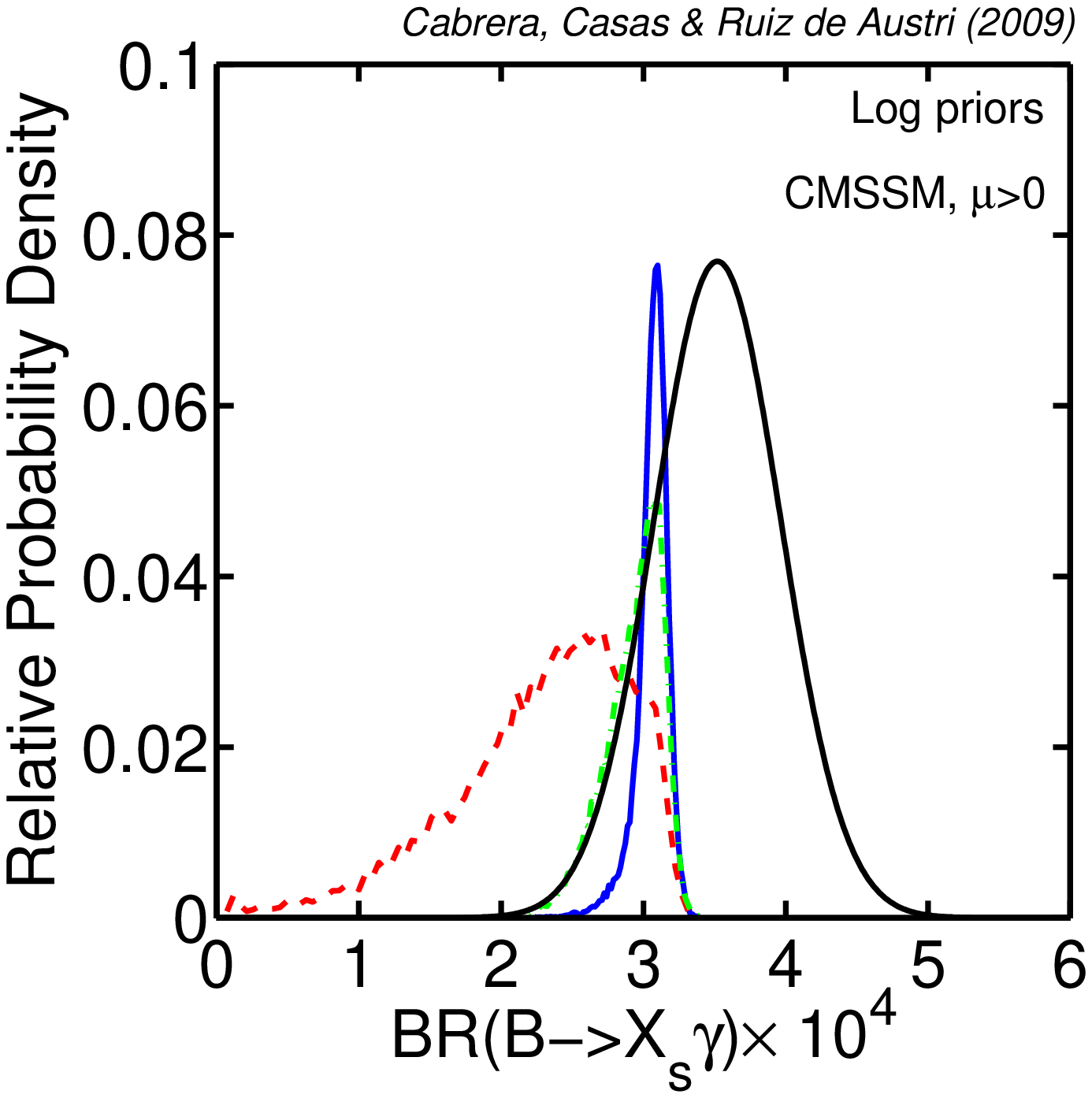} \hspace{1.2cm}
\includegraphics[angle=0,width=0.4\linewidth]{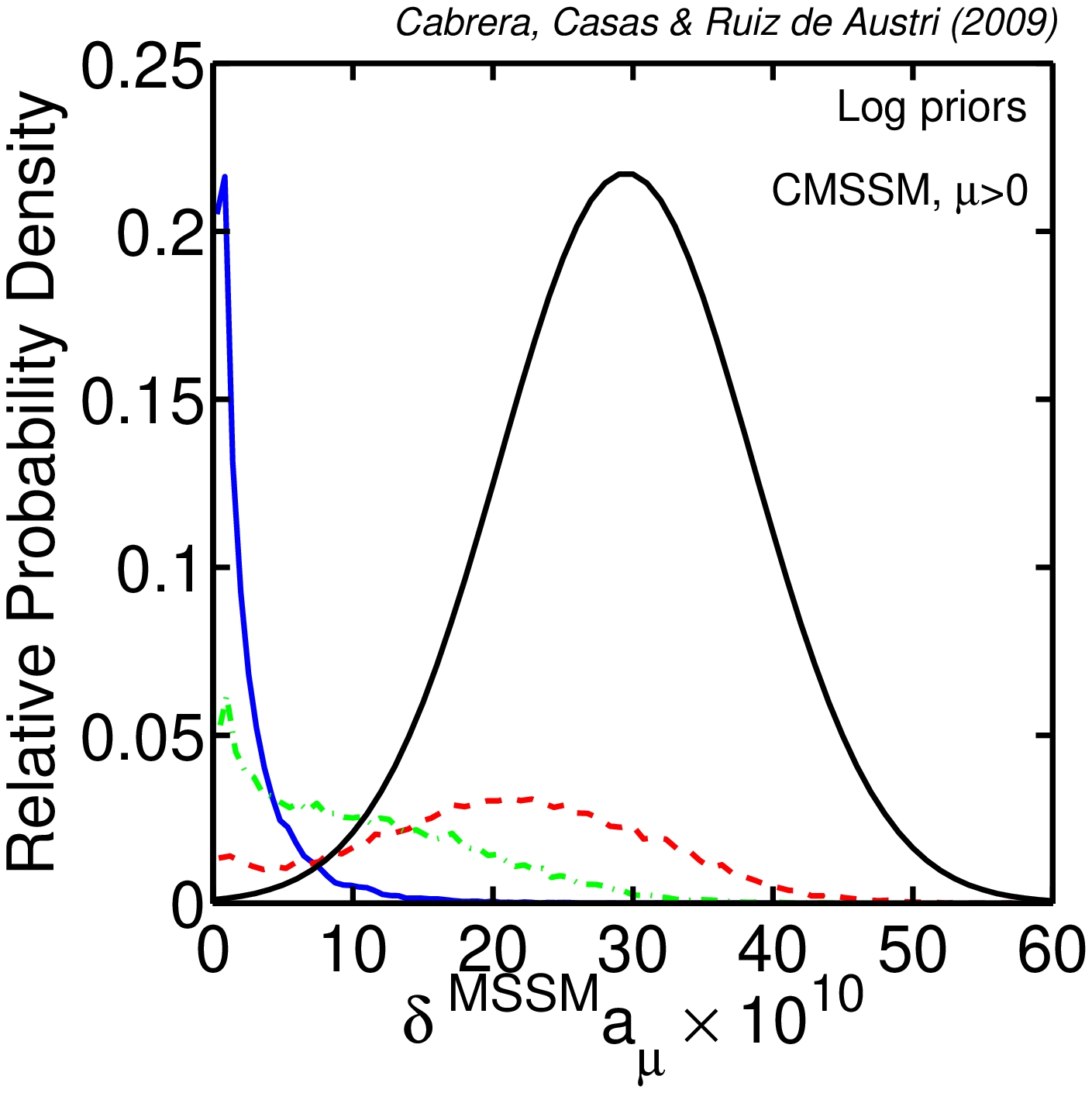} 
\caption{Non-normalized 1D marginalized posterior probability distribution for $BR(B\rightarrow X_s \gamma)$ (left panel) and for $\delta^{\rm MSSM} a_\mu$ (right panel).The code for the lines is as in Fig. 6. Besides, the black (solid) gaussians represent the experimental likelihood.}
\end{center}
\end{figure}

Let us also remark that, if the Higgs mass turns out to be ${\cal O}$(10) GeV above the present experimental limit, the tension between the Higgs mass and $a^{\rm exp}_\mu$ would be dramatic and could not be reconciled: $m_h$ ($a^{\rm exp}_\mu$) would require too large (small) soft masses, see the discussion in subsect. 4.1.

Fig.8 shows the probability distribution in the $\{M,m\}$ and $\{\tan\beta,M\}$ planes, as in Fig.5, once the $a_\mu$ constraint (based on $e^+e^-$ data) is included. Comparison with Fig. 5 clearly shows the big push of the soft terms into the low-energy region. Actually, most of the probability falls now within the LHC reach (even in the short term), which is great news for the potential discovery of SUSY (if the $a_\mu$ discrepancy is really there).

\begin{figure}[t]
\begin{center}
\label{}
\includegraphics[angle=0,width=0.4\linewidth]{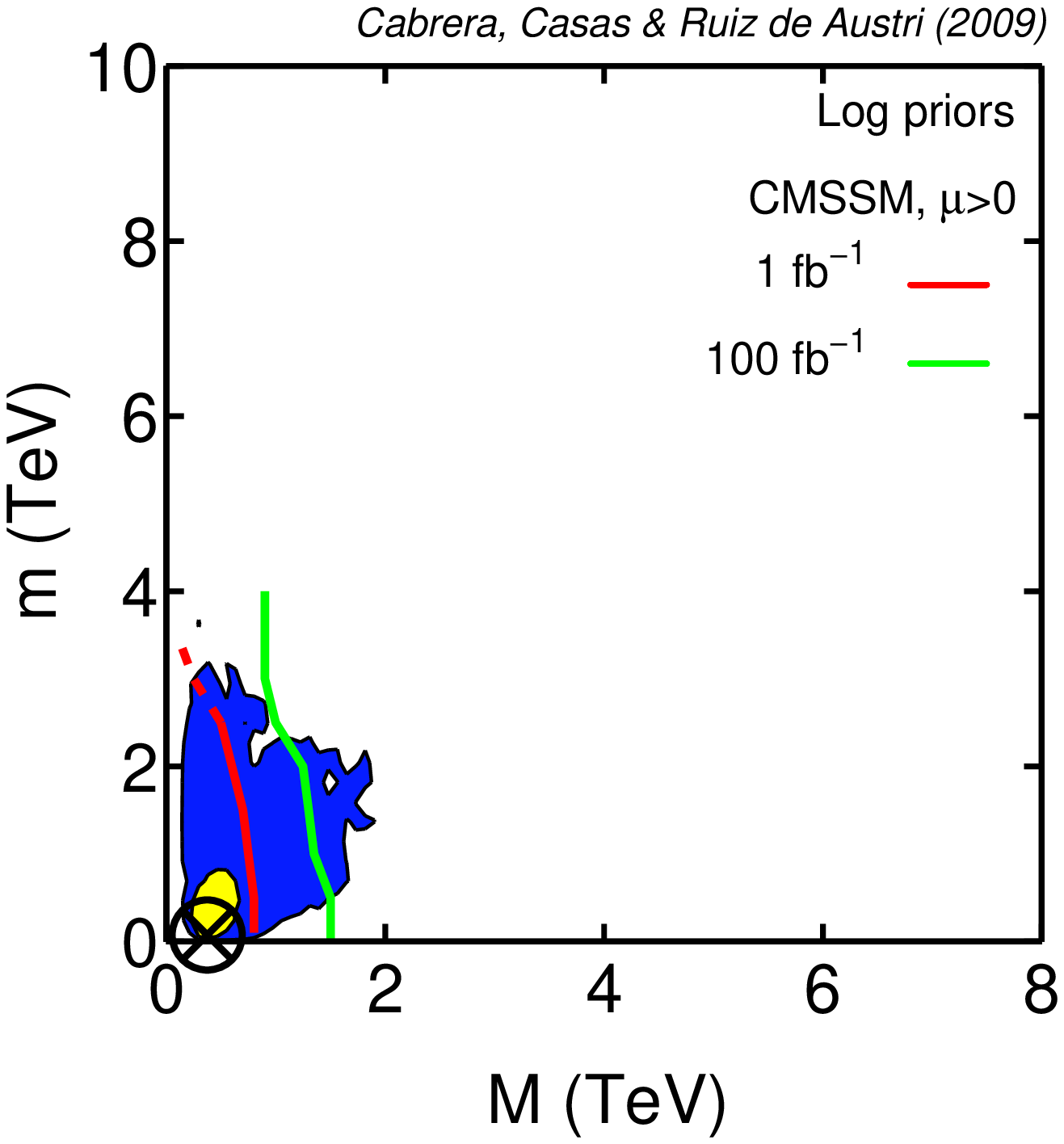} \hspace{1.2cm}
\includegraphics[angle=0,width=0.4\linewidth]{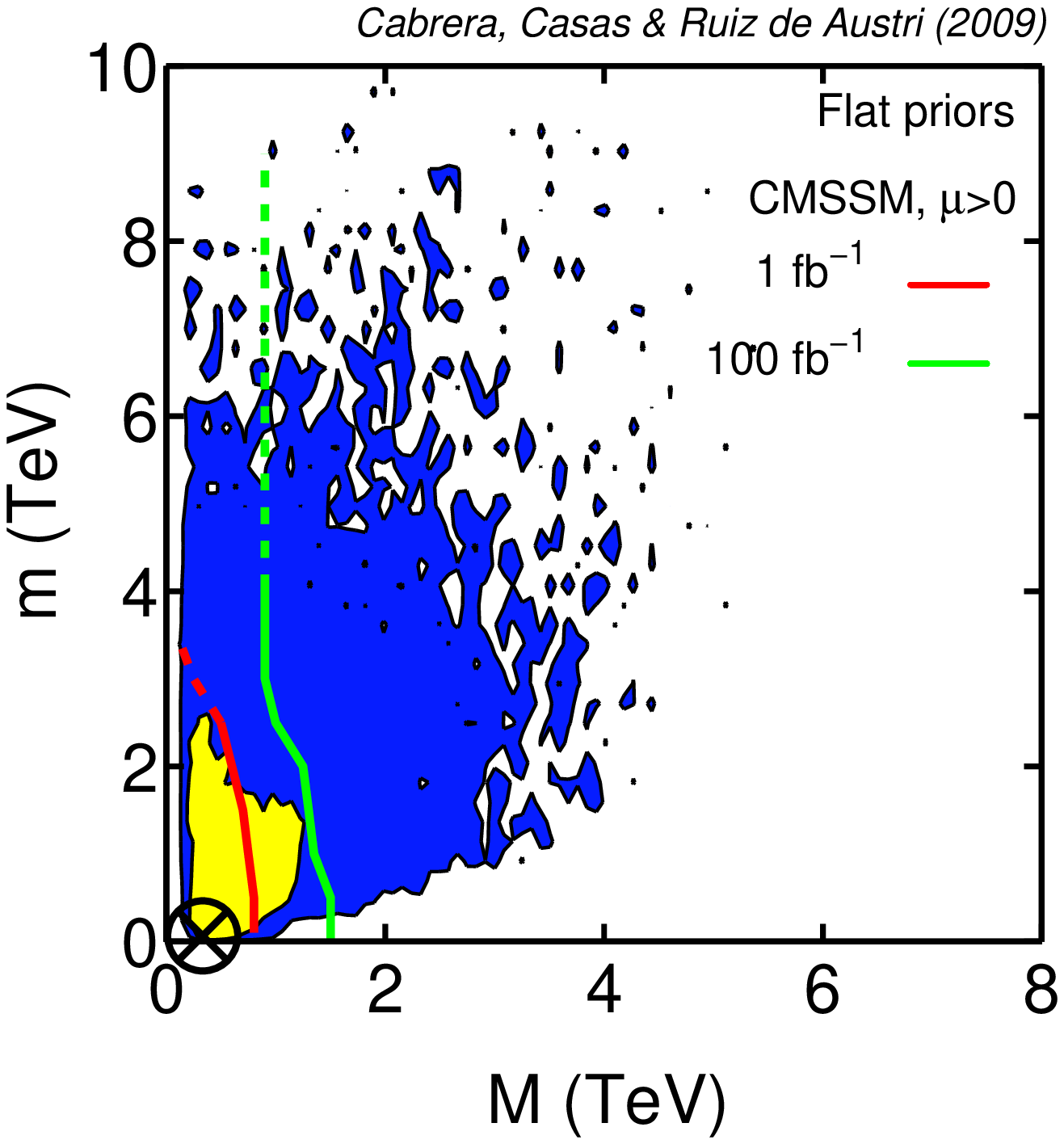} \\ \vspace{1.0cm}
\includegraphics[angle=0,width=0.4\linewidth]{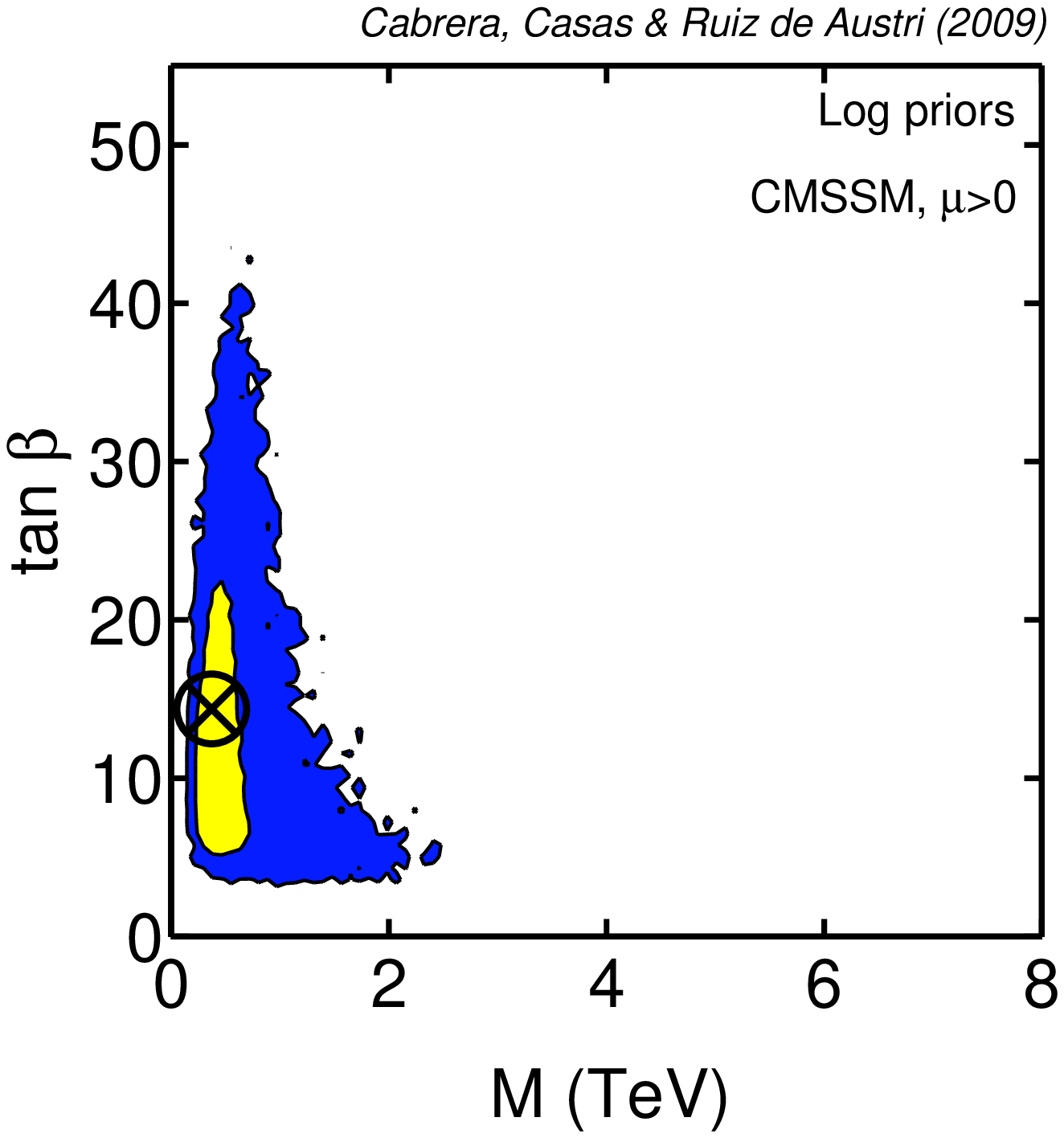} \hspace{1.2cm}
\includegraphics[angle=0,width=0.4\linewidth]{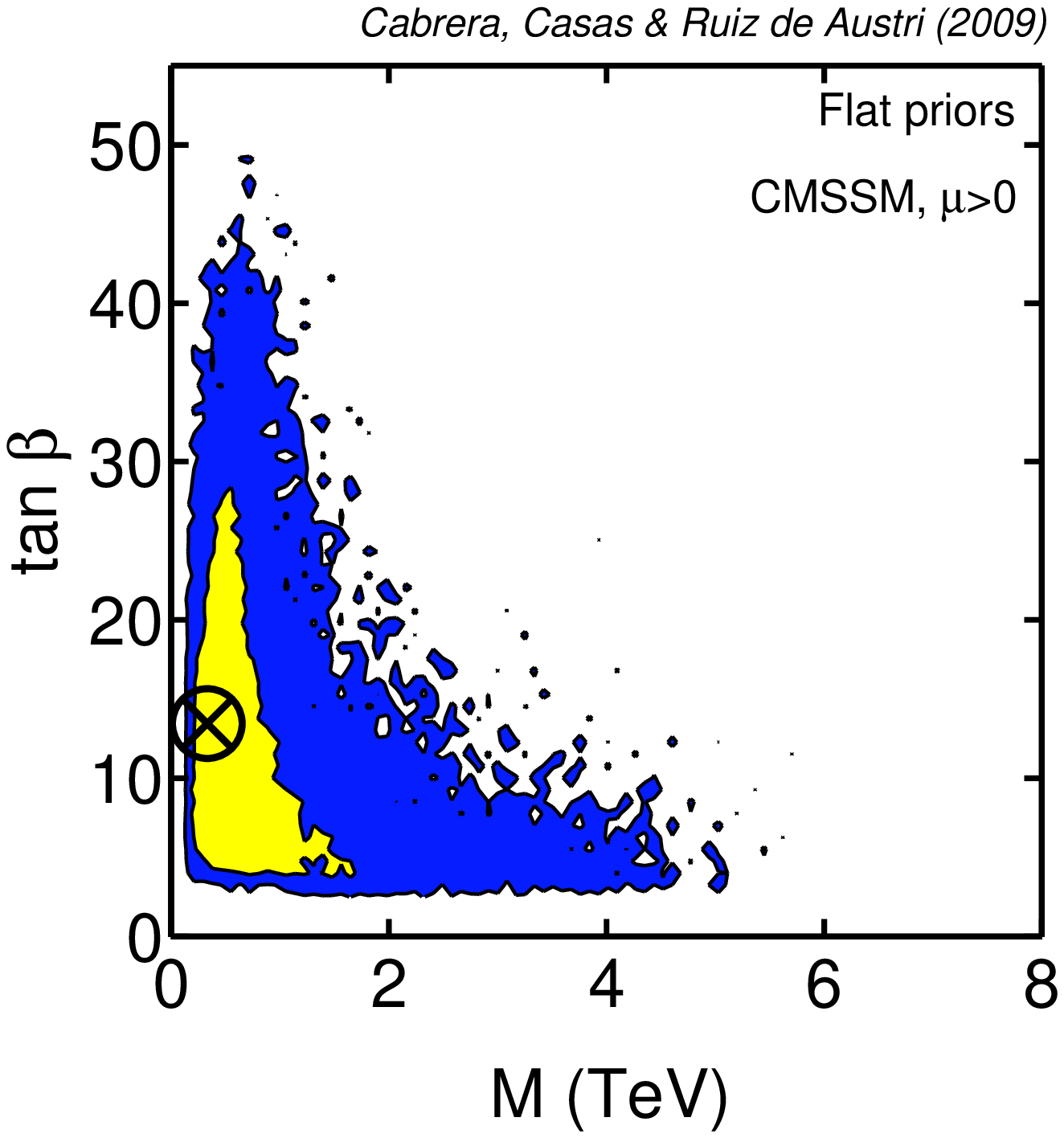}
\caption{As in Fig. 5 but with the additional constraint from $a_\mu$, based on $e^+e^-$ data.}
\end{center}
\end{figure}

\vspace{0.2cm}
\noindent {\bf Using $\delta_{\rm had}^{\rm SM}a_\mu$ from $\tau$ data}
\vspace{0.2cm}

In this case, there is no big discrepancy between $a_\mu^{\rm SM}$ and $a_\mu^{\rm exp}$, so $\delta^{\rm MSSM} a_\mu $ does not need to be large. Consequently, the probability distributions are essentially unchanged by the inclusion of the $a_\mu$ constraint, and are very similar to those shown in subsect. 4.1 (Figs. 2--5).

Consequently, if $a_\mu$ is not a signal of new physics, the size of EW breaking continues to be the only piece of data that brings SUSY to scales accessible to LHC (apart from Dark Matter considerations, which we examine next).

\subsection{Constraints from Dark Matter}

There are different astrophysical and cosmological observations that offer impressive evidence of the existence of Dark Matter (DM) in the universe (see Table 2 for a recent determination of $\Omega_{\rm DM}$). On the other hand, the consistency with the observed large structure of the universe favours cold dark matter (CDM), i.e. non-relativistic matter at the beginning of galaxy formation. This leads to the hypothesis of a weakly interacting massive particle (WIMP) as the component of CDM. 

Supersymmetry offers a good candidate for such a WIMP, namely the LSP, which is stable in the standard (R-parity conserving) SUSY formulations (for a review see \cite{Jungman:1995df}). Although, depending on the models, there are several possibilities for the SUSY WIMP, the most popular and natural candidate is the lightest neutralino, $\chi^0$, which is the LSP in most of the CMSSM parameter space. However the calculations show that typically too many neutralinos are produced after inflation. Therefore some efficient annihilation mechanism is required in order to bring $\Omega_{\rm DM}$ down to the allowed range.
In the context of CMSSM there are four such mechanisms known, which take place in four different regions of the parameter space:
\vspace{0.2cm}

\noindent
{\em Bulk region:} Neutralinos can be annihilated (into leptons) via sleptons if the masses of the latter are not high. This requires rather small $m$ and $M$ soft parameters, in potential conflict with the Higgs mass bound.
\vspace{0.2cm}

\noindent
{\em Focus Point region:} For moderate or large values of $\tan{\beta}$ the electroweak
scale is quite insensitive to the variation of $m$. For large enough values of $m$, 
the $\mu$ parameter decreases, which drives the LSP to get a significant Higgsino-component, making its annihilations (into vector bosons) more efficient. 
\vspace{0.2cm}

\noindent
{\em Co-annihilation region:} If the mass of the second lightest supersymmetric particle (NLSP) is close to that of the LSP, the annihilation of the latter is enhanced through co-annihilation processes. In the CMSSM this mechanism takes place typically with an stau NLSP. In the parameter space this corresponds to a
rather narrow region with $M>m$.
\vspace{0.2cm}

\noindent
{\em Higgs funnel region:} When the mass of the pseudoscalar $A^0$-boson becomes close to twice the neutralino LSP and $\tan\beta$ is large, the annihilation occurs quite efficiently through the $A^0$ resonance.

\begin{figure}[t]
\begin{center}
\label{}
\includegraphics[angle=0,width=0.4\linewidth]{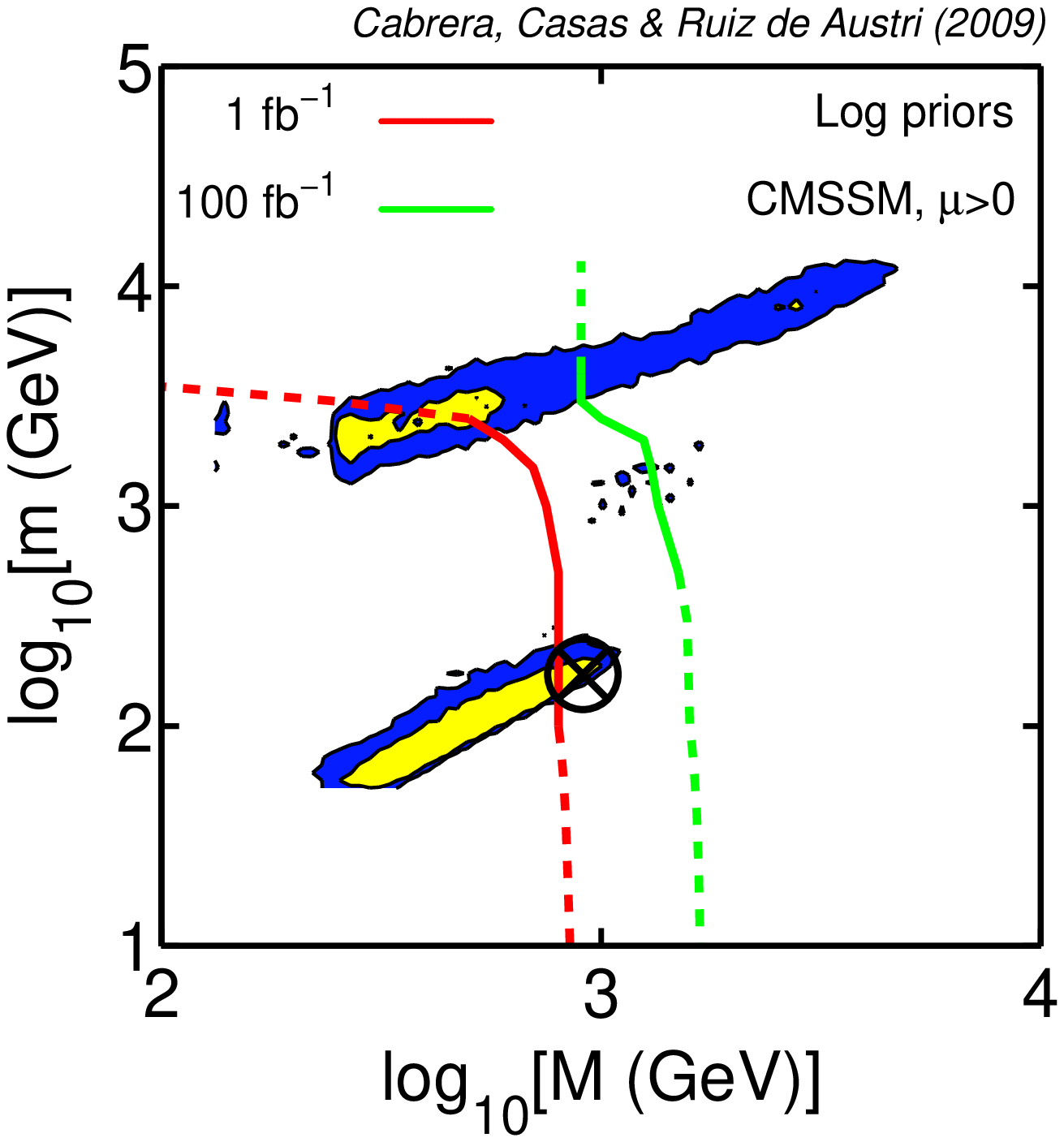} \hspace{1.2cm}
\includegraphics[angle=0,width=0.4\linewidth]{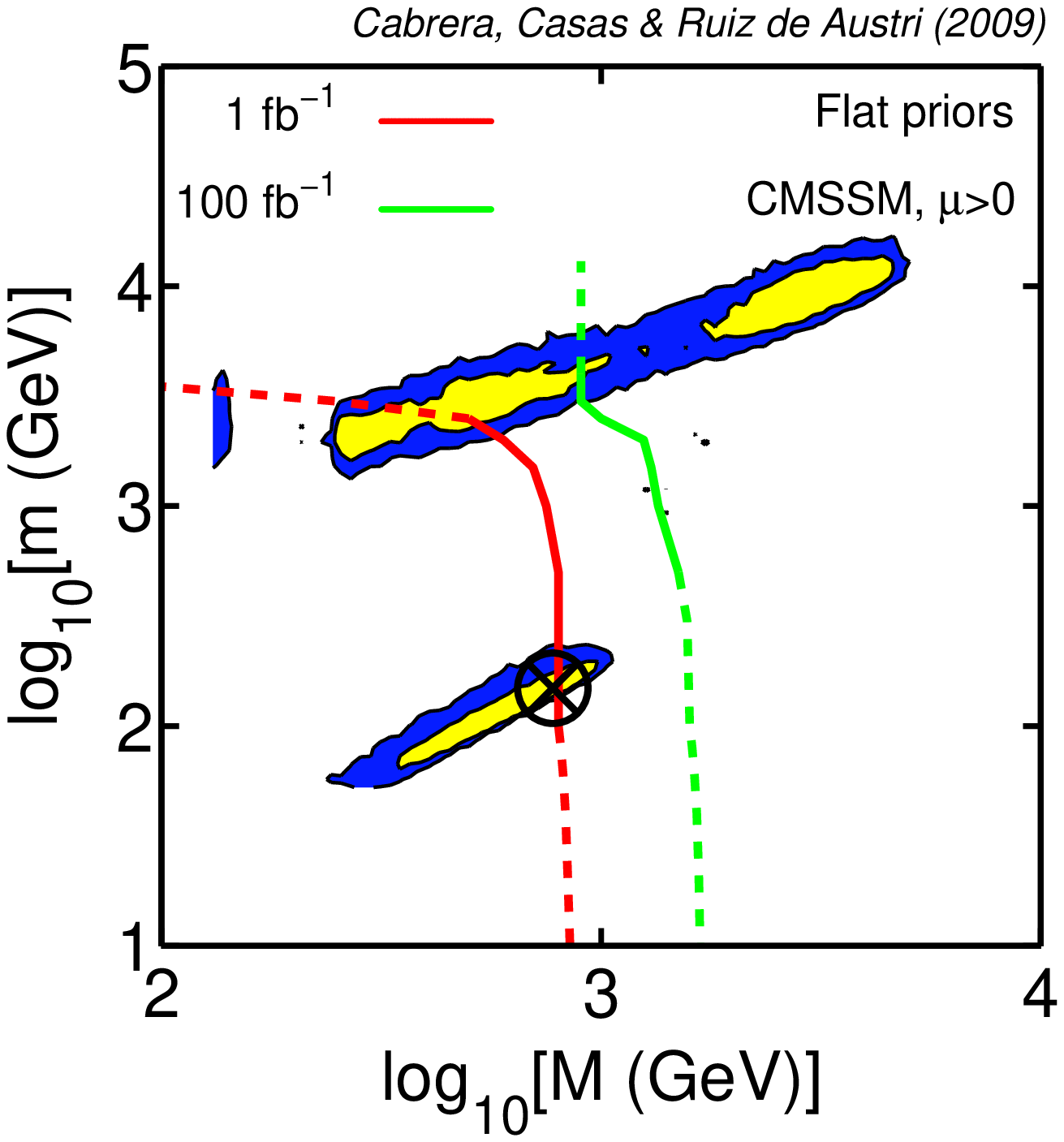} \\ \vspace{1.0cm}
\includegraphics[angle=0,width=0.4\linewidth]{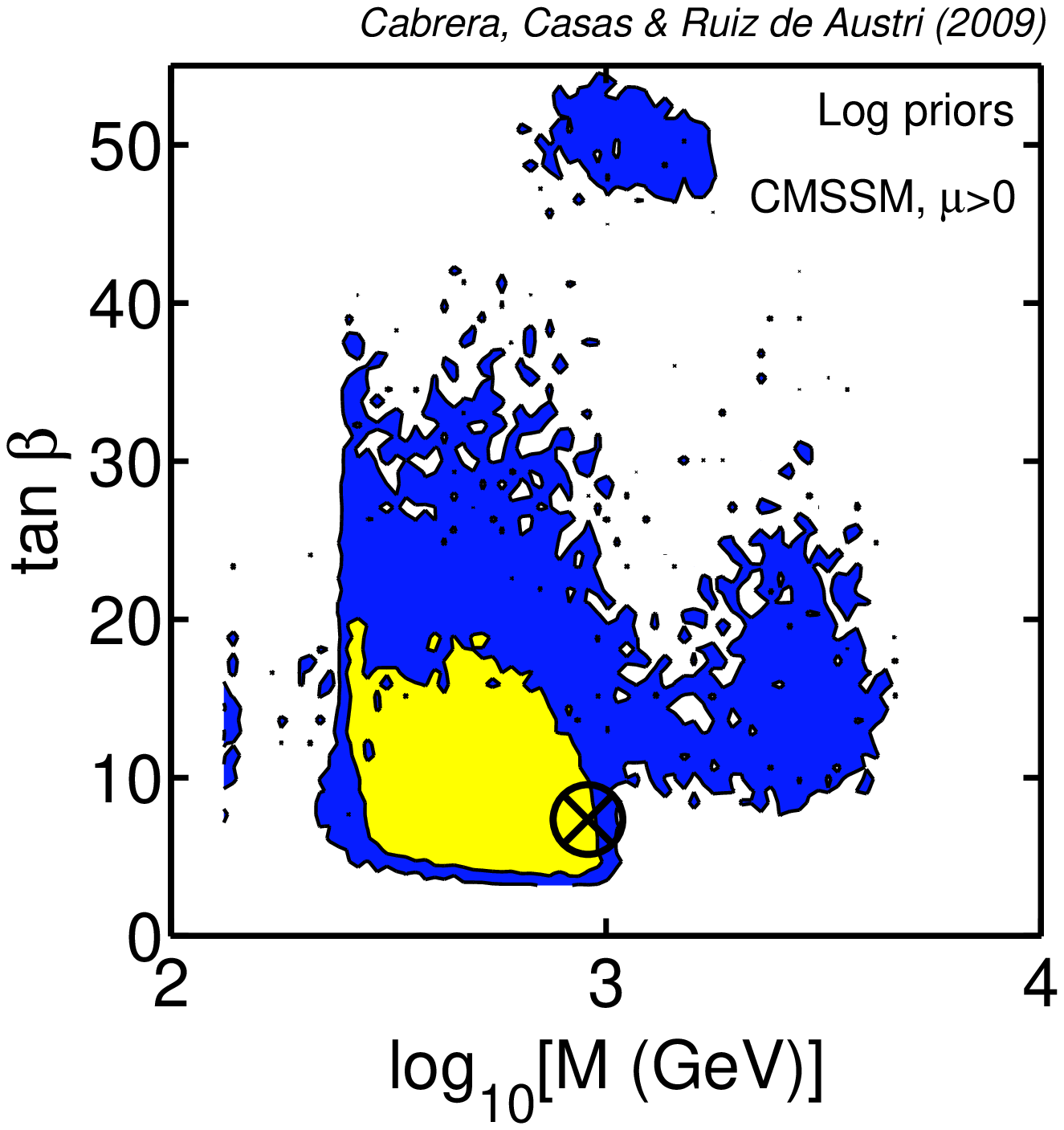} \hspace{1.2cm}
\includegraphics[angle=0,width=0.4\linewidth]{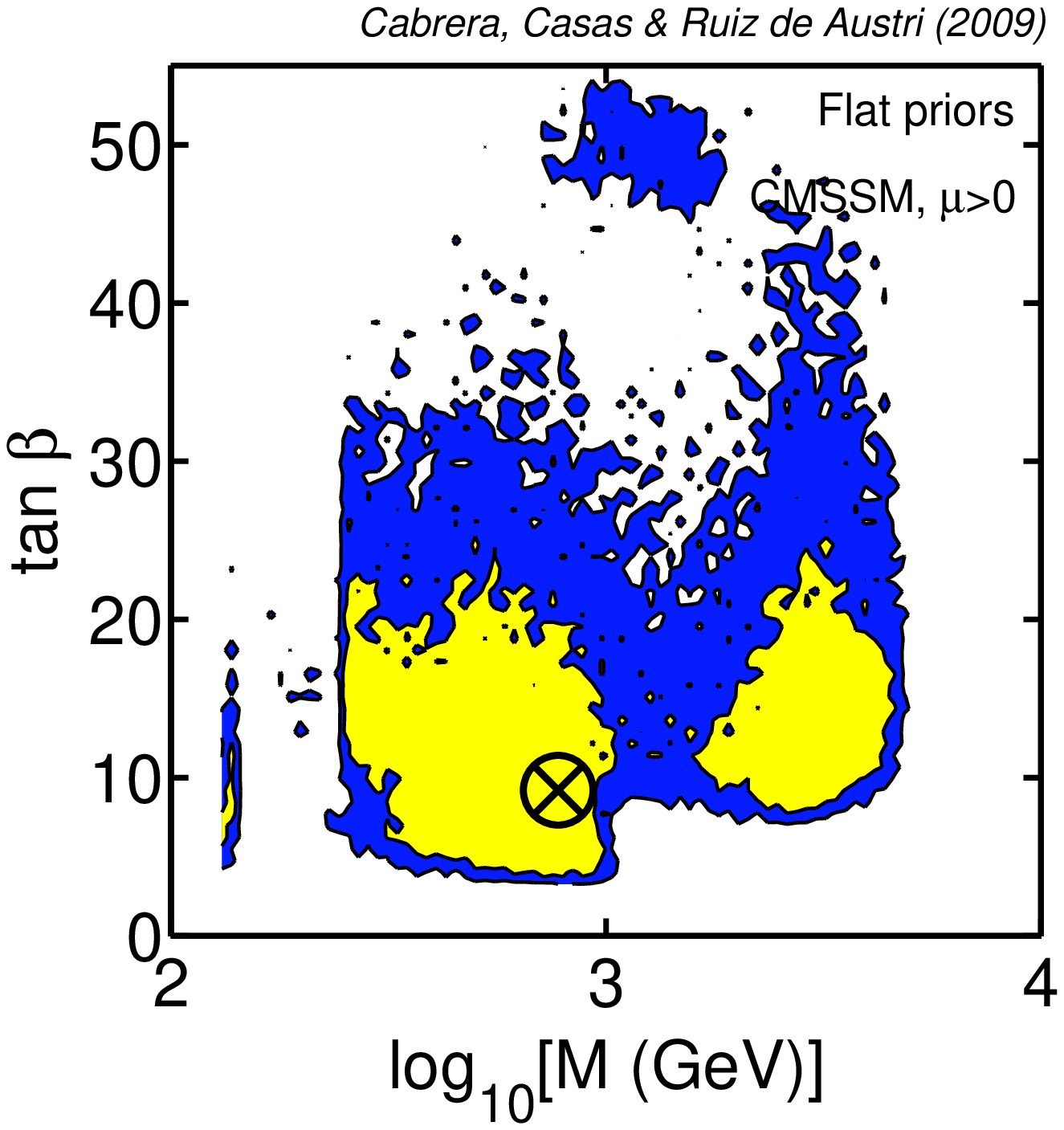}
\caption[text]{As in Fig. 5 but with an additional constraint from the WMAP CDM abundance. }
\end{center}
\end{figure}

\vspace{0.2cm}
\noindent
In order to evaluate the viability of supersymmetric CDM in each point of the CMSSM parameter space, we use the \texttt{MicrOMEGAs} code \cite{micromegas} integrated into \texttt{SuperBayes}. The corresponding likelihood, assuming that all the CDM is made up of neutralinos, is then incorporated to the pdf in the Bayesian scan.
Fig.9 shows the resulting probability distribution in the $\{M,m\}$ and $\{M, \tan\beta\}$ planes (i.e. when all the parameters but two are marginalized) for logarithmic and flat priors. In these figures we have {\em not} included the information about $a_\mu$. The $\{M,m\}$--plane plots show a kind of blurring with respect to usual plots in the literature, due to the integration in the variables $A, \tan\beta$. Still, the above-mentioned four viable regions are visible in Fig.9. On the other hand, the $\{M, \tan\beta\}$ plots show two big preferred regions. The largest one ocurrs at $M < 1$ TeV and contains (mixed) the Co-annihilation, Bulk and part of the Focus Point regions. The second one occurs at $M > 1$ TeV and corresponds to the part of the Focus Point region that needs moderate to large values of $\tan\beta$. Besides, the 
very small island around $M=200$ GeV (also visible in the $\{M, m\}$ plots) corresponds also to the Focus Point region. Finally, the Higgs funnel region, which becomes significant for very large values of $\tan\beta$, is located around $\tan\beta = 50$. Let us remark that, since 
some of the previous regions require large $\tan\beta$, the latter becomes more probable than before-including CDM constraints.

Although the favoured regions are qualitatively similar for logarithmic and flat priors, quantitatively the area of highest probability is extended into larger (even inaccessible to LHC) soft masses in the case of flat prior. This is because the DM constraints, though quite severe, do not select a unique region of the parameter space but several ones, located in different zones of the CMSSM parameter space, as discussed above. Consequently, the prior assumed for the parameter space plays a relevant role when comparing the relative probability of these regions.

Regarding the impact on the LHC potential of discovery, roughly speaking, including DM constraints the low-energy gets favoured and therefore the detection of SUSY at the LHC, as can be seen by comparing Figs. 5 and 9. However, there survive large (though less probable, especially for log prior) high-energy areas out of the LHC reach. Consequently, again, if we are unlucky, even if DM is supersymmetric, it could escape LHC detection (especially if the Higgs mass is not close to its experimental limit).

In any case, again, one should be cautious at interpreting these results as a robust constraint on the CMSSM. Certainly, they are so with an ``standard" cosmology. However, it could happen that other regions of the MSSM parameter space are cosmologically viable if, e.g. the overproduction of CDM is diluted by electroweak baryogenesis. Admittedly, the latter is not a most natural or popular scenario of inflation, but mechanisms for it have been explored \cite{Knox:1992iy}. Alternatively, the LSP could be unstable assuming tiny violations of R-parity, see e.g. \cite{Ibarra:2008jk}. In these cases the observed dark matter should be provided by other candidate, e.g. an axion. But this is not a drawback for the model.
Of course, CDM constraints are extremely interesting and they have to be taken into account. But it seems sensible not to put them at the same level as e.g. electroweak observables.

\begin{figure}[t]
\begin{center}
\label{}
\includegraphics[angle=0,width=0.4\linewidth]{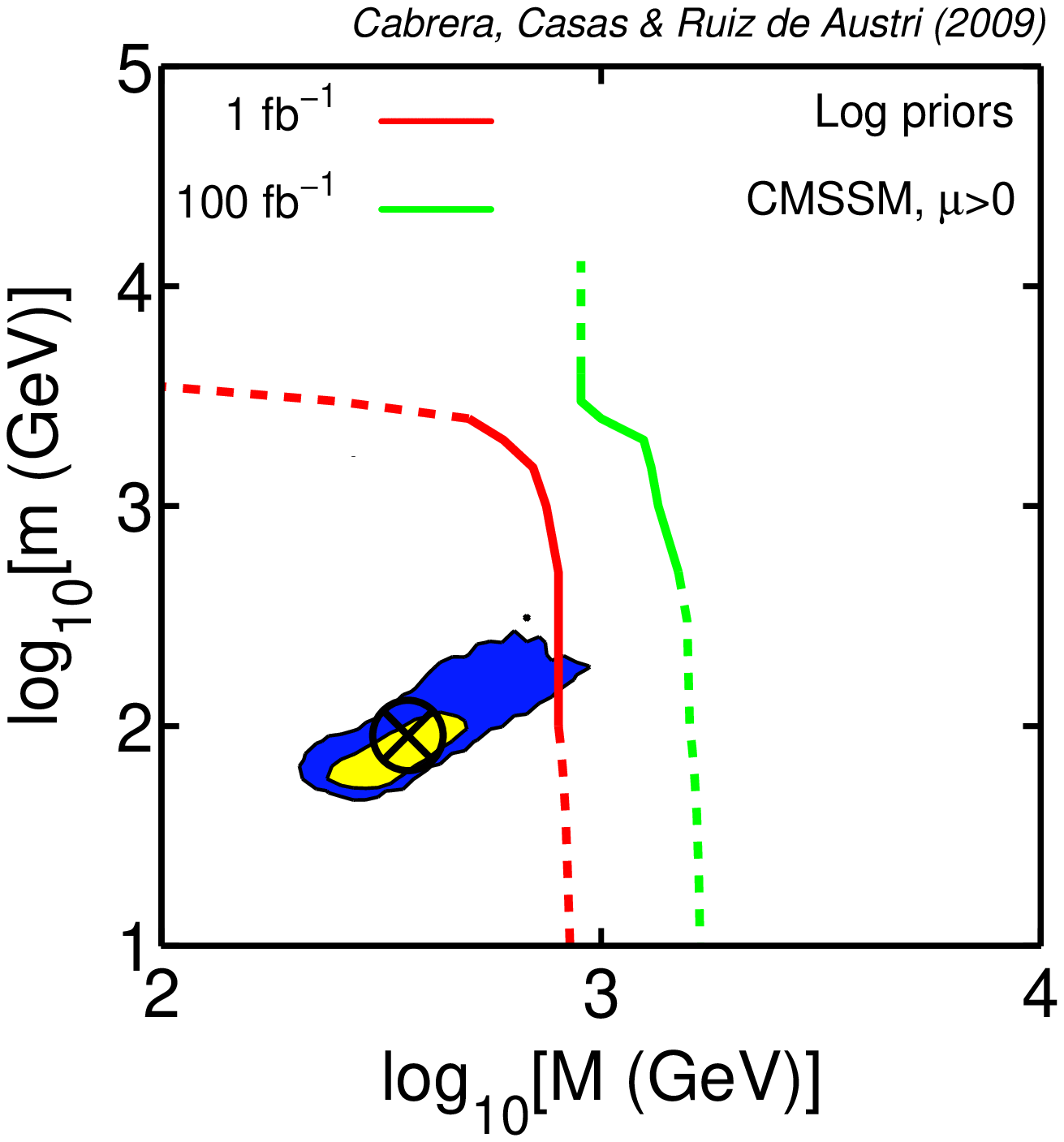} \hspace{1.2cm}
\includegraphics[angle=0,width=0.4\linewidth]{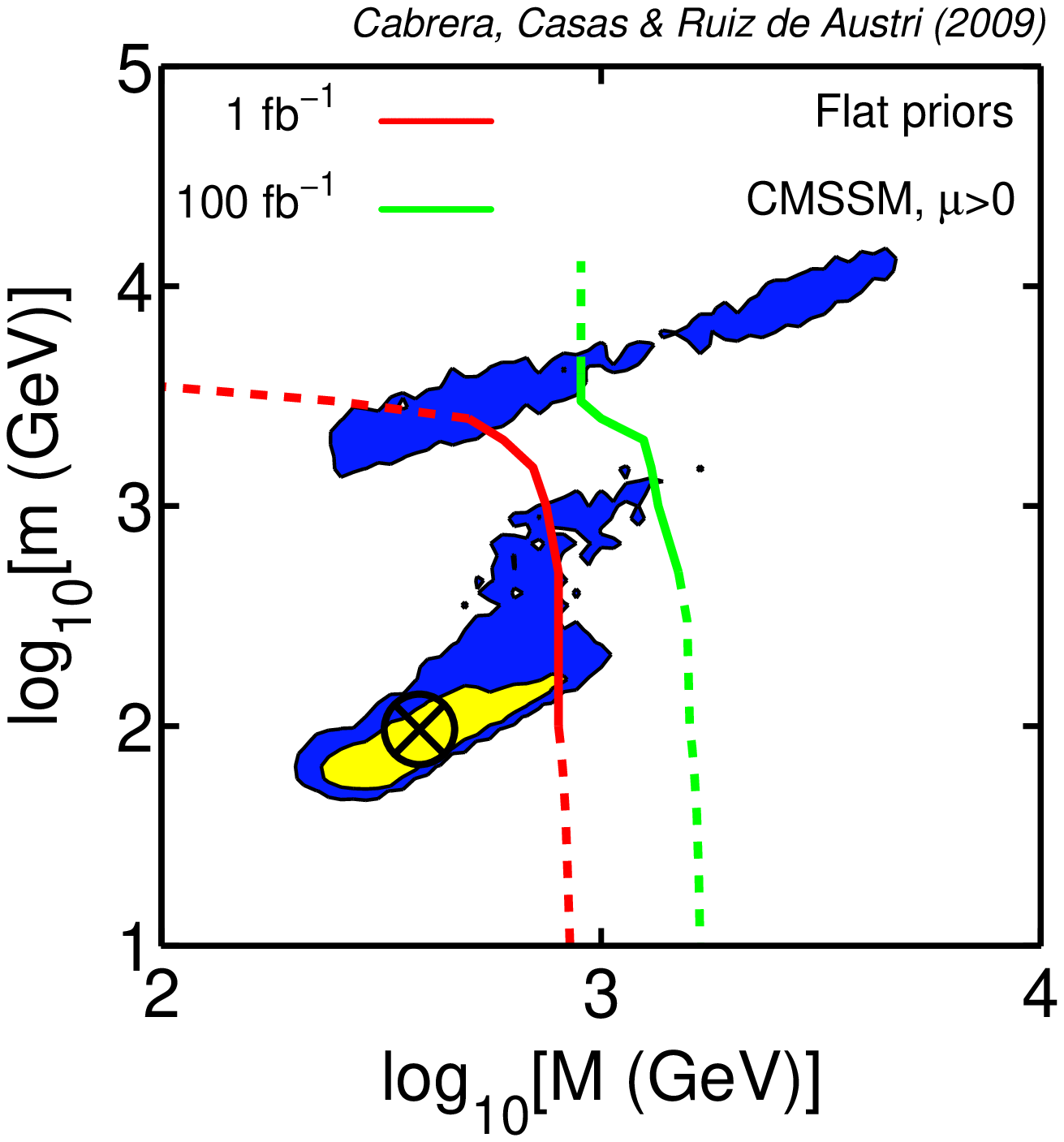} \\ \vspace{1.0cm}
\includegraphics[angle=0,width=0.4\linewidth]{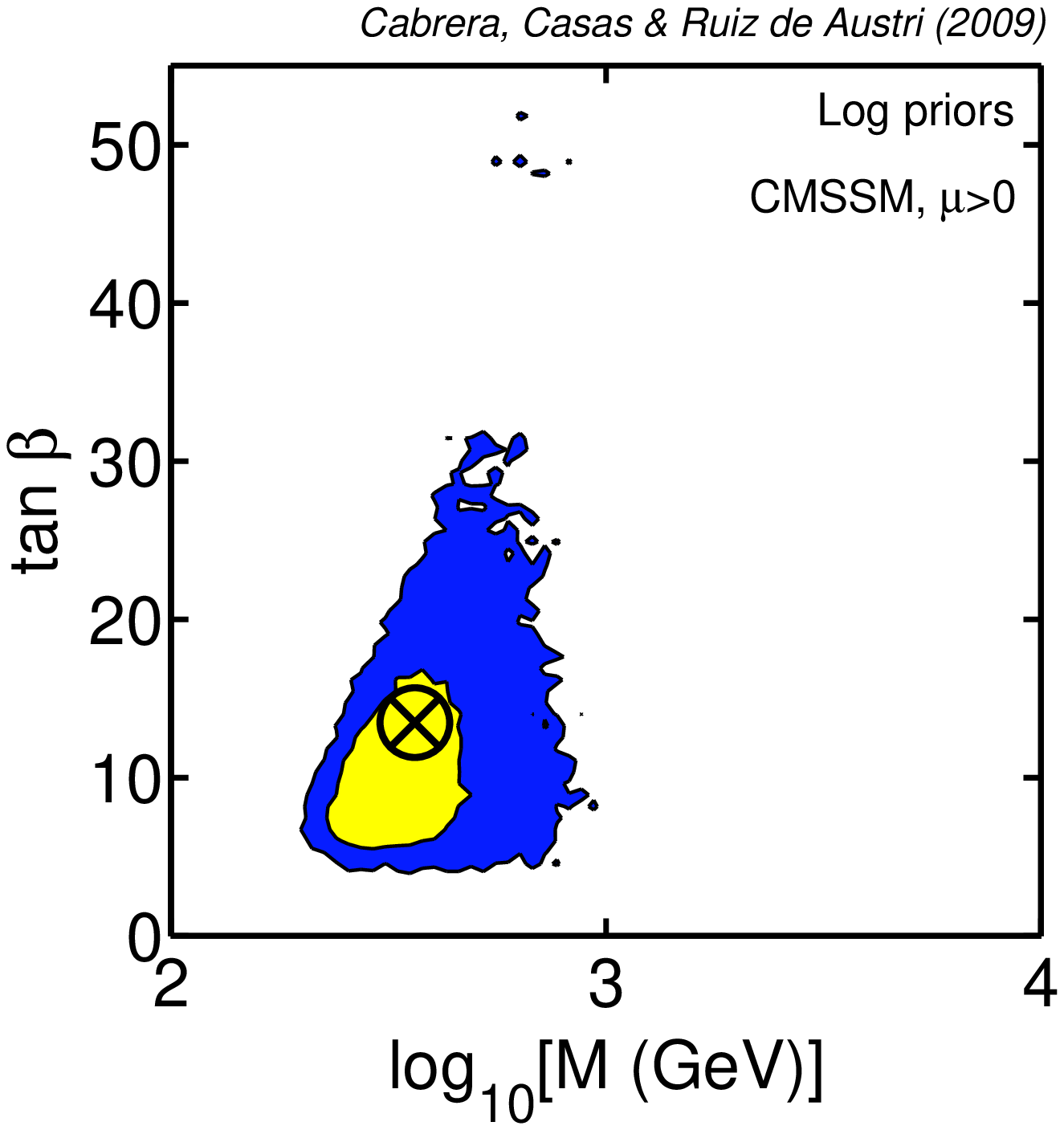} \hspace{1.2cm}
\includegraphics[angle=0,width=0.4\linewidth]{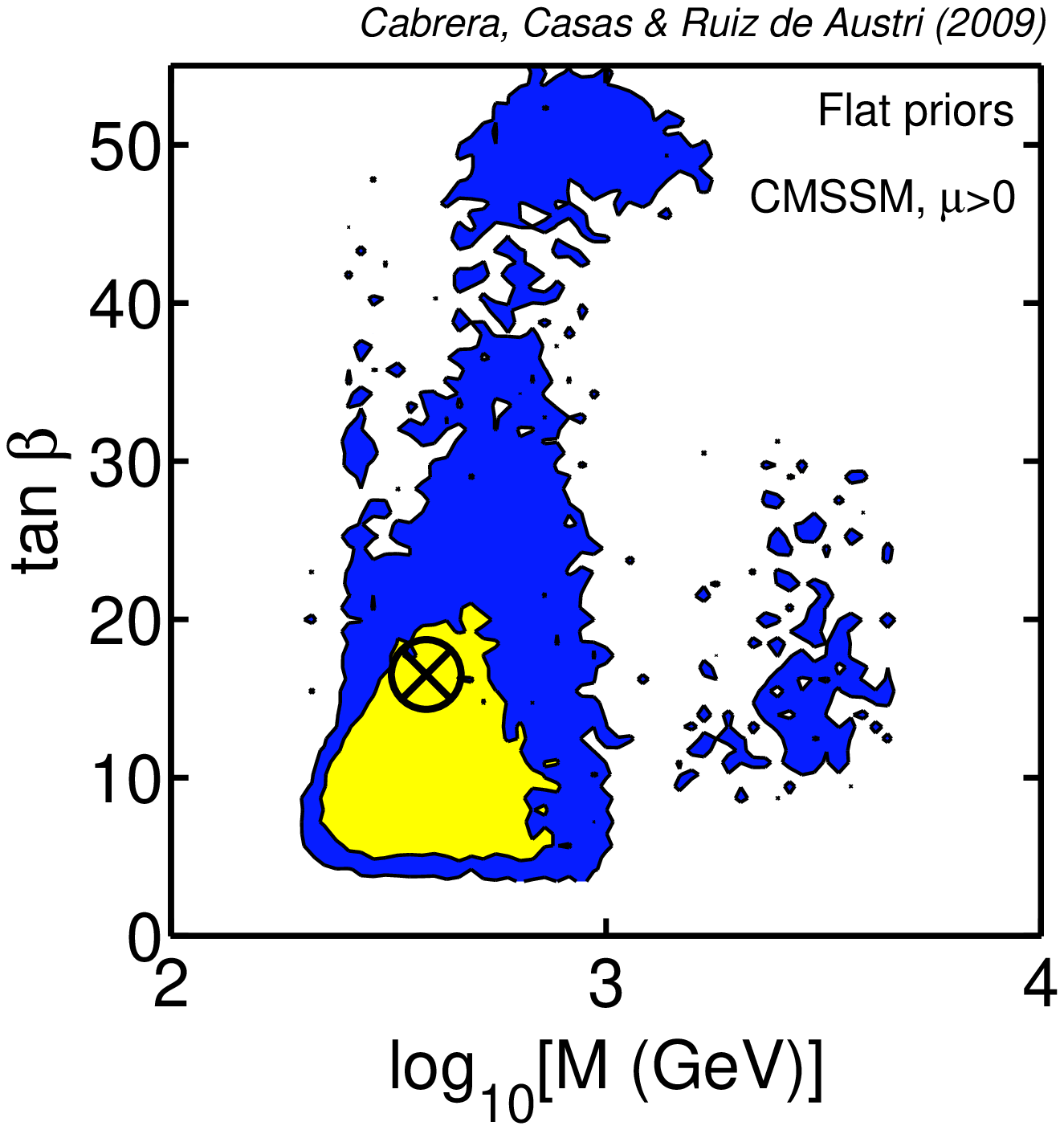}
\caption{As in Fig. 5 but with an additional constraint from the WMAP CDM abundance.}
\end{center}
\end{figure}

Finally, Fig.~10 shows the $\{M,m\}$--plane plots when the $a_\mu$ constraint (based on $e^+e^-$ data) is incorporated to the analysis as well. Clearly, the regions with ``too large" soft masses (to reproduce the $a_\mu^{\rm exp}$) are now suppressed, leaving a quite definite region at low-energy. More precisely, the bulk and co-annihilation regions are now clearly selected amongst the various possibilities to obtain $\Omega_{DM}$.
We stress, however, that in this case one should be cautious about both the $\Omega_{DM}$ and the $a_\mu$ constraints. Note in particular that, if the $a_\mu$ constraint is based on $\tau$ data, it does not produce relevant restrictions and, consequently,
the corresponding plots are quite similar to those of Fig. 9.

\section{Negative sign of $\mu$}

So far all the results and plots presented correspond to $\mu>0$. The analysis 
for $\mu<0$ is completely similar. The most worth-mentioning difference is that with $\mu<0$ the MSSM contributions to $a_\mu$ have negative sign and thus become useless to reconcile theory and experiment (a discrepancy that is only present {\em if} $\delta_{\rm had}^{\rm SM}a_\mu$ is evaluated using $e^+ e^- \rightarrow $ had data). On the other hand, the contributions to $b\rightarrow s, \gamma$ have now positive sign, which is the ``right" sign to push the theoretical result closer to the experimental value (see Table 1). This effect, however, has less impact than $a_\mu$ in the distribution of probability.

\subsection{Results}

The results for $\mu<0$ are summarized in Figs. 11, 12, 13 and 14, which are as previous Figs. 5,8,9 and 10, but with opposite sign of $\mu$.

\begin{figure}[t]
\begin{center}
\label{}
\includegraphics[angle=0,width=0.4\linewidth]{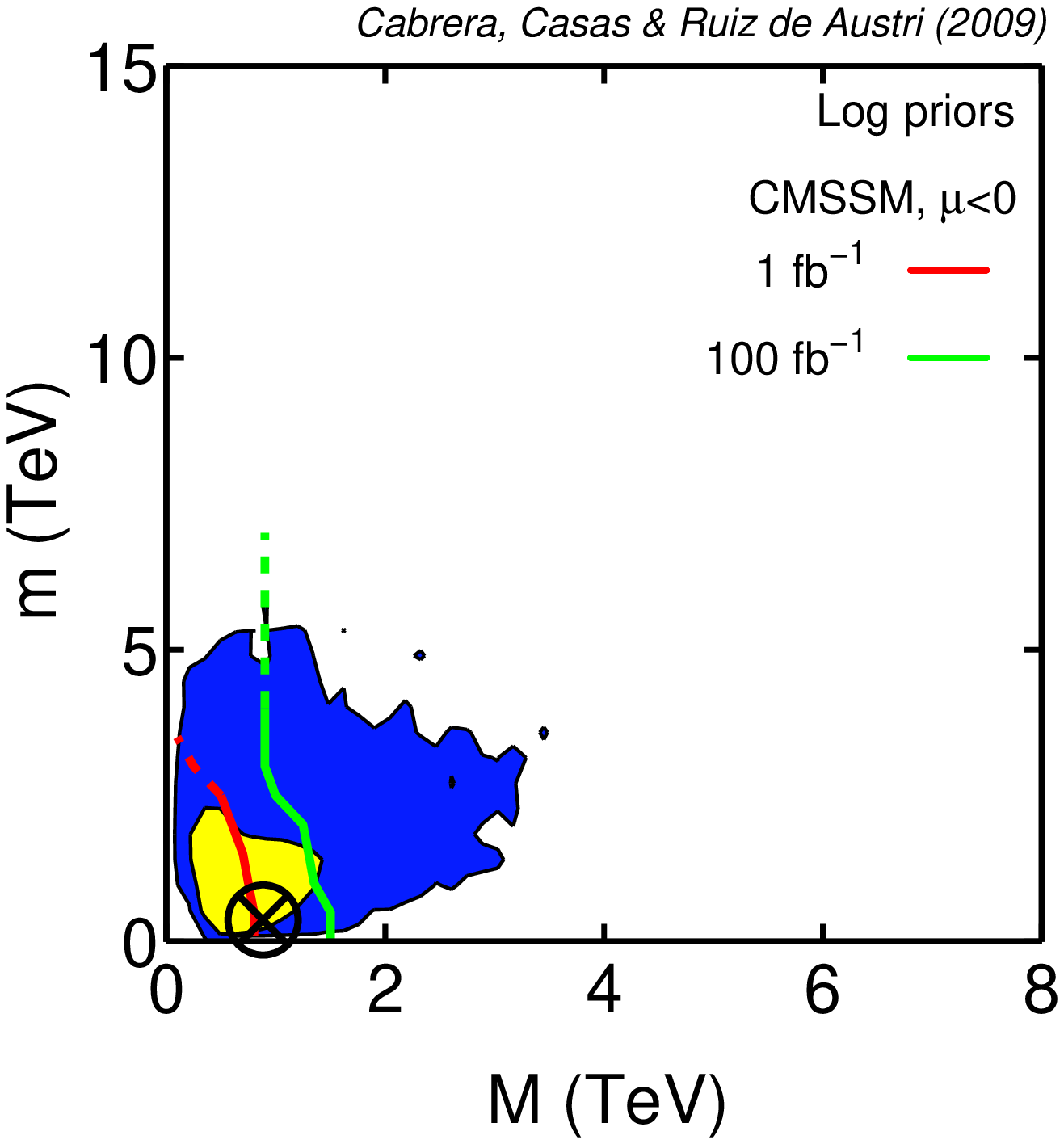} \hspace{1.2cm}
\includegraphics[angle=0,width=0.4\linewidth]{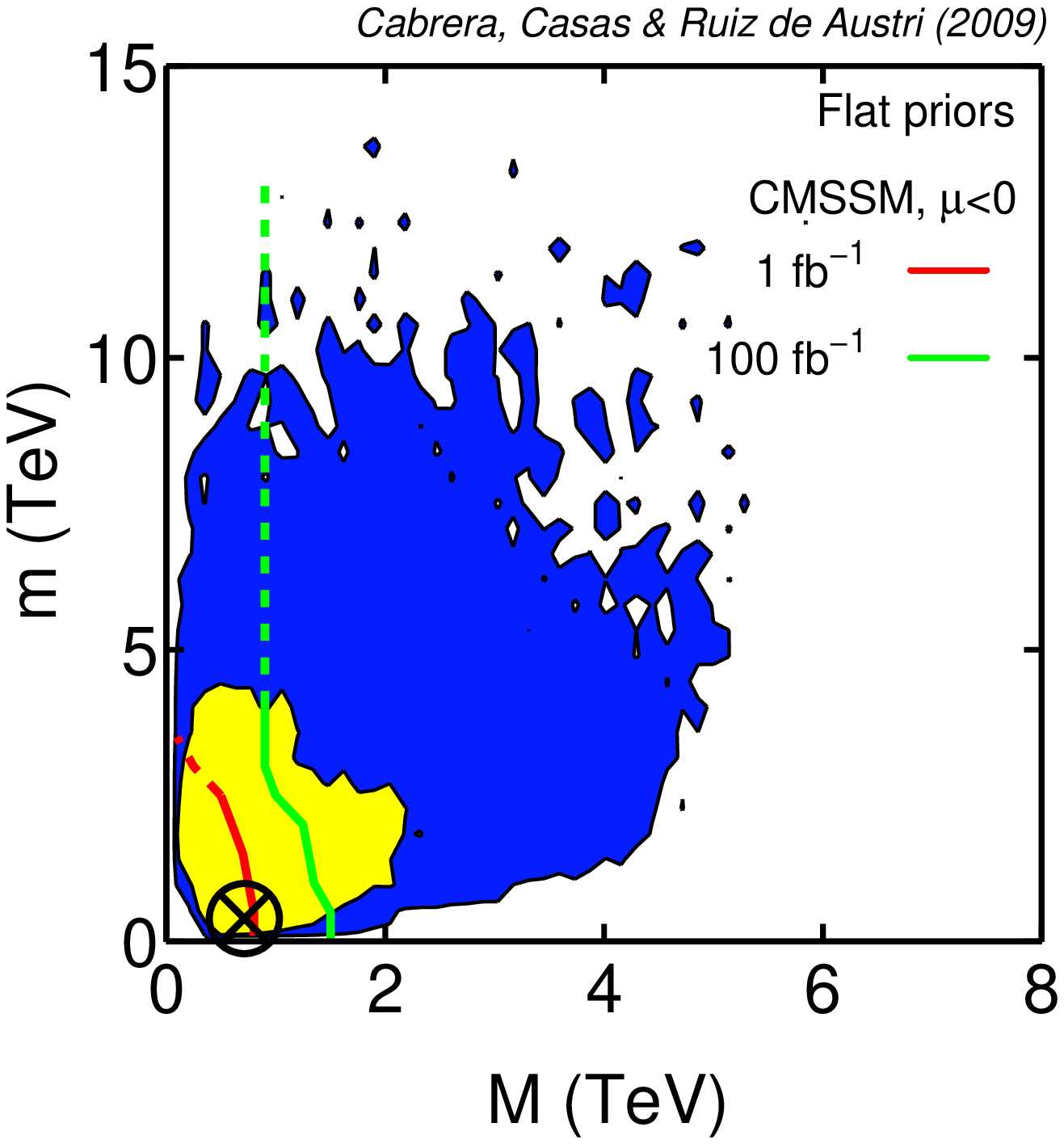} \\ \vspace{1.0cm}
\includegraphics[angle=0,width=0.4\linewidth]{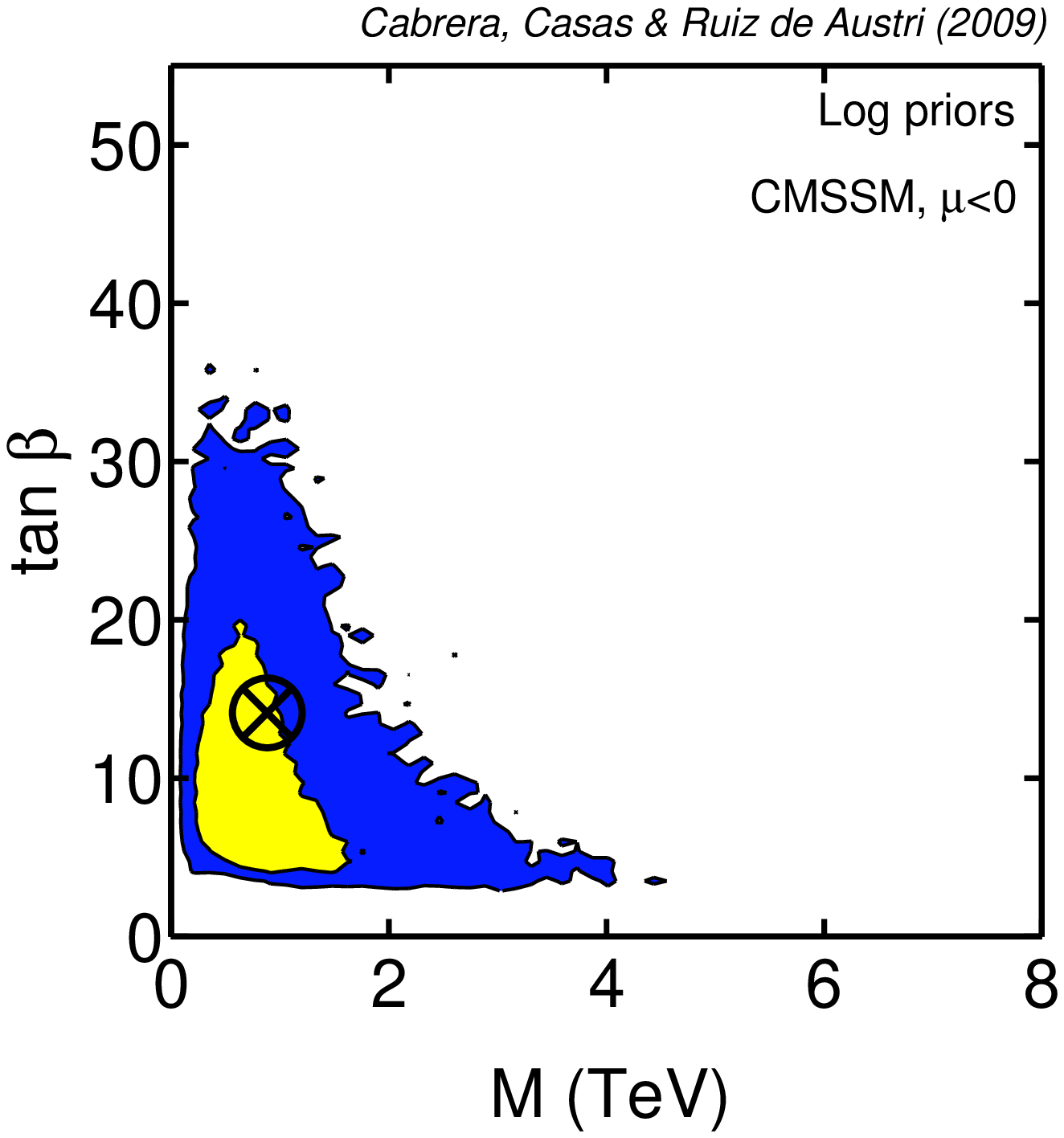} \hspace{1.2cm}
\includegraphics[angle=0,width=0.4\linewidth]{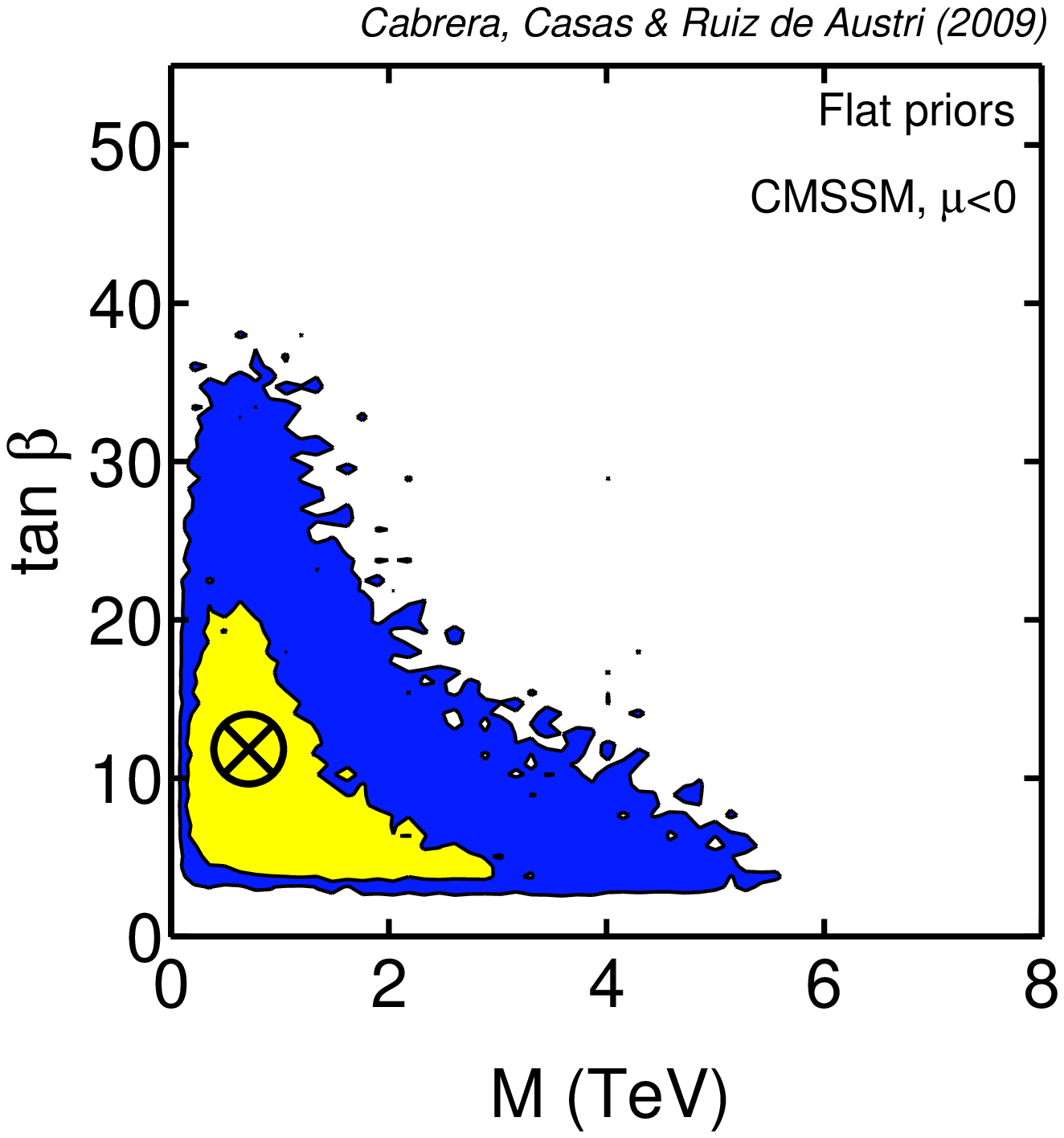}
\caption{As in Fig. 5 but with $\mu<0$.}
\end{center}
\end{figure}

Fig. 11 shows the posterior distribution function when only the most robust set of data (EW and B(D)-physics observables, and limits on particle masses) are taken into account. Because of the above-mentioned $b\rightarrow s, \gamma$ observable, the distribution is now slightly shifted to smaller soft masses (now ``it pays" to have a moderately sizeable SUSY contribution to this process), as it is clear from comparison with Fig. 5. The effect is welcome, as it pushes SUSY towards regions of the parameter space more accessible to LHC. However the impact is far from dramatic.

Fig. 12 shows the posterior when $a_\mu$ (evaluated using $e^+e^-$ data) is included in the analysis. Now the difference with the analogue for positive $\mu$ (Fig. 8) is really dramatic. Recall that now the SUSY contributions to $a_\mu$ have the wrong sign, so it does not pay to have smaller soft masses.  Consequently, Fig. 12 is similar to Fig. 11 (i.e. before including $a_\mu$ constraints) and, actually, the soft masses are pushed to slightly higher values. 

\begin{figure}[tbh!]
\begin{center}
\label{}
\includegraphics[angle=0,width=0.4\linewidth]{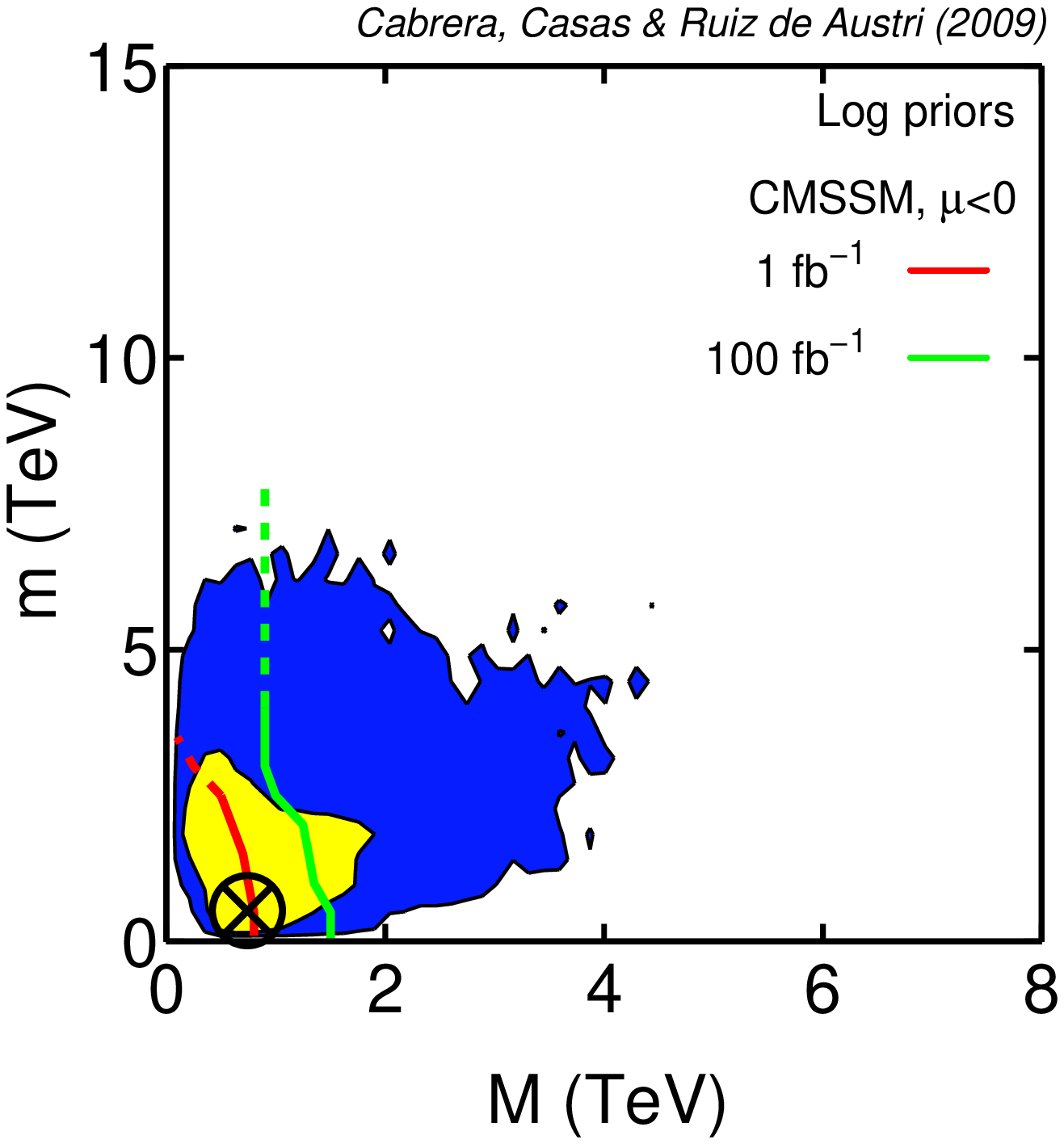} \hspace{1.2cm}
\includegraphics[angle=0,width=0.4\linewidth]{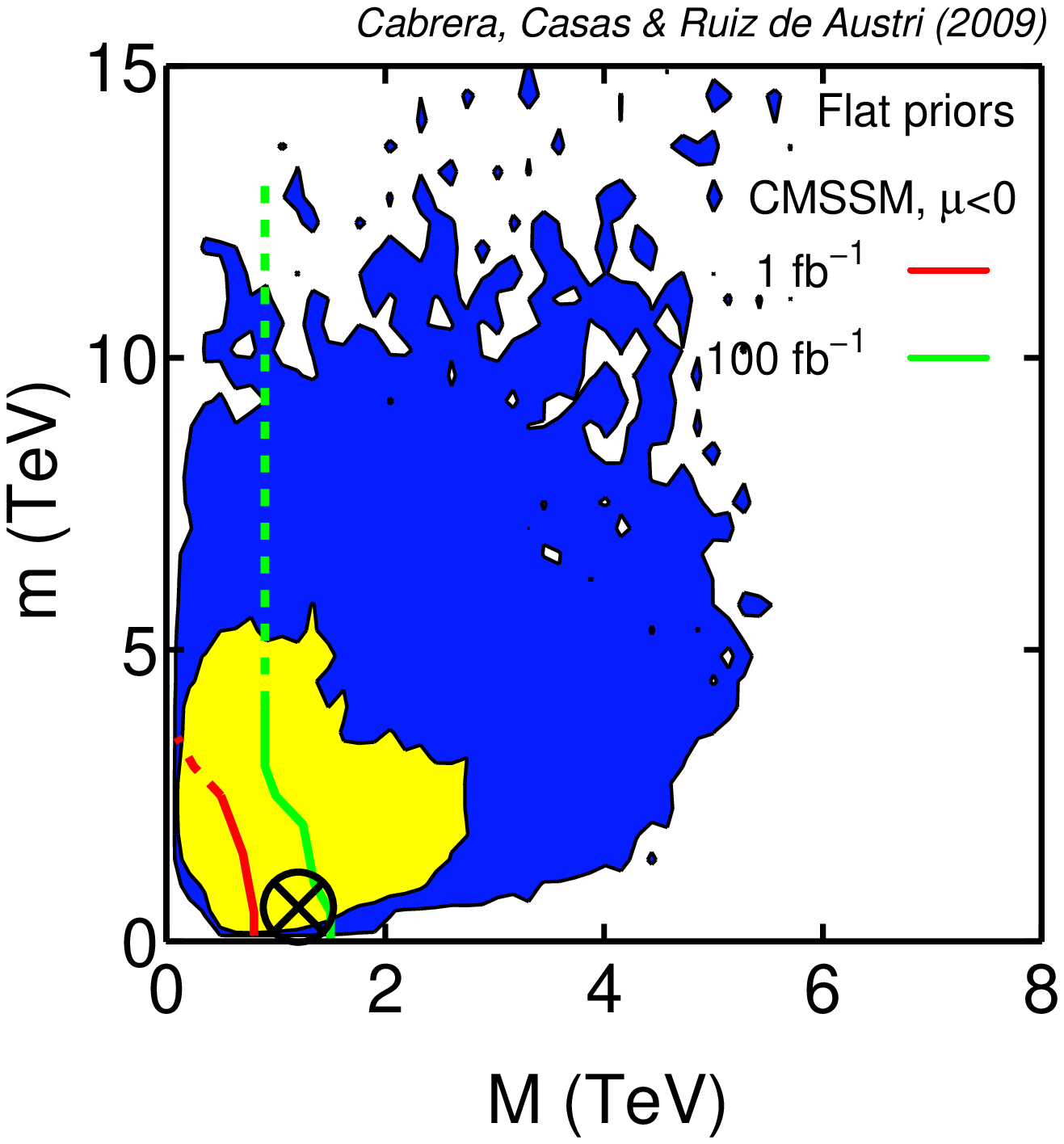} \\ \vspace{1.0cm}
\includegraphics[angle=0,width=0.4\linewidth]{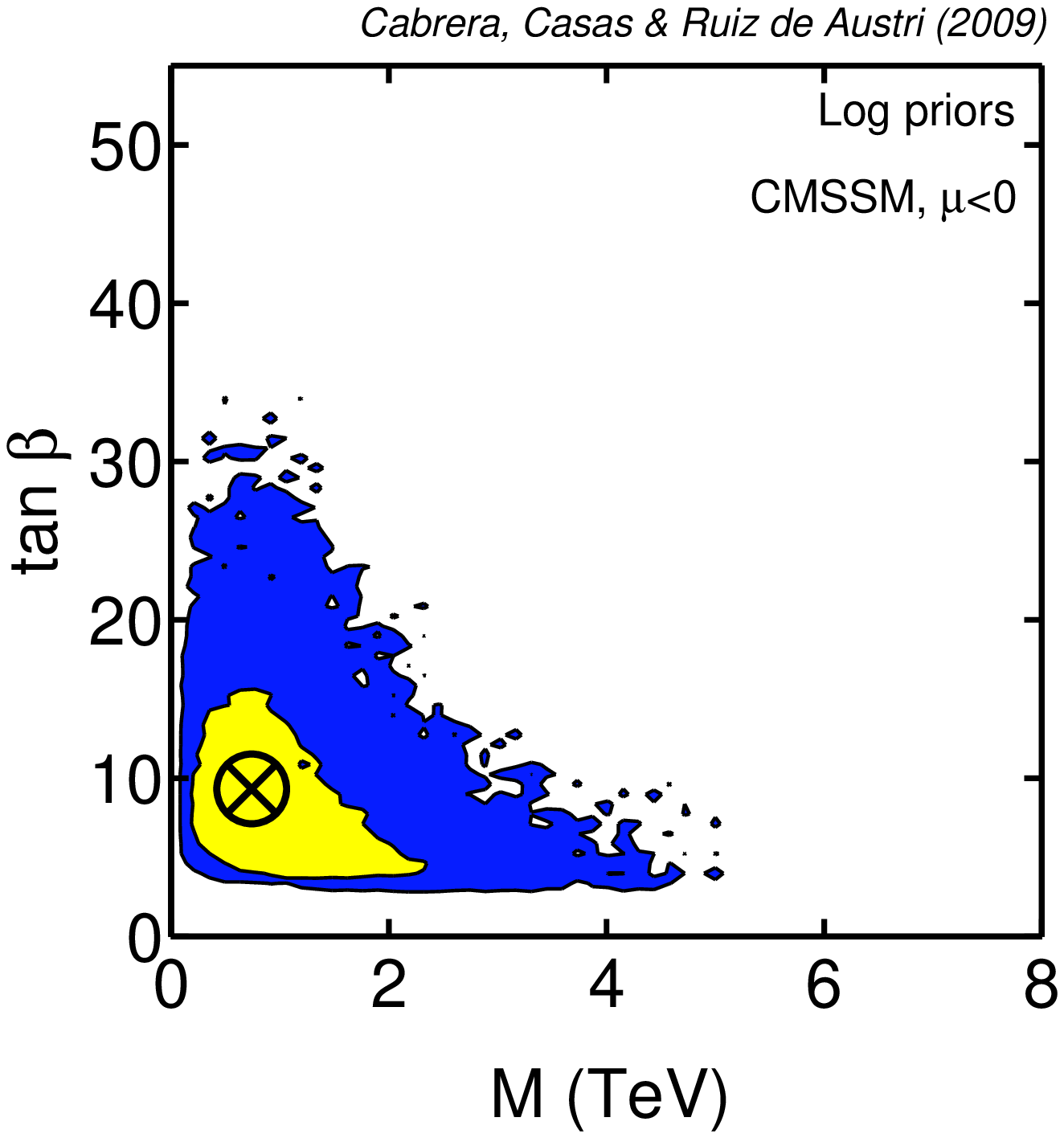} \hspace{1.2cm}
\includegraphics[angle=0,width=0.4\linewidth]{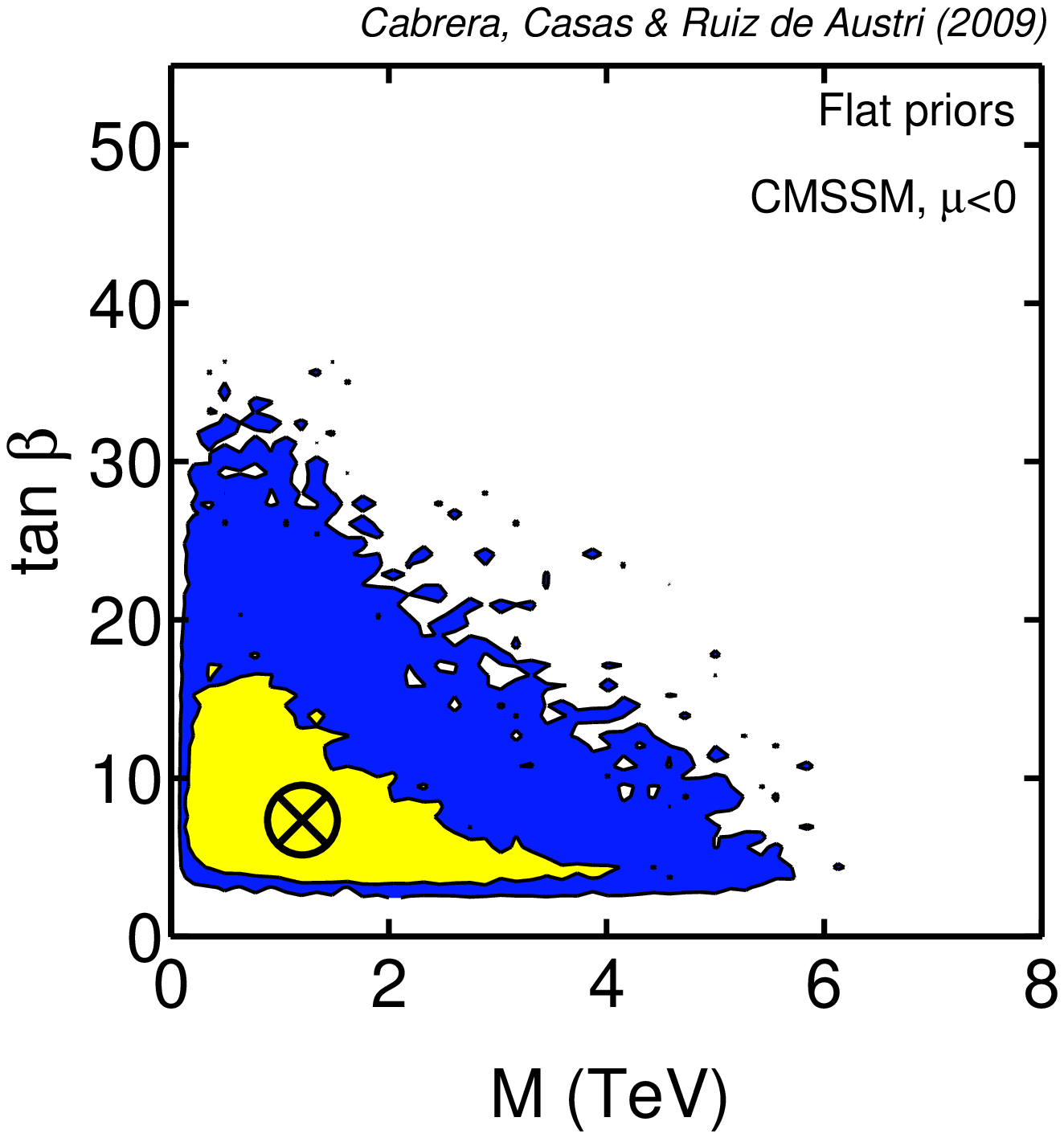}
\caption{As in Fig. 8 but with $\mu<0$.}
\end{center}
\end{figure}

Fig. 13 shows the posterior when one considers the previous robust set of data (not including $a_\mu$) plus the constraints from Dark Matter. Since Dark Matter has a great potential to select preferred regions in the parameter space, the results are quite similar to those for $\mu>0$, Fig. 9; and the same comments hold here.

\begin{figure}[t]
\begin{center}
\label{}
\includegraphics[angle=0,width=0.4\linewidth]{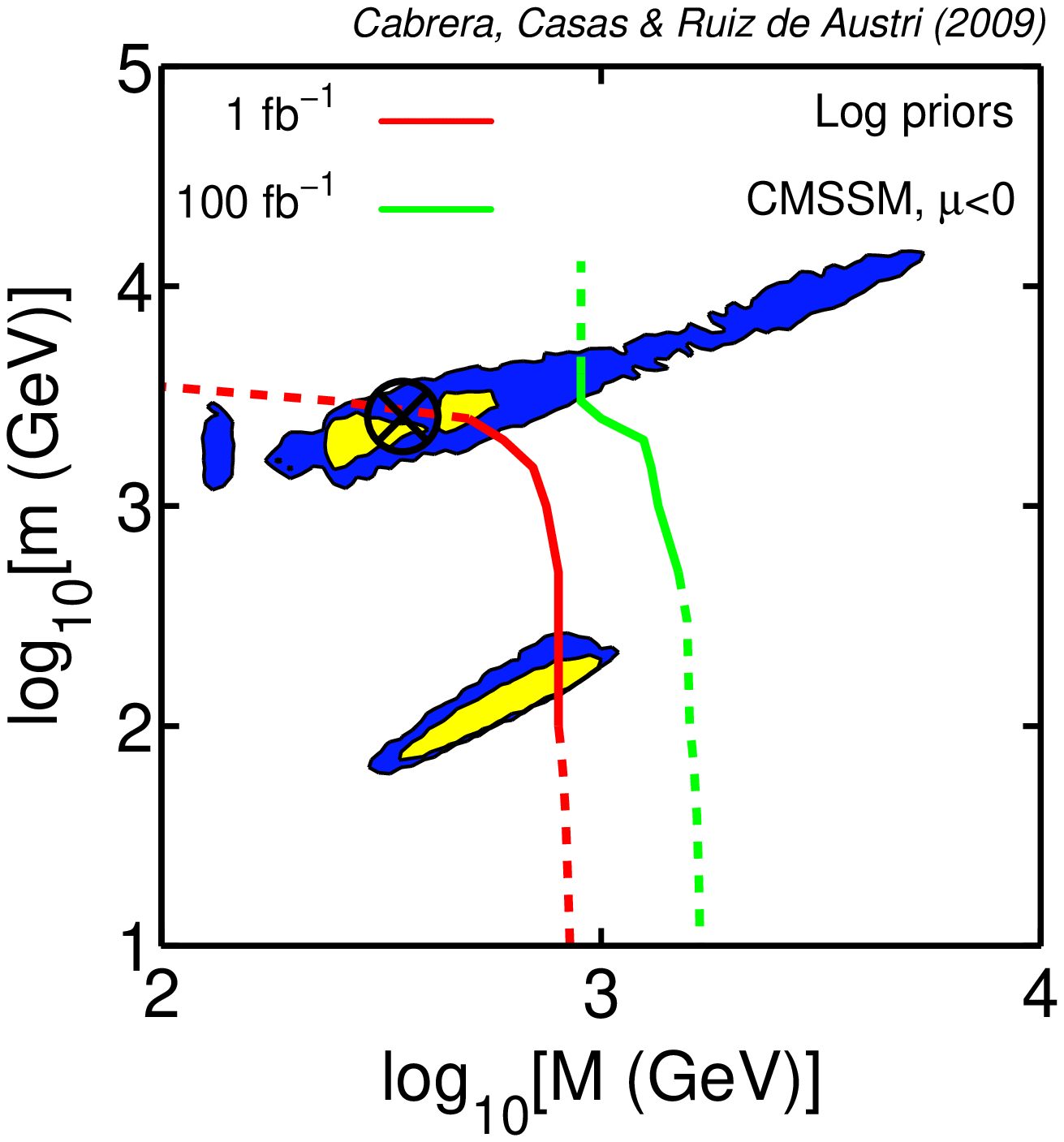} \hspace{1.2cm}
\includegraphics[angle=0,width=0.4\linewidth]{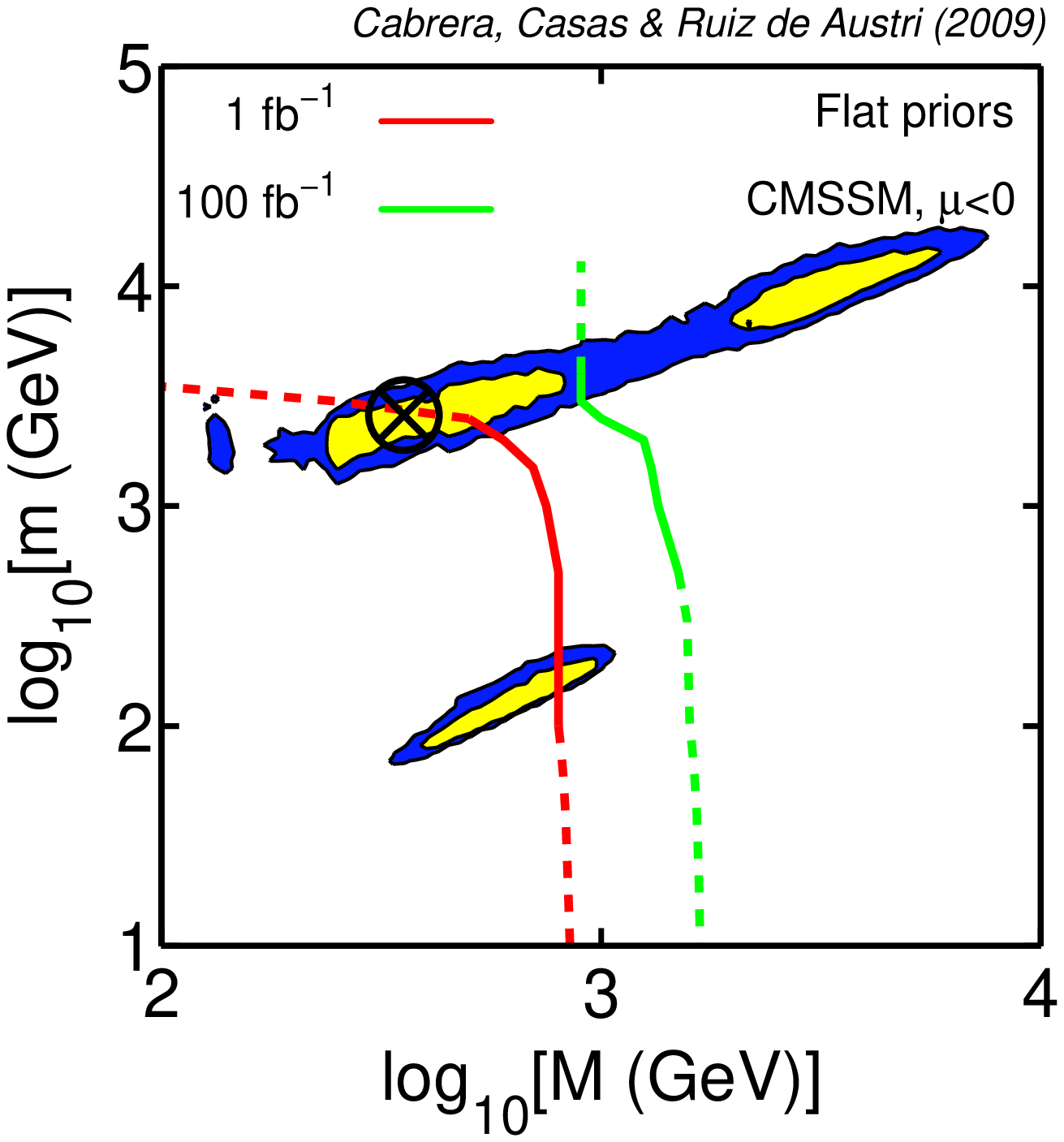} \\ \vspace{1.0cm}
\includegraphics[angle=0,width=0.4\linewidth]{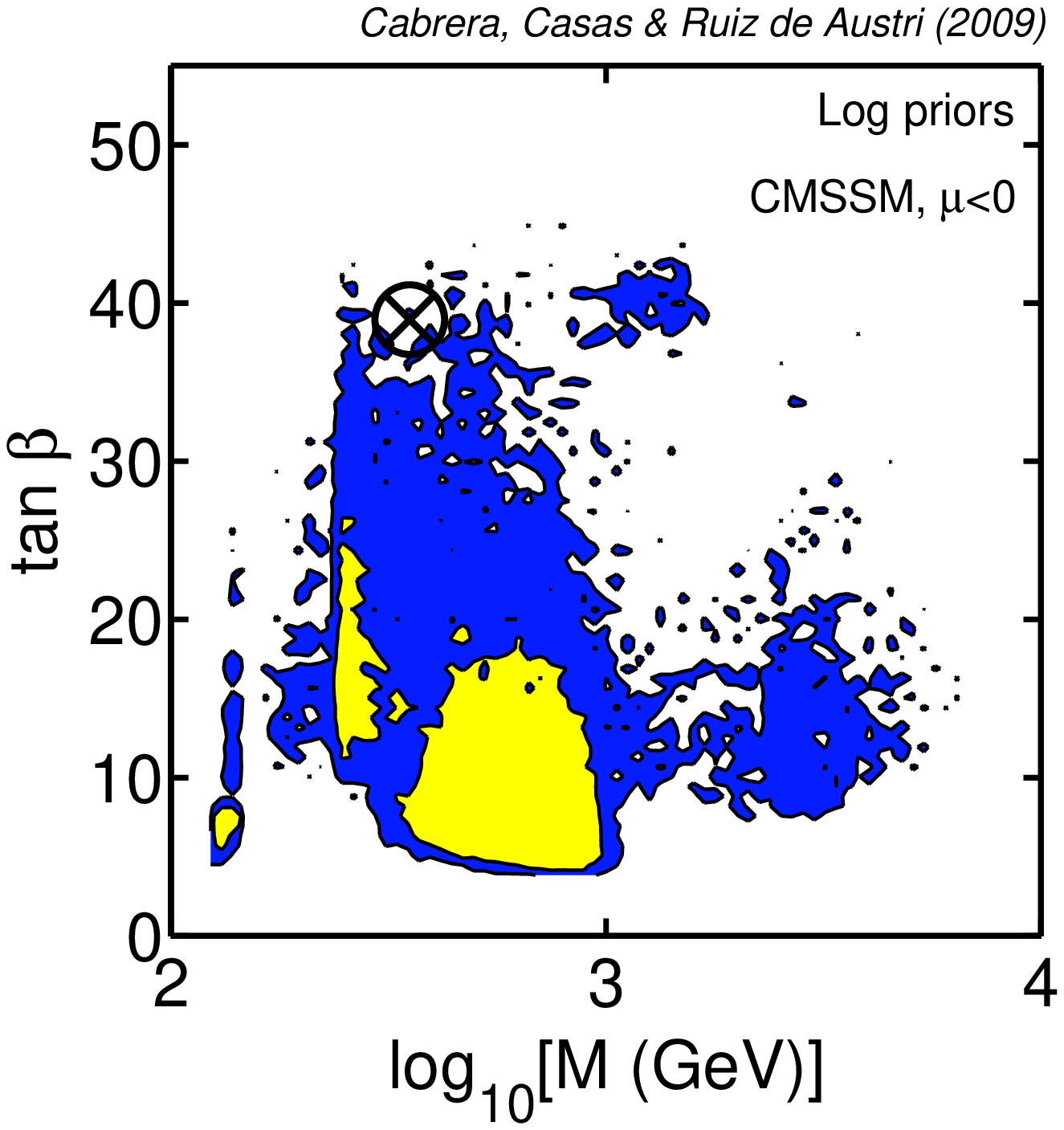} \hspace{1.2cm}
\includegraphics[angle=0,width=0.4\linewidth]{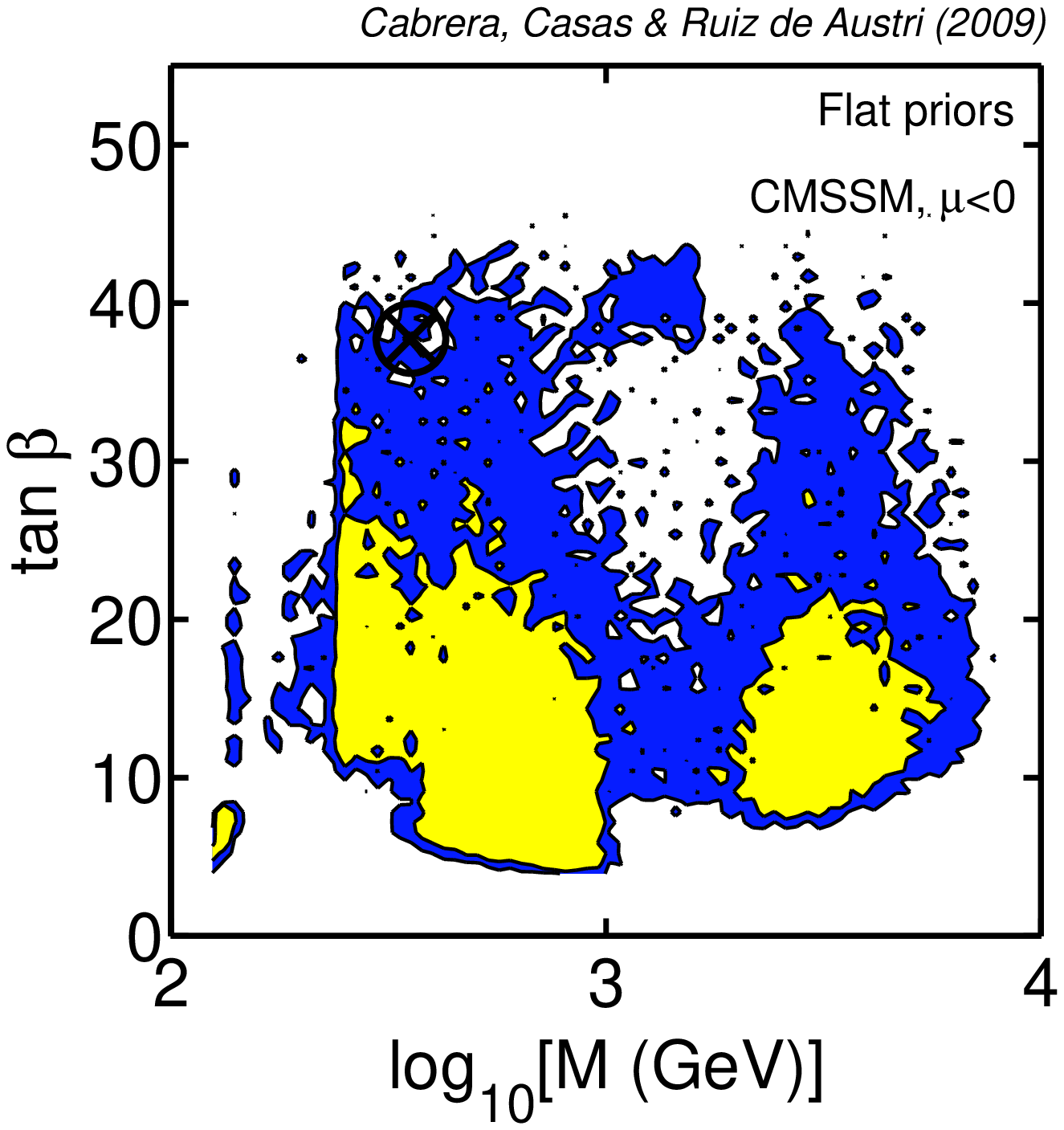}
\caption{As in Fig. 9 but with $\mu<0$.}
\end{center}
\end{figure}

Finally, Fig. 14 shows the posterior when all the experimental information, including $a_\mu$ (evaluated using $e^+e^-$ data) is taken into account. Similarly to our above discussion of Fig. 12, the results do not change much after the inclusion of the
$a_\mu$ constraint. In consequence, Fig. 14 is quite similar to Fig. 13, with a certain
penalization of too small soft masses. Again, this is in strong contrast with the $\mu>0$ case, where the low-energy (bulk and co-annihilation) regions were preferred (see Fig. 10).

\begin{figure}[t]
\begin{center}
\label{}
\includegraphics[angle=0,width=0.4\linewidth]{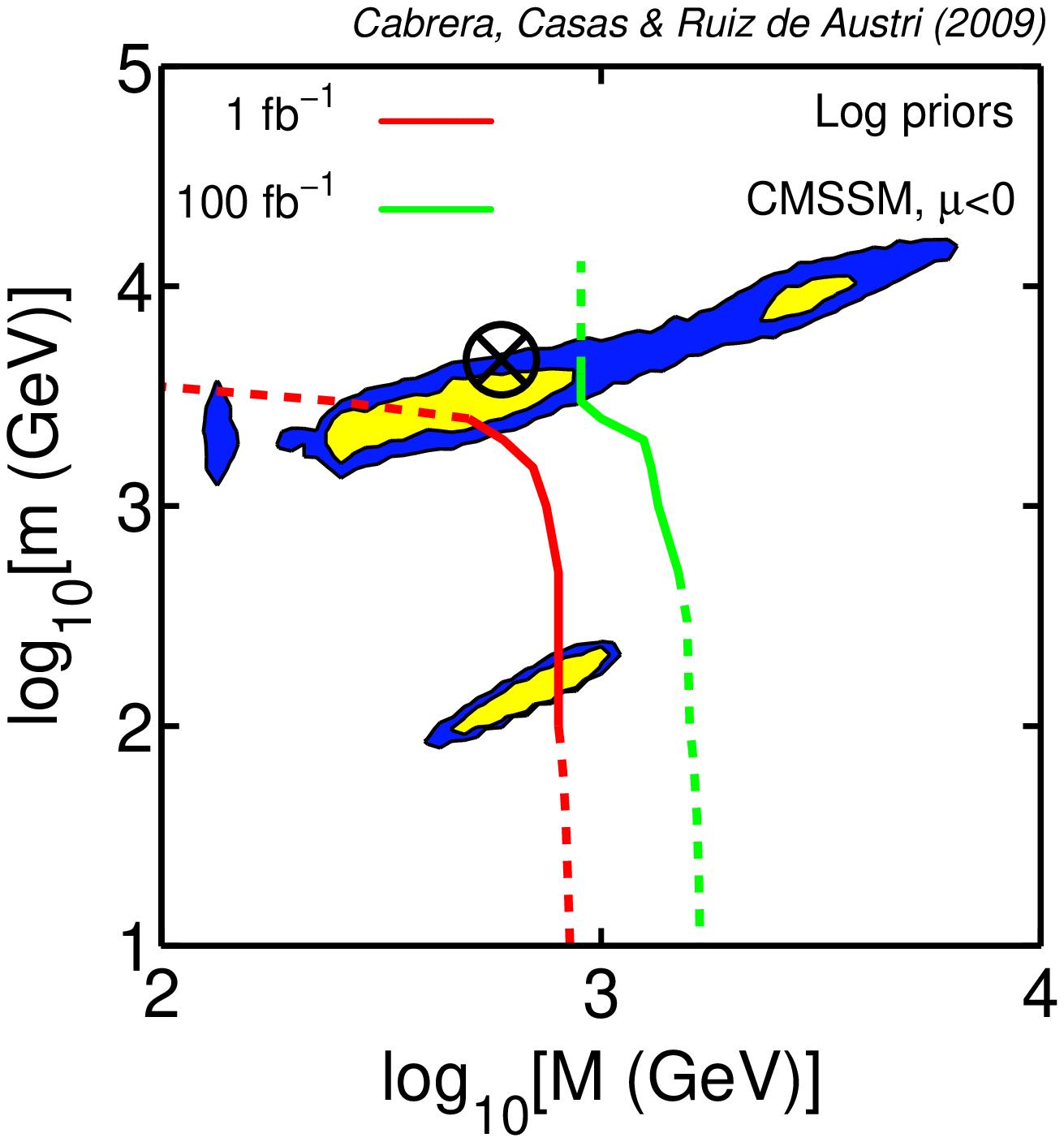} \hspace{1.2cm}
\includegraphics[angle=0,width=0.4\linewidth]{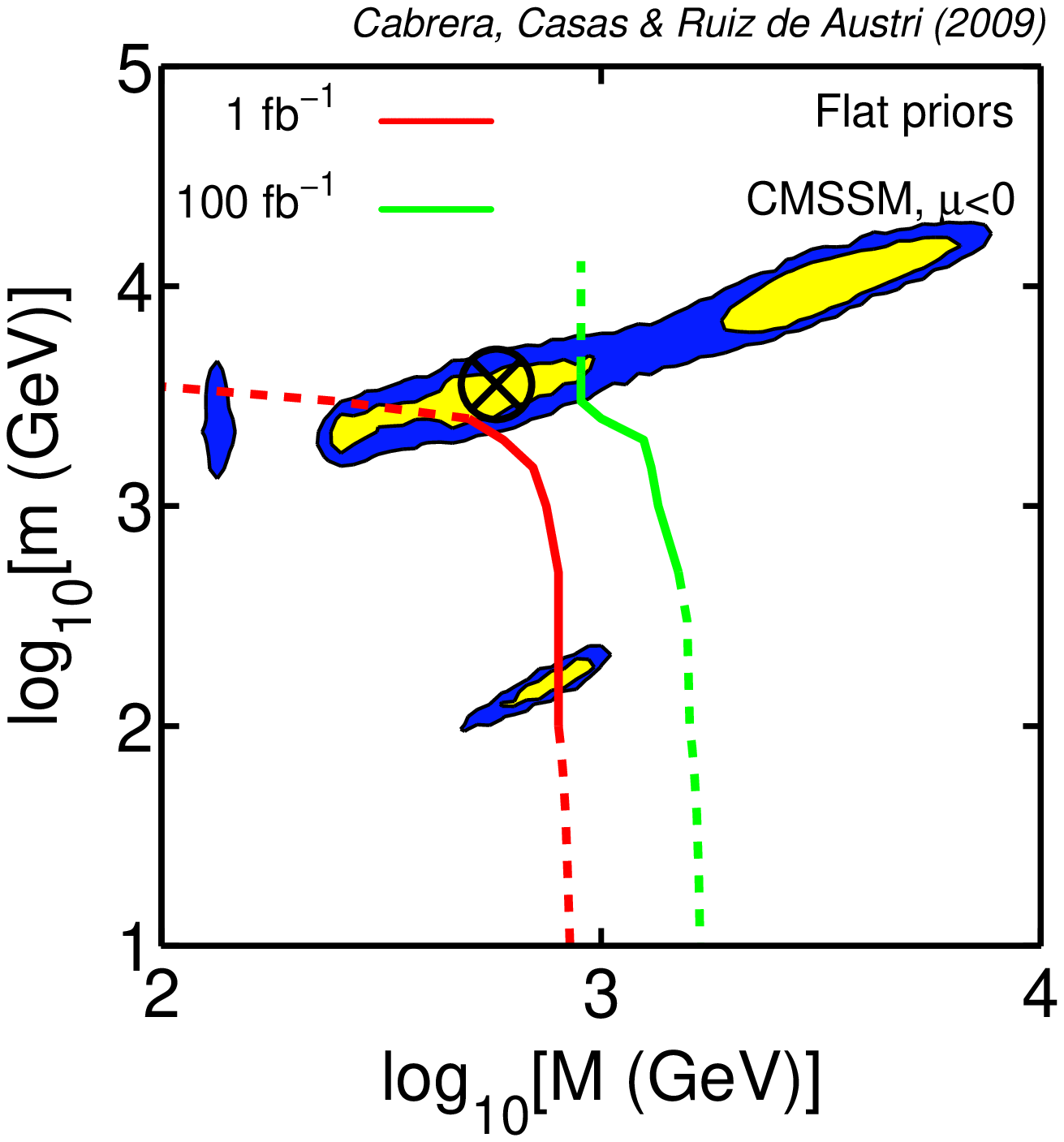} \\ \vspace{1.0cm}
\includegraphics[angle=0,width=0.4\linewidth]{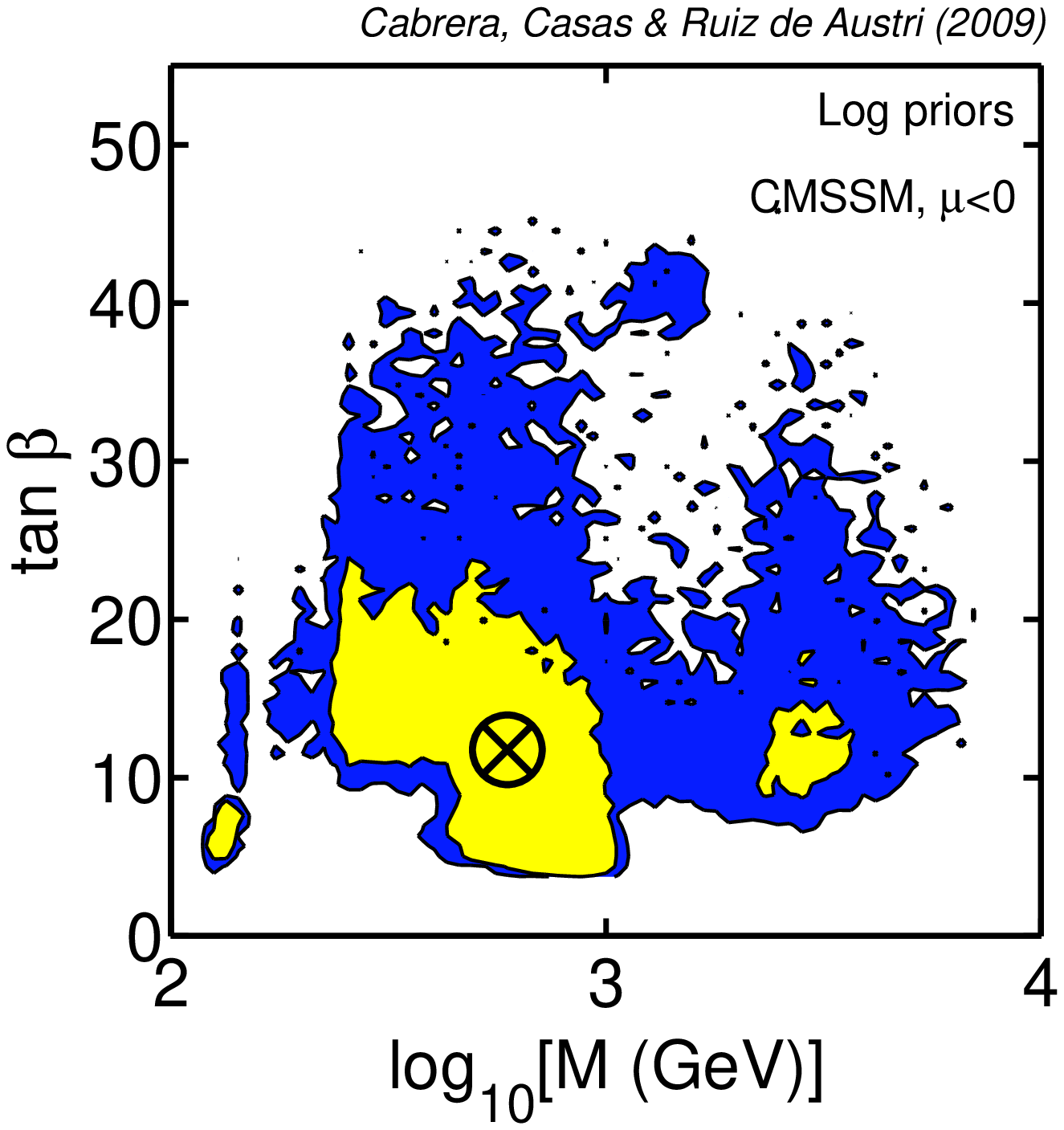} \hspace{1.2cm}
\includegraphics[angle=0,width=0.4\linewidth]{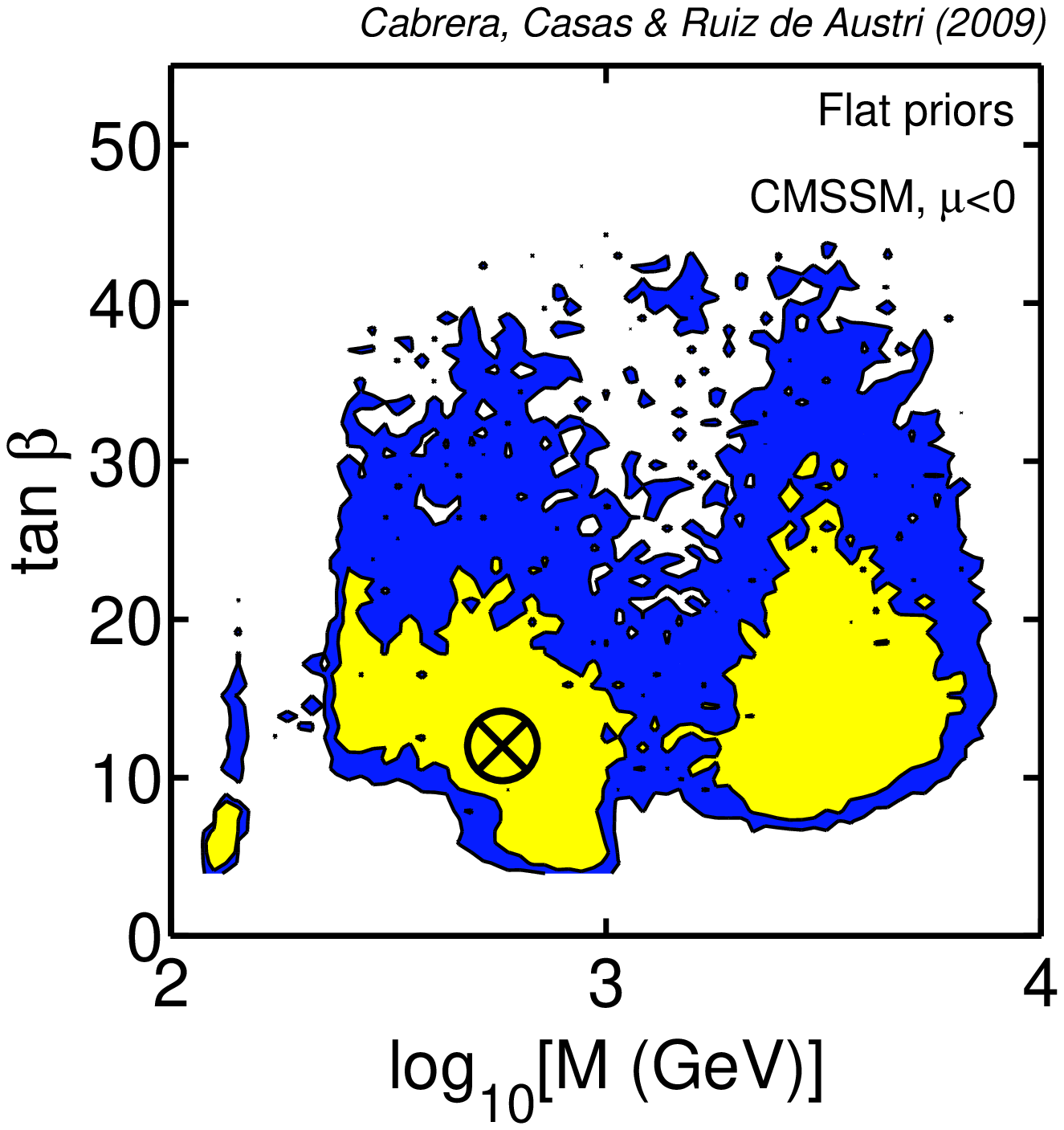}
\caption{As in Fig. 10 but with $\mu<0$.}
\end{center}
\end{figure}

\subsection{Positive versus negative $\mu$}

In order to compare the relative probability of the $\mu>0$ and $\mu<0$ branches, one has to evaluate the Bayesian evidences of both cases. We recall that the Bayesian evidence is the piece in the denominator of eq.(\ref{Bayes}), i.e. $p({\rm data})$, sometimes called ${\cal Z}$. Note that, integrating both sides of eq.(\ref{Bayes}), and using the fact that $p(\theta_i|{\rm data})$ must be correctly normalized, one simply obtains 
\bea
\label{BayesEvid}
{\cal Z}\ \equiv\ p({\rm data})\ =\ \int d\theta_1\cdots\ d\theta_N \  p({\rm data}|\theta_i)\ p(\theta_i)\ ,
\eea
i.e. the evidence is the integral of the likelihood times the prior, and therefore it is a measure of the global probability of the model.
Note that in doing parameter inference within a given model, ${\cal Z}$ plays the role 
of a normalization factor and can be (and it is) usually ignored. However, it is the key quantity to compare the relative probability of two different models.
Let ${\cal M}_1$, ${\cal M}_2$ be two models with prior probabilities $p({\cal M}_1)$, $p({\cal M}_2)$. Then the relative posterior probability of the two models, given a set of data, is simply
\begin{equation}   
\frac{p({\cal M}_{1}|data)}{p({\cal M}_{2}|data)}
=\frac{\mathcal{Z}_{1}\ p({\cal M}_{1})}{\mathcal{Z}_{2}\ p({\cal M}_{2})}
=B_{12}\frac{p({\cal M}_{1})}{p({\cal M}_{2})},
\label{bfactor}
\end{equation}
where $B_{12} \equiv {\cal Z}_1/{\cal Z}_2$ is called the Bayes factor and 
$p({\cal M}_{1})/p({\cal M}_{2})$ is the prior factor, often set to unity. 

The natural logarithm of the Bayes factor provides a useful indication of the different performance of two models. In Table~\ref{tab:Jeffreys}, we summarize the translation of the Bayes factor to relative probabilities and a conventional interpretation of them \cite{Jeffreys}, which we follow in this paper.

\begin{table}
\begin{center}
\begin{tabular}{|c|c|c|c|}
\hline
$|\ln B_{10}|$ & Odds & Probability & Strength of evidence \\ 
\hline\hline
$<1.0$ & $\lesssim 3:1$ & $<0.750$ & Inconclusive \\
$1.0$ & $\sim 3:1$ & $0.750$ & Weak Evidence \\
$2.5$ & $\sim 12:1$ & $0.923$ & Moderate Evidence \\
$5.0$ & $\sim 150:1$ & $0.993$ & Strong Evidence \\ \hline
\end{tabular}
\end{center}
\caption{The scale we use for the interpretation of model probabilities.}
\label{tab:Jeffreys}
\end{table}

The evaluation of the Bayesian evidence is in general a numerically challenging 
task, as it involves a multidimensional integral over the whole parameter space.
In addition the likelihood is often multi-modal, or it has strong degeneracies 
that confine the posterior to thin sheets in parameter space. 
Standard techniques as thermodynamic integration \cite{O'Ruanaidh} have been 
proposed for a reliably estimation of the evidence, but they are extremely 
computationally-expensive.
Perhaps the most elegant algorithm proposed up to now is the
{\em nested sampling} as referred to in Sec.~3. This technique has greatly 
reduced the computational cost of model selection.

Let us now come back to our task, namely comparing the evidences for the positive and negative $\mu$ branches, considered here as different models with equal prior probabilities. We have applied the MultiNest algorithm to obtain the Bayes factor of these two models, $B_{+-}={\cal Z}_+/{\cal Z}_-$.
Of course the results depend on the experimental information considered, which enters the likelihood piece in eq.(\ref{BayesEvid}). The results are given in Table~\ref{tab:evidences}. 

\begin{table}
\begin{center}
\begin{tabular}{|c|c|c|}
\hline
%$\Delta ln Z$& \multicolumn{3}{c|}{symmetric $\mathcal{L}_{\rm DM}$} &
%\multicolumn{5}{c|}{asymmetric $\mathcal{L}_{\rm DM}$} \\ \hline
Observables  &  $\ln \, B_{+-}$ (flat) & $\ln \, B_{+-}$ (log)    \\ \hline \hline
EW + Bounds + B-physics        &  $-0.32 \pm 0.048$ &   $-0.48 \pm 0.049$     \\ 
EW + Bounds + B-physics + $(g-2)_\mu$    &   $0.81 \pm  0.043$   & $1.73 \pm  0.052$        \\ 
EW + Bounds + B-physics + $\Omega_{\rm DM}$   & $-0.31 \pm 0.068$   &  $-0.66 \pm 0.066$  \\
EW + Bounds + B-physics + $(g-2)_\mu$ + $\Omega_{\rm DM}$   & $1.9 \pm 0.065$  & $3.71 \pm  0.068$        \\
\hline 
\end{tabular}
\caption{The natural log of the Bayes factor ($\ln \, B_{+-}$) for $\mu > 0$ 
and $\mu < 0$. A positive (negative) value indicates a preference 
for $\mu > 0$ ($\mu < 0$).}
\label{tab:evidences}
\end{center}
\end{table}

The first column of Table~\ref{tab:evidences}
indicates the set of experimental data taken into account, the notation is self-explanatory and corresponds to the different cases previously defined. The discussion of subsect.~5.1 allows to understand the numbers of the table. When only the most robust pieces of experimental information are used (first row), the performance of both models is similar. The $\mu<0$ branch is slightly favoured, due to its capability to reproduce the central value of $b\rightarrow s, \gamma$, but the effect is not really significant, as is shown by a value of 
$|\ln \, B_{+-}|$ well below 0.75, see Table~\ref{tab:Jeffreys}. This holds when $\Omega_{DM}$ constraints are incorporated into the analysis (third row of Table~\ref{tab:evidences}). On the other hand, $(g-2)_\mu$ constraints (when evaluated using $e^+e^- \rightarrow $had data) clearly favour the $\mu>0$ branch, as discussed above, which is reflected in
the numbers of the second and fourth rows of Table~\ref{tab:evidences}. 
Using the conventions of Table~\ref{tab:Jeffreys}, we see that the global evidence in favour of positive $\mu$ is weak-to-moderate (not strong but already significant).
Note that this effect is stronger for log prior, since in that case the high-energy region (the preferred one for $\mu<0$) gets an additional penalization. Likewise, when $\Omega_{DM}$ constraints are included (at the same time as $(g-2)_\mu$), the preference for positive $\mu$ gets even stronger. This is because, $\Omega_{DM}$ constraints favours (in terms of statistical weight) the low-energy region of the parameter space, and this is the region strongly preferred (penalized) by $a_\mu$ constraints for positive (negative) $\mu$.

\section{Comparison to previous work} 

Some of the previous work in this subject has been collected in refs.
\cite{Allanach:2005kz, Allanach:2006jc, deAustri:2006pe, Allanach:2007qk, Roszkowski:2007fd, Buchmueller:2008qe, Trotta:2008bp, Ellis:2008di, AbdusSalam:2009qd, Buchmueller:2009fn}
 All of them are Bayesian analyses, except refs.\cite{Buchmueller:2008qe, Buchmueller:2009fn}. However, a fair comparison with our work is tricky, since these articles often make assumptions very different from us about the priors and ranges of the initial parameters (and even about which are the initial parameters). Also they may include different pieces of experimental information. The last point is dramatic regarding $(g-2)_\mu$, as is clear from subsect. 4.2. Nevertheless it is interesting to compare our work with this previous literature, to make clearer how all these differences do affect the results and conclusions. For the sake of concreteness, we have considered five previous representative works, corresponding to refs.\cite{Allanach:2005kz, deAustri:2006pe, Allanach:2007qk, AbdusSalam:2009qd, Buchmueller:2009fn}.

In ref.\cite{Allanach:2005kz}, which was pioneering in MSSM Bayesian analyses, $\tan\beta$ was considered as an initial parameter, with flat prior. As a result, there is no penalization of the large $\tan\beta$ region, which thus becomes even favoured by experimental data (probably because of Dark Matter constraints, see below). Besides, the authors include always the experimental data concerning $(g-2)_\mu$ (based on $e^+e^-$ data) and Dark Matter constraints. Finally, the priors for the soft terms are taken as flat, with ranges bound by 2 TeV.
Hence their Fig.2 would correspond to our Fig. 10 (they are based on essentially the same experimental data). Actually, the $\{M,m\}$ plots of the two figures are not very different, although ours favour more clearly the low-energy region (due to the incorporation of the electroweak scale, as discussed in sect. 2). This effect would have been more impressive if in ref.\cite{Allanach:2005kz} they had unplugged the $(g-2)_\mu$ and Dark Matter data. And much more if, besides, they had widened the allowed range of the soft terms. On the other hand, their $\{M,\tan\beta\}$ plots favour more clearly the region of very large $\tan\beta$ (Higgs Funnel region). In our opinion this effect is not realistic, since $\tan\beta$ is clearly a derived parameter, and this fact introduces a Jacobian factor in the associated probability distribution, penalizing large $\tan\beta$.

In ref.\cite{deAustri:2006pe} the initial assuptions were similar to those of ref.\cite{Allanach:2005kz} (and the results are consistent with each other). Therefore the comparison with our results is also similar. In this case, however, the authors tried two different classes of ranges for the soft parameters (up to 2 TeV and 4 TeV) and, also, they probed to disconnect $(g-2)_\mu$. From their Fig. 16, it is clear that, by unplugging $(g-2)_\mu$, the preferred region for $M$ goes from 0.5--1 TeV to 1--1.5 TeV. Comparing to our Fig.9 (which is now the corresponding one), we see that in our analysis the high-energy region is more penalized, which is not surprising.

In ref.\cite{Allanach:2007qk}, a refined version of the analysis of ref.\cite{Allanach:2005kz} was presented. In this case, $\tan\beta$ was considered a derived parameter (which introduces a Jacobian factor). Also, $M_Z$ was marginalized, as in our case (for a detailed comparison between the two procedures see ref.\cite{Cabrera:2008tj}). Therefore, the initial set up of ref.\cite{Allanach:2007qk} is the most similar one to ours. Their priors, however, are quite different and somewhat arbitrary (though reasonable). They would correspond more or less to our logarithmic priors, allowing very large ranges for the parameters. In their results the authors observed indeed a penalization of the high-energy region, which they attributed to the choice of the priors. We think, however, that it is mainly a consequence of the marginalization of $M_Z$, and the effective penalization of fine-tuning that it entails (something that is far from obvious at first sight). In their Fig. 3 they compare their results with those of ref.\cite{Allanach:2005kz}. There one can clearly see the extra penalization of the high-energy region. The $\{M,m\}$ and $\{M,\tan\beta\}$ plots of that figure correspond to the (log prior) plots of our Fig. 10. Indeed, both figures are quite consistent (theirs are even more tilted towards low energy, probably due to the additional effect of their choice of priors). Unfortunately, they do not explore unplugging $(g-2)_\mu$ and Dark Matter data, so a comparison with other results and plots of our paper is not possible.

In ref.\cite{AbdusSalam:2009qd} a Bayesian analysis of the so-called pMSSM ("phenomenological MSSM") was presented. This model has many initial parameters ($\sim 20)$, all of them defined at low-energy. Apart from that, the set up of the analysis was similar to that of ref.\cite{Allanach:2005kz}.
In particular they took $\tan\beta$ as an initial parameter, and considered flat priors and finite ranges for the soft parameters ($< 4 $TeV), including 
$(g-2)_\mu$ and Dark Matter experimental data in all instances. In order to make any comparison with our work, one has to focus on particular quantities. A good example is the gluino mass, $M_{\tilde g}$, which for mSUGRA is $\sim 2.5 M$. From their Fig. 3 the peak of the probability distribution of $M_{\tilde g}$ is around 2--3 TeV, which would correspond to $M\sim 1$ TeV. This should be compared with our Fig. 10 (the one based on a similar set of experimental data). In our case the peak of the distribution is around 400 GeV, showing that we get an extra penalization of the high-energy region, as explained in this paper. Unplugging $(g-2)_\mu$ and Dark Matter, the differences would have been more dramatic, especially if the allowed ranges of the parameters were stretched.

Finally, in ref.\cite{Buchmueller:2009fn} a frequentist analysis of the MSSM was presented. This is a point of view complementary to the Bayesian approach, followed here. The authors of ref.\cite{Buchmueller:2009fn} perform a scan of the parameter space of the CMSSM (and also of the so-called NUHM1 model), evaluating the likelihood (based on the $\chi^2$). This leads to zones of estimated probability (inside contours of constant $\chi^2$) around the best fit points in the parameter space. Their Fig. 1 (($\{M,m\}$ plane) corresponds to our Fig. 10. However, notice that, in their case, the unplotted variables are optimized to obtain the best $\chi^2$, whereas in our case they are marginalized. Nevertheless, it is remarkable that the two figures are quite similar (especially comparing with our log-prior plot). This is an encouraging result. Indeed, the frequentist and Bayesian approaches must converge when the quality of data increases. This means that the bulk of the probability is centered around the best-fit points. This coincidence is also observed when the authors probe to unplug $(g-2)_\mu$ (compare their Fig. 2 to our Fig. 9). Since they do not explore to unplug Dark Matter, it is not possible to make further comparisons. It is likely, that in that case their $68\%$ and $95\%$ c.l. regions become much more extended in the parameter space, thus taking up a large portion of the high-energy (non-accesible to LHC) region. Notice that a frequentist approach cannot penalize those regions from fine-tuning arguments. Fine-tuning has to do with statistical weight (see subsect. 2.1) and a frequentist analysis is based in likelihood, i.e. the ability to reproduce the experiment. Without $(g-2)_\mu$ and Dark Matter data, the experimental reasons to stick to low-energy are much less powerful. In other words, without $(g-2)_\mu$ and Dark Matter it is likely that the convergence between frequentist and Bayesian approaches is still weak.

\section{Summary and conclusions}

The idea of an MSSM forecast for the LHC is to use all the present (theoretical and experimental) information available to determine the relative probability of the different regions of the MSSM parameter space. This includes theoretical constraints (and perhaps prejudices) and experimental constraints. 
An appropriate framework to perform such forecast is 
the Bayesian approach, which allows a sensible statistical analysis, clearly identifying the objective and subjective pieces of information. The latter are incorporated in the {\em prior}, which represents the ``theoretical" probability density that is assigned a priori to each point in the parameter space. Ignoring the prior factor is not necessarily the most reasonable or ``free of prejudices" attitude. Such procedure is equivalent to 
a completely flat prior in the parameters. But one needs some theoretical basis to establish, at least, the parameters whose prior can be reasonably taken as flat. Besides, a choice for the allowed ranges of the various parameters is necessary in order to make statistical statements. The Bayesian approach allows to keep track of the influence of the prior (whether explicit or implicit) upon the results.

In this paper we have performed a Bayesian analysis of the MSSM with universal soft terms at high energy (a scenario sometimes denoted CMSSM or MSUGRA). We have improved
previous studies by means of a careful handling of the various pieces of information: 

\begin{itemize}

\item
First, we do not incorporate ad hoc measures of the fine-tuning to penalize regions of the parameter space with large soft terms. Such penalization arises from the Bayesian analysis itself when the experimental value of $M_Z$ is considered on the same foot as the rest of experimental information (and not as a constraint of the model). Nicely, this permits to scan the whole parameter space, allowing arbitrarily large soft terms. Still, the low-energy region is statistically favoured (even before including dark matter or $a_\mu$ constraints). 
Incidentally, this statistical argument supports low-energy supersymmetry breaking (in the observable sector), even in a landscape scenario. On the other hand, in a frequentist analysis (thus ignoring the prior factor), the high-energy region is essentially non-disfavoured at all (before including dark matter or $a_\mu$ constraints), since it works as well as the ordinary SM. Using fine-tuning arguments to penalize it, would take us back to the choice of an implicit (and non-trivial) prior.

\item 
We have done a rigorous treatment of the nuisance variables, in particular Yukawa couplings, showing that the usual practice of taking the Yukawas "as required" to reproduce the fermion masses, approximately corresponds to taking logarithmically flat priors in the Yukawa couplings. We argue that this is a most reasonable choice.

\item 

Although we start with the usual MSSM initial parameters, $\{m,M,A,B,\mu\}$ (plus Yukawa couplings and other nuisance variables) we use an efficient (and actually quite common)
set of variables to scan the MSSM parameter space. Besides trading $\mu$ by $M_Z$ and the Yukawa couplings by the fermion masses, it is extremely convenient to trade
$B$ by $\tan\beta$. These changes introduce a global Jacobian factor in the density probability when working in the new (and more suitable) parameters for the scan. Once the information about $M_Z^{\rm exp}$ is incorporated (by marginalizing $M_Z$) the {\em effective prior} in the new variables inherits the Jacobian factor, as is explicit in eq.(\ref{eff_prior}). A quite accurate analytical expression for it is given in eq.(\ref{approx_eff_prior}), which is valid for any MSSM (not just the CMSSM). This effective prior contains inside the above-mentioned penalization of fine-tuned regions, but we stress that the latter has not been introduced by hand. Actually, these expressions for the effective prior contain no ad hoc constraints or prejudices, since the prior in the initial variables is still undefined.

\item

We have developed a sensible prior in the initial variables. Our basic  assumption has been that the soft-breaking terms share a common origin and, hence, it is logical to assume that their sizes are also similar. So, it is reasonable to assume that a particular soft term can get any value (with essentially flat probability) of the order of the typical size of the soft terms in the observable sector, $M_S\sim F/\Lambda$, or below it. 
Then, concerning the prior in $M_S$ itself, we have taken the two basic choices: a flat prior and a logarithmic prior (i.e. flat in the magnitude of $M_S$). We perform the analysis for the two priors, even though we think the logarithmic one is more realistic, taking into account what we know about mechanisms of SUSY breaking. This allows us to quantify the dependence of the results on the choice of the prior.

\end{itemize}

The second part of the paper (sections 3-5) is devoted to incorporate all the important experimental constrains to the analysis, obtaining a map of the probability density of the MSSM parameter space, i.e. the MSSM forecast for the LHC. Since not all the experimental information is equally robust, 
we perform separate analyses depending on the group of observables used. The main results are the following:

\begin{itemize}

\item First we include only the most robust experimental data: E.W. and B(D)-physics observables, and lower bounds on the masses of supersymmetric particles and the Higgs mass. Then, the favoured region of the MSSM parameter space lies at low-energy, but there is a significant portion out of the LHC reach. This effect is more prominent in the case of flat prior, but it is visible both for flat and logarithmic prior. The main responsible for this situation is the lower bound on the Higgs mass: to become consistent with experiment the Higgs mass needs sizeable radiative corrections, which require larger soft terms. We show that increasing the Higgs mass in a few GeV affects substantially the amount of parameter space within the LHC reach. In consequence, if we wish to detect SUSY at LHC, let us hope that the Higgs mass is close to the present experimental limit. We also show the present preferred credibility interval for the Higgs mass (Table 3).

\item Then we add the information about $a_\mu$. As is well-known, the impact of this observable depends dramatically on the way one computes the SM hadronic contribution. Using $e^+ e^-\rightarrow$ had data (the most common choice in the literature), 
the soft terms are dramatically pushed into the low-energy region (for $\mu>0$), well inside the LHC reach. Actually, the push is so strong that the predictions for other observables, in particular $b\rightarrow s\ \gamma$, start to be too large. Furthermore, if the Higgs mass turns out to be ${\cal O}$(10) GeV above the present experimental limit, the tension between the Higgs mass and $a_\mu$ would be dramatic and could not be reconciled: $m_h$ ($a_\mu$) would require too large (small) soft masses. 

Using $\tau$-decay data, instead $e^+ e^-\rightarrow$ had, there is no big discrepancy between $a_\mu^{\rm SM}$ and $a_\mu^{\rm exp}$, so the SUSY contribution does not need to be large. Consequently, the probability distributions are essentially unchanged by the inclusion of the $a_\mu$ constraint. Although the more direct $e^+e^-$ data are usually preferred to evaluate $a_\mu^{\rm SM}$, this discrepancy is warning us to be cautious about this procedure.

\item We then consider the impact of Dark Matter (DM) constraints (unplugging $a_\mu$ constraints), namely we require that the WIMP responsible for $\Omega_{\rm DM}$ is the supersymmetric LSP (typically the lightest neutralino). As is known, this selects four regions in the parameter space: Bulk, Focus point, Co-annihilation and Higgs-funnel. When all the SUSY parameters, but the universal scalar and gaugino mass ($m$ and $M$ respectively), are marginalized, these regions appear clearly in the $m-M$ plots. One can observe a certain blurring with respect to usual plots in the literature, due to the integration in the other variables.

Roughly speaking, including DM constraints the low-energy gets favoured and therefore the detection of SUSY at the LHC. However, there survive large high-energy areas out of the LHC reach. Consequently, again, if we are unlucky, even if DM is supersymmetric, it could escape LHC detection (especially if the Higgs mass is not close to its present experimental limit). In any case, we have stressed that one should be cautious at interpreting these results as a robust constraint on the CMSSM, discussing possible ways-out for the (in principle) disfavoured regions.

\item Finally we consider all constraints at the same time (including $a_\mu$ and $\Omega_{\rm DM}$). The bulk and co-annihilation regions are now clearly selected (for $\mu>0$) amongst the various possibilities to obtain $\Omega_{DM}$. Again, these results have to be taken with caution.

\item We perform a similar analysis for $\mu<0$. The most important change in the results is that when $a_\mu$ (using $e^+e^-$ data) is taken into account, this scenario becomes unfavoured, as it cannot reproduce the experimental measure. Besides, there is no push of the MSSM parameters into the low energy region (actually, the opposite is true, but in a mild way).

We also perform a Bayesian comparison of both scenarios, showing quantitatively how better the $\mu>0$ case performs than the $\mu<0$ one. Actually, the advantage of the positive-$\mu$ case only occurs when $a_\mu$ constraints (using $e^+e^-$ data) are taken into account.

\end{itemize}

In summary, LHC offers an exciting horizon for SUSY discovery, but there is still a non-negligible possibility that it escapes detection, especially if the Higgs mass is not close to its present experimental value. So we should cross fingers.

\section*{Acknowledgements}

We thank P. Slavich for interesting discussions and suggestions. 
M. Eugenia Cabrera thanks Ben Allanach for useful discussions and the hospitality of DAMTP (Cambridge) during a two months visit.
J.A. Casas thanks A. Ibarra for useful discussions and the support and hospitality of TUM (Munich) during a one month visit.
This work has been partially supported by the MICINN, Spain, under contract FPA 2007--60252, the Comunidad de Madrid through Proyecto HEPHACOS S-0505/ESP--0346, and by the European Union through the UniverseNet (MRTN--CT--2006--035863).
M.~E. Cabrera acknowledges the 
financial support of the CSIC through a predoctoral research grant (JAEPre 07 00020).
The work of R. Ruiz de Austri has been supported in part by MEC (Spain) 
under grant FPA2007-60323, by Generalitat Valenciana under grant 
PROMETEO/2008/069 and by the Spanish Consolider-Ingenio 2010 Programme 
CPAN (CSD2007-00042).
The use of the ciclope cluster of the IFT-UAM/CSIC is also acknowledged.


\begin{thebibliography}{99}
%

%\cite{Allanach:2005kz}
\bibitem{Allanach:2005kz}
  B.~C.~Allanach and C.~G.~Lester,
  %``Multi-Dimensional mSUGRA Likelihood Maps,''
  Phys.\ Rev.\  D {\bf 73} (2006) 015013
  [arXiv:hep-ph/0507283].
  %%CITATION = PHRVA,D73,015013;%%
%
%\cite{Allanach:2006jc}
\bibitem{Allanach:2006jc}
  B.~C.~Allanach,
  %``Naturalness priors and fits to the constrained minimal supersymmetric
  %standard model,''
  Phys.\ Lett.\  B {\bf 635} (2006) 123
  [arXiv:hep-ph/0601089].
  %%CITATION = PHLTA,B635,123;%%
%
%\cite{de Austri:2006pe}
\bibitem{deAustri:2006pe}
  R.~R.~de Austri, R.~Trotta and L.~Roszkowski,
  %``A Markov chain Monte Carlo analysis of the CMSSM,''
  JHEP {\bf 0605} (2006) 002
  [arXiv:hep-ph/0602028].
  %%CITATION = JHEPA,0605,002;%%
%
%\cite{Allanach:2007qk}
\bibitem{Allanach:2007qk}
  B.~C.~Allanach, K.~Cranmer, C.~G.~Lester and A.~M.~Weber,
  %``Natural Priors, CMSSM Fits and LHC Weather Forecasts,''
  JHEP {\bf 0708} (2007) 023
  [arXiv:0705.0487 [hep-ph]].
  %%CITATION = JHEPA,0708,023;%%
%
%\cite{Roszkowski:2007fd}
\bibitem{Roszkowski:2007fd}
  L.~Roszkowski, R.~Ruiz de Austri and R.~Trotta,
  %``Implications for the Constrained MSSM from a new prediction for b to s
  %gamma,''
  JHEP {\bf 0707}, 075 (2007)
  [arXiv:0705.2012 [hep-ph]].
  %%CITATION = JHEPA,0707,075;%%
%\cite{Buchmueller:2008qe}
\bibitem{Buchmueller:2008qe}
  O.~Buchmueller {\it et al.},
  %``Predictions for Supersymmetric Particle Masses in the CMSSM using Indirect
  %Experimental and Cosmological Constraints,''
  JHEP {\bf 0809} (2008) 117
  [arXiv:0808.4128 [hep-ph]].
  %%CITATION = JHEPA,0809,117;%%
%
%\cite{Trotta:2008bp}
\bibitem{Trotta:2008bp}
  R.~Trotta, F.~Feroz, M.~P.~Hobson, L.~Roszkowski and R.~Ruiz de Austri,
  %``The impact of priors and observables on parameter inferences in the
  %Constrained MSSM,''
  arXiv:0809.3792 [hep-ph].
  %%CITATION = ARXIV:0809.3792;%%
%
%\cite{Ellis:2008di}
\bibitem{Ellis:2008di}
  J.~Ellis,
  %``Prospects for Discovering Supersymmetry at the LHC,''
  Eur.\ Phys.\ J.\  C {\bf 59} (2009) 335
  [arXiv:0810.1178 [hep-ph]].
  %%CITATION = EPHJA,C59,335;%%
%
%\cite{AbdusSalam:2009qd}
\bibitem{AbdusSalam:2009qd}
  S.~S.~AbdusSalam, B.~C.~Allanach, F.~Quevedo, F.~Feroz and M.~Hobson,
  %``Fitting the Phenomenological MSSM,''
  arXiv:0904.2548 [hep-ph].
  %%CITATION = ARXIV:0904.2548;%%
%
%\cite{Buchmueller:2009fn}
\bibitem{Buchmueller:2009fn}
  O.~Buchmueller {\it et al.},
  %``Likelihood Functions for Supersymmetric Observables in Frequentist Analyses
  %of the CMSSM and NUHM1,''
  Eur.\ Phys.\ J.\  C {\bf 64} (2009) 391
  [arXiv:0907.5568 [hep-ph]].
  %%CITATION = EPHJA,C64,391;%%
%
%\cite{D'Agostini:1995fv}
\bibitem{D'Agostini:1995fv}
  G.~D'Agostini,
  %``Probability and Measurement Uncertainty in Physics - a Bayesian Primer,''
  arXiv:hep-ph/9512295;
  %%CITATION = HEP-PH/9512295;%%
%%\cite{Trotta:2008qt}
%\bibitem{Trotta:2008qt}
  R.~Trotta,
  %``Bayes in the sky: Bayesian inference and model selection in cosmology,''
  Contemp.\ Phys.\  {\bf 49} (2008) 71
  [arXiv:0803.4089 [astro-ph]].
  %%CITATION = CTPHA,49,71;%%
%
%\cite{Martin:1997ns}
\bibitem{Martin:1997ns}
  For a review see S.~P.~Martin,
  %``A supersymmetry primer,''
  arXiv:hep-ph/9709356.
  %%CITATION = HEP-PH/9709356;%%
%
%\cite{Ellis:1986yg}
\bibitem{Ellis:1986yg}
  J.~R.~Ellis, K.~Enqvist, D.~V.~Nanopoulos and F.~Zwirner,
  %``Observables In Low-Energy Superstring Models,''
  Mod.\ Phys.\ Lett.\  A {\bf 1} (1986) 57.
  %%CITATION = MPLAE,A1,57;%%
%
%\cite{Barbieri:1987fn}
\bibitem{Barbieri:1987fn}
  R.~Barbieri and G.~F.~Giudice,
  %``Upper Bounds On Supersymmetric Particle Masses,''
  Nucl.\ Phys.\  B {\bf 306} (1988) 63.
  %%CITATION = NUPHA,B306,63;%%
%
%\cite{de Carlos:1993yy}
\bibitem{deCarlos:1993yy}
  B.~de Carlos and J.~A.~Casas,
  %``One loop analysis of the electroweak breaking in supersymmetric models and
  %the fine tuning problem,''
  Phys.\ Lett.\  B {\bf 309} (1993) 320
  [arXiv:hep-ph/9303291];
  %%CITATION = PHLTA,B309,320;%%
%\cite{Anderson:1994dz}
%\bibitem{Anderson:1994dz}
  G.~W.~Anderson and D.~J.~Castano,
  %``Measures of fine tuning,''
  Phys.\ Lett.\  B {\bf 347} (1995) 300
  [arXiv:hep-ph/9409419];
  %%CITATION = PHLTA,B347,300;%%
%\cite{Athron:2007ry}
%\bibitem{Athron:2007ry}
  P.~Athron and D.~J.~Miller,
  %``A New Measure of Fine Tuning,''
  Phys.\ Rev.\  D {\bf 76} (2007) 075010
  [arXiv:0705.2241 [hep-ph]];
  %%CITATION = PHRVA,D76,075010;%%
%\cite{Ciafaloni:1996zh}
%\bibitem{Ciafaloni:1996zh}
  P.~Ciafaloni and A.~Strumia,
  %``Naturalness upper bounds on gauge mediated soft terms,''
  Nucl.\ Phys.\  B {\bf 494} (1997) 41
  [arXiv:hep-ph/9611204];
  %%CITATION = NUPHA,B494,41;%%
%\cite{Casas:2004gh}
%\bibitem{Casas:2004gh}
  J.~A.~Casas, J.~R.~Espinosa and I.~Hidalgo,
  %``Implications for new physics from fine-tuning arguments. I: Application  to
  %SUSY and seesaw cases,''
  JHEP {\bf 0411} (2004) 057
  [arXiv:hep-ph/0410298];
  %%CITATION = JHEPA,0411,057;%%
%
%\cite{Casas:2005ev}
%\bibitem{Casas:2005ev}
%  J.~A.~Casas, J.~R.~Espinosa and I.~Hidalgo,
  %``Implications for new physics from fine-tuning arguments. II: Little  Higgs
  %models,''
[ibid],   
JHEP {\bf 0503} (2005) 038
  [arXiv:hep-ph/0502066].
  %%CITATION = JHEPA,0503,038;%%
%
%\cite{Giusti:1998gz}
\bibitem{Giusti:1998gz}
  L.~Giusti, A.~Romanino and A.~Strumia,
  %``Natural ranges of supersymmetric signals,''
  Nucl.\ Phys.\  B {\bf 550} (1999) 3
  [arXiv:hep-ph/9811386].
  %%CITATION = NUPHA,B550,3;%%
%
%\cite{Strumia:1999fr}
\bibitem{Strumia:1999fr}
  A.~Strumia,
  %``Naturalness of supersymmetric models,''
  arXiv:hep-ph/9904247;
  %%CITATION = HEP-PH/9904247;%%
%\cite{Feruglio:2002af}
%\bibitem{Feruglio:2002af}
  F.~Feruglio, A.~Strumia and F.~Vissani,
  %``Neutrino oscillations and signals in beta and 0nu 2beta experiments,''
  Nucl.\ Phys.\  B {\bf 637} (2002) 345
  [Addendum-ibid.\  B {\bf 659} (2003) 359]
  [arXiv:hep-ph/0201291].
  %%CITATION = NUPHA,B637,345;%%
%
\bibitem{Cabrera:2008tj}
M.~E.~Cabrera, J.~A.~Casas and R.~Ruiz de Austri,
%``Bayesian approach and Naturalness in MSSM analyses for the LHC,''
JHEP {\bf 0903}, 075 (2009)
[arXiv:0812.0536 [hep-ph]].
%%CITATION = JHEPA,0903,075;%%
%
\bibitem{Berger:1999}
J.~O.~Berger, B.~Liseo and R.~L.~Wolpert,
%``Integrated likelihood methods for eliminating nuisance variables"
Statistical Science 1999, Vol. 14, No. 1, 1-28
%
\bibitem{softsusy}
  B.~C.~Allanach,
%  {\it SOFTSUSY: a C++ program for calculating supersymmetric spectra}, 
  Comput. \ Phys. \ Commun. {\bf 143} (2002) 305  
  [arXiv:hep-ph/0104145].
%
%\cite{Ibanez:1983di}
\bibitem{Ibanez:1983di}
  L.~E.~Ibanez and C.~Lopez,
  %``N=1 Supergravity, The Weak Scale And The Low-Energy Particle Spectrum,''
  Nucl.\ Phys.\  B {\bf 233} (1984) 511.
  %%CITATION = NUPHA,B233,511;%%
%
\bibitem{Hall:1993gn}
  L.~J.~Hall, R.~Rattazzi and U.~Sarid,
  %``The Top quark mass in supersymmetric SO(10) unification,''
  Phys.\ Rev.\  D {\bf 50} (1994) 7048
  [arXiv:hep-ph/9306309].
  %%CITATION = PHRVA,D50,7048;%%
%
\bibitem{Nelson:1993vc}
  A.~E.~Nelson and L.~Randall,
  %``Naturally large tan BETA,''
  Phys.\ Lett.\  B {\bf 316} (1993) 516
  [arXiv:hep-ph/9308277].
  %%CITATION = PHLTA,B316,516;%%

%\cite{Kaplunovsky:1993rd}
\bibitem{Kaplunovsky:1993rd}
  V.~S.~Kaplunovsky and J.~Louis,
  %``Model independent analysis of soft terms in effective supergravity and in
  %string theory,''
  Phys.\ Lett.\  B {\bf 306} (1993) 269
  [arXiv:hep-th/9303040].
  %%CITATION = PHLTA,B306,269;%%
%
%\cite{Giudice:1988yz}
\bibitem{Giudice:1988yz}
  G.~F.~Giudice and A.~Masiero,
  %``A Natural Solution to the mu Problem in Supergravity Theories,''
  Phys.\ Lett.\  B {\bf 206} (1988) 480.
  %%CITATION = PHLTA,B206,480;%%
%
\bibitem{Feroz:2007kg}
F. Feroz and  M.~P. Hobson  
%{\it  Multimodal nested sampling: an efficient and robust alternative to 
% MCMC methods for astronomical data analysis},
Mon. Not. Roy. Astron. Soc. \textbf{384} 449 (2008).
%
\bibitem{superbayes} Available from: \texttt{http://superbayes.org}
%
\bibitem{SkillingNS}
J. Skilling, Nested sampling, in R.~Fischer, R.~Preuss and U.~\protect{von
Toussaint} (Eds)  
%{\itshape Bayesian Inference and Maximum Entropy Methods in
%Science and Engineering}
, 735  (Amer.\ Inst.\ Phys.\, conf.\ proc.\, 2004),
pp. 395--405.
%
\bibitem{Skilling:2006}
J. Skilling, Bayesian Analysis \textbf{1} 833--861 (2006).
%
\bibitem{Schael:2006cr}
  S.~Schael {\it et al.}  [ALEPH Collaboration and DELPHI Collaboration and
                  L3 Collaboration and ],
  %``Search for neutral MSSM Higgs bosons at LEP,''
  Eur.\ Phys.\ J.\  C {\bf 47} (2006) 547
  [arXiv:hep-ex/0602042].
  %%CITATION = EPHJA,C47,547;%%
%
\bibitem{Casas:1998vh}
  J.~A.~Casas, J.~R.~Espinosa and H.~E.~Haber,
  %``The Higgs mass in the MSSM infrared fixed point scenario,''
  Nucl.\ Phys.\  B {\bf 526} (1998) 3
  [arXiv:hep-ph/9801365];
  %%CITATION = NUPHA,B526,3;%%
  H.~Baer, C.~h.~Chen, M.~Drees, F.~Paige and X.~Tata,
  %``Supersymmetry reach of Tevatron upgrades: The large tan(beta) case,''
  Phys.\ Rev.\  D {\bf 58} (1998) 075008
  [arXiv:hep-ph/9802441];
  %%CITATION = PHRVA,D58,075008;%%
  M.~Jurcisin and D.~I.~Kazakov,
  %``Infrared quasi fixed points and mass predictions in the MSSM. II: Large
  %tan(beta) scenario,''
  Mod.\ Phys.\ Lett.\  A {\bf 14} (1999) 671
  [arXiv:hep-ph/9902290].
  %%CITATION = MPLAE,A14,671;%%
%
\bibitem{Ellis:2001msa}
  J.~R.~Ellis, T.~Falk, G.~Ganis, K.~A.~Olive and M.~Srednicki,
  %``The CMSSM Parameter Space at Large tan beta,''
  Phys.\ Lett.\  B {\bf 510} (2001) 236
  [arXiv:hep-ph/0102098].
  %%CITATION = PHLTA,B510,236;%%
%
%\cite{Ball:2007zza}
\bibitem{Ball:2007zza}
  G.~L.~Bayatian {\it et al.}  [CMS Collaboration],
  %``CMS technical design report, volume II: Physics performance,''
  J.\ Phys.\ G {\bf 34} (2007) 995;
  %%CITATION = JPHGB,G34,995;%%
%\cite{Aad:2009wy}
%\bibitem{Aad:2009wy}
  G.~Aad {\it et al.}  [The ATLAS Collaboration],
  %``Expected Performance of the ATLAS Experiment - Detector, Trigger and
  %Physics,''
  arXiv:0901.0512 [hep-ex].
  %%CITATION = ARXIV:0901.0512;%%
%
\bibitem{Baer:2009dn}
H.~Baer, V.~Barger, A.~Lessa and X.~Tata,
%``Supersymmetry discovery potential of the LHC at $\sqrt{s}=$10 and 14 TeV
%without and with missing $E_T$,''
JHEP {\bf 0909} (2009) 063
[arXiv:0907.1922 [hep-ph]].
%%CITATION = JHEPA,0909,063;%%
%
%\cite{Susskind:2004uv}
\bibitem{Susskind:2004uv}
  L.~Susskind,
  %``Supersymmetry breaking in the anthropic landscape,''
  arXiv:hep-th/0405189.
  %%CITATION = HEP-TH/0405189;%%
%
\bibitem{topmass:mar08}
By CDF Collaboration and D0 Collaboration, 
%{\it A Combination of CDF and D0 Results on the Mass of the Top Quark}
arXiv:0803.1683 [hep-ex].
%
\bibitem{pdg07}
W.-M.~Yao \etal, 
%{\it The Review of Particle Physics}, 
{\em J. Phys.} {\bf G33} (2006) 1 and 2007 partial update for the 2008 edition.
%
\bibitem{Hagiwara:2006jt}
K.~Hagiwara, A.~D.~Martin, D.~Nomura and T.~Teubner, 
%{\it Improved predictions for g-2 of the muon and $\alpha_{\rm QED}(M_Z^2)$},
\plb{649}{2007}{173}.% [hep-ph/0611102].
%
%
\bibitem{Dedes:2003km}
  A.~Dedes, G.~Degrassi and P.~Slavich,
  %``On the two-loop Yukawa corrections to the MSSM Higgs boson masses at  large
  %tan(beta),''
  Nucl.\ Phys.\  B {\bf 672} (2003) 144
  [arXiv:hep-ph/0305127].
  %%CITATION = NUPHA,B672,144;%%
%\cite{Barbieri:2000gf}
%
\bibitem{awramik-acfw04}
M.~Awramik, M.~Czakon, A.~Freitas and G.~Weiglein,
%{\it Precise prediction for the W boson mass in the standard model},
\prd{69}{2004}{053006} [hep-ph/0311148]; 
%{\it Complete two-loop electroweak fermionic corrections to
%$\sineff$ and indirect determination of the
%Higgs boson mass}, 
\prl{93}{2004}{201805}  [hep-ph/0407317].
%
\bibitem{dghhjw97}
A.~Djouadi, P.~Gambino, S.~Heinemeyer, W.~Hollik,
C.~Junger and G.~Weiglein, 
%{\it Leading QCD corrections to scalar
%quark contributions to electroweak precision observables},
\prd{57}{1998}{4179} [hep-ph/9710438].

\bibitem{deBoer:2003xm}
  W.~de Boer and C.~Sander,
  %``Global electroweak fits and gauge coupling unification,''
  Phys.\ Lett.\  B {\bf 585} (2004) 276
  [arXiv:hep-ph/0307049].
  %%CITATION = PHLTA,B585,276;%%

\bibitem{Heinemeyer:2004gx}
  S.~Heinemeyer, W.~Hollik and G.~Weiglein,
  %``Electroweak precision observables in the minimal supersymmetric  standard
  %model,''
  Phys.\ Rept.\  {\bf 425} (2006) 265
  [arXiv:hep-ph/0412214].
  %%CITATION = PRPLC,425,265;%%
%\cite{Barbieri:2000gf}
\bibitem{Barbieri:2000gf}
  R.~Barbieri and A.~Strumia,
  %``The 'LEP paradox',''
  arXiv:hep-ph/0007265.
  %%CITATION = HEP-PH/0007265;%%
%\cite{Barbieri:2004qk}
%\bibitem{Barbieri:2004qk}
  R.~Barbieri, A.~Pomarol, R.~Rattazzi and A.~Strumia,
  %``Electroweak symmetry breaking after LEP-1 and LEP-2,''
  Nucl.\ Phys.\  B {\bf 703} (2004) 127
  [arXiv:hep-ph/0405040].
  %%CITATION = NUPHA,B703,127;%%
%

\bibitem{Degrassi:2007kj}
G.~Degrassi, P.~Gambino and P.~Slavich,
%``SusyBSG: a fortran code for BR[B -> Xs gamma] in the MSSM with Minimal
%Flavor Violation,''
Comput.\ Phys.\ Commun.\  {\bf 179} (2008) 759
[arXiv:0712.3265 [hep-ph]].
%%CITATION = CPHCB,179,759;%%
%
\bibitem{Degrassi:2006eh}
G.~Degrassi, P.~Gambino and P.~Slavich,
%``QCD corrections to radiative B decays in the MSSM with minimal flavor
%violation,''
Phys.\ Lett.\  B {\bf 635} (2006) 335
[arXiv:hep-ph/0601135].
%%CITATION = PHLTA,B635,335;%%

%\cite{D'Ambrosio:2002ex}
\bibitem{D'Ambrosio:2002ex}
G.~D'Ambrosio, G.~F.~Giudice, G.~Isidori and A.~Strumia,
%``Minimal flavour violation: An effective field theory approach,''
Nucl.\ Phys.\  B {\bf 645}, 155 (2002)
[arXiv:hep-ph/0207036].
%%CITATION = NUPHA,B645,155;%%

\bibitem{for1}
J.~Foster, K.~Okumura and L.~Roszkowski, 
%{\it New Higgs effects in
%B--physics in supersymmetry with general flavour mixing},
\plb{609}{2005}{102} [hep-ph/0410323].

\bibitem{Mahmoudi:2008tp}
F.~Mahmoudi,
%``SuperIso v2.3: A Program for calculating flavor physics observables in
%Supersymmetry,''
Comput.\ Phys.\ Commun.\  {\bf 180}, 1579 (2009)
[arXiv:0808.3144 [hep-ph]].
%%CITATION = CPHCB,180,1579;%%

\bibitem{lepwwg}
See \texttt{http://lepewwg.web.cern.ch/LEPEWWG}.

\bibitem{Miller:2007kk}
J.~P.~Miller, E.~de Rafael and B.~L.~Roberts,
%``Muon g-2: Review of Theory and Experiment,''
Rept.\ Prog.\ Phys.\  {\bf 70}, 795 (2007)
[arXiv:hep-ph/0703049].
%%CITATION = RPPHA,70,795;%%


\bibitem{Davier:2009ag}
M.~Davier {\it et al.},
%``The Discrepancy Between tau and e+e- Spectral Functions Revisited and the
%Consequences for the Muon Magnetic Anomaly,''
arXiv:0906.5443 [hep-ph].
%%CITATION = ARXIV:0906.5443;%%
%
\bibitem{cdf-deltambs}
The CDF Collaboration,
%{\it Measurement of the  $B_s-\bar{B}_s$ oscillation frequency}, 
\prl{97}{2006}{062003}.% [hep-ex/0606027]
and 
%{\it Observation of $B_s-\bar{B}_s$ oscillations}, 
\prl{97}{2006}{242003}.% [hep-ex/0609040].
%

%
\bibitem{hfag} Heavy Flavor Averaging
Group (HFAG) (E. Barberio \etal), 
%{\it Averages of b-hadron properties at the end of 2007}, 
arXiv:0808.1297 [hep-ex].
%

%
\bibitem{:2008cy}
B.~Aubert {\it et al.}  [BABAR Collaboration],
%``Measurement of Branching Fractions and CP and Isospin Asymmetries in $B \to
%K^{*} \gamma$,''
arXiv:0808.1915 [hep-ex].
%%CITATION = ARXIV:0808.1915;%%
%%
\bibitem{Aubert:2007dsa}
B.~Aubert {\it et al.}  [BABAR Collaboration],
%``Observation of the semileptonic decays $B \to D^{*} \tau^{-} \bar{\nu}$(
%$\tau^{)}$ and evidence for $B \to D \tau^{-} \bar{\nu}$( $\tau^{)}$,''
Phys.\ Rev.\ Lett.\  {\bf 100} (2008) 021801
[arXiv:0709.1698 [hep-ex]].
%%CITATION = PRLTA,100,021801;%%
%
\bibitem{Antonelli:2008jg}
M.~Antonelli {\it et al.}  [FlaviaNet Working Group on Kaon Decays],
%``Precision tests of the Standard Model with leptonic and semileptonic kaon
%decays,''
arXiv:0801.1817 [hep-ph].
%%CITATION = ARXIV:0801.1817;%%
%
%
\bibitem{Akeroyd:2009tn}
A.~G.~Akeroyd and F.~Mahmoudi,
%``Constraints on charged Higgs bosons from D(s)+- -> mu+- nu and D(s)+- ->
%tau+- nu,''
JHEP {\bf 0904} (2009) 121
[arXiv:0902.2393 [hep-ph]].
%%CITATION = JHEPA,0904,121;%%
%
%
%
\bibitem{wmap5yr}
J.~Dunkley  \etal\ [The WMAP Collaboration],
%{\it Five-year Wilkinson Microwave Anisotropy Probe (WMAP) Observations: Likelihoods and parameters from the WMAP data}, 
arXiv:0803.0586 [astro-ph].
%
%
%
\bibitem{cdf-bsmumu}
The CDF Collaboration,
%{\it Search for $B_s\to\mu^+\mu^-$ and
% $B_d\to\mu^+\mu^-$ decays with 2 fb$^{-1}$ of $p\bar{p}$ collisions}, 
\prl{97}{2006}{242003}.
%
%
\bibitem{lhwg}
The LEP Higgs Working Group,
\texttt{http://lephiggs.web.cern.ch/LEPHIGGS};\\ G.~Abbiendi \etal\ [the ALEPH Collaboration, the DELPHI
 Collaboration, the L3 Collaboration and the OPAL Collaboration, The
 LEP Working Group for Higgs Boson Searches], 
%{\it Search for the standard model Higgs boson at LEP},
\plb{565}{2003}{61}.% [hep-ex/0306033].
%
\bibitem{rrt2}
L.~Roszkowski, R.~Ruiz de Austri and R.~Trotta, 
%{\it  On the
%detectability of the CMSSM light Higgs boson at the Tevatron},
\jhep{0704}{2007}{084} [hep-ph/0611173].
%\cite{Ellis:1991zd}
\bibitem{Ellis:1991zd}
  J.~R.~Ellis, G.~Ridolfi and F.~Zwirner,
  %``On radiative corrections to supersymmetric Higgs boson masses and their
  %implications for LEP searches,''
  Phys.\ Lett.\  B {\bf 262} (1991) 477;
  %%CITATION = PHLTA,B262,477;%%
%\cite{Okada:1990vk}
%\bibitem{Okada:1990vk}
  Y.~Okada, M.~Yamaguchi and T.~Yanagida,
  %``Upper bound of the lightest Higgs boson mass in the minimal supersymmetric
  %standard model,''
  Prog.\ Theor.\ Phys.\  {\bf 85} (1991) 1;
  %%CITATION = PTPKA,85,1;%%
%\cite{Haber:1990aw}
%\bibitem{Haber:1990aw}
  H.~E.~Haber and R.~Hempfling,
  %``Can the mass of the lightest Higgs boson of the minimal supersymmetric
  %model be larger than m(Z)?,''
  Phys.\ Rev.\ Lett.\  {\bf 66} (1991) 1815;
  %%CITATION = PRLTA,66,1815;%%
%\cite{Barbieri:1990ja}
%\bibitem{Barbieri:1990ja}
  R.~Barbieri, M.~Frigeni and F.~Caravaglios,
  %``The Supersymmetric Higgs for heavy superpartners,''
  Phys.\ Lett.\  B {\bf 258} (1991) 167.
  %%CITATION = PHLTA,B258,167;%%
%
%\cite{Davier:2009zi}
\bibitem{Davier:2009zi}
  M.~Davier, A.~Hoecker, B.~Malaescu, C.~Z.~Yuan and Z.~Zhang,
  %``Reevaluation of the hadronic contribution to the muon magnetic anomaly
  %using new e+e- -> pi+pi- cross section data from BABAR,''
  arXiv:0908.4300 [hep-ph].
  %%CITATION = ARXIV:0908.4300;%%
%
%
\bibitem{Heinemeyer:2004yq}
S.~Heinemeyer, D.~Stockinger and G.~Weiglein,
%``Electroweak and supersymmetric two-loop corrections to (g-2)(mu),''
Nucl.\ Phys.\  B {\bf 699}, 103 (2004)
[arXiv:hep-ph/0405255].
%%CITATION = NUPHA,B699,103;%%
%
%

\bibitem{Marchetti:2008hw}
S.~Marchetti, S.~Mertens, U.~Nierste and D.~Stockinger,
%``Tan(beta)-enhanced supersymmetric corrections to the anomalous magnetic
%moment of the muon,''
Phys.\ Rev.\  D {\bf 79} (2009) 013010
[arXiv:0808.1530 [hep-ph]].
%%CITATION = PHRVA,D79,013010;%%
%
%
\bibitem{Feroz:2009dv}
F.~Feroz, M.~P.~Hobson, L.~Roszkowski, R.~Ruiz de Austri and R.~Trotta,
%``Are BR(b->s gamma) and (g-2)_muon consistent within the Constrained
%MSSM?,''
arXiv:0903.2487 [hep-ph].
%%CITATION = ARXIV:0903.2487;%%
%
%
%\cite{Jungman:1995df}
\bibitem{Jungman:1995df}
  G.~Jungman, M.~Kamionkowski and K.~Griest,
  %``Supersymmetric dark matter,''
  Phys.\ Rept.\  {\bf 267} (1996) 195
  [arXiv:hep-ph/9506380].
  %%CITATION = PRPLC,267,195;%%
%
%
\bibitem{micromegas}
G.~Belanger, F.~Boudjema, A.~Pukhov and A.~Semenov,
%{\it MicrOMEGAs: A program for calculating the relic density in the MSSM},
Comput.\ Phys.\ Commun.\  {\bf 149} (2002) 103 [hep-ph/0112278]; 
%{\it MicrOMEGAs: Version 1.3}, 
Comput.\ Phys.\ Commun.\  {\bf 174}, 577 (2006) [hep-ph/0405253].
%
%
%\cite{Knox:1992iy}
\bibitem{Knox:1992iy}
  L.~Knox and M.~S.~Turner,
  %``Inflation at the electroweak scale,''
  Phys.\ Rev.\ Lett.\  {\bf 70} (1993) 371
  [arXiv:astro-ph/9209006];
  %%CITATION = PRLTA,70,371;%%
%\cite{GarciaBellido:1999sv}
%\bibitem{GarciaBellido:1999sv}
  J.~Garcia-Bellido, D.~Y.~Grigoriev, A.~Kusenko and M.~E.~Shaposhnikov,
  %``Non-equilibrium electroweak baryogenesis from preheating after
  %inflation,''
  Phys.\ Rev.\  D {\bf 60} (1999) 123504
  [arXiv:hep-ph/9902449].
  %%CITATION = PHRVA,D60,123504;%%
%
%\cite{Ibarra:2008jk}
\bibitem{Ibarra:2008jk}
  A.~Ibarra and D.~Tran,
  %``Decaying Dark Matter and the PAMELA Anomaly,''
  JCAP {\bf 0902} (2009) 021
  [arXiv:0811.1555 [hep-ph]].
  %%CITATION = JCAPA,0902,021;%%


%%%%%%%%%% roberto's refs %%%%%%%%%%%%%%%%%%%%%


\bibitem{Jeffreys}
H. Jeffreys, Theory of probability, 3rd edn, Oxford Classics 
series (reprinted 1998) (Oxford University Press, Oxford, UK, 1961).


\bibitem{O'Ruanaidh}
J. ´O Ruanaidh and W. Fitzgerald, 
Numerical Bayesian Methods Applied to Signal Processing. 
Springer Verlag: New York, 1996.


\end{thebibliography}
\end{document}